\documentclass[a4paper,aps,prd,twocolumn,nofootinbib] {revtex4}
\usepackage[T1]{fontenc}
\usepackage[brazil,english]{babel}
\usepackage[usenames,dvipsnames]{color}
  \definecolor{darkblue}{RGB}{0,0,150}
\usepackage{amsmath,amssymb}
\usepackage{relsize}
\usepackage{graphicx}
\usepackage{nicefrac}

\usepackage[unicode,bookmarks=false,colorlinks=true,linkcolor=blue,
urlcolor =blue, citecolor=blue, anchorcolor=blue]{hyperref}

\usepackage[caption=false]{subfig}

\usepackage{bigints}
\usepackage{geometry}
\usepackage{stackrel}
\usepackage{appendix}

\begin{document}

\title{Stability of electrically charged stars, regular black holes,
quasiblack holes, and quasinonblack holes}

\author{Angel D. D. Masa$^{1,}$\footnote{angel.masa@ufabc.edu.br},
Jos\'{e} P. S.  Lemos$^{2,}$\footnote{joselemos@ist.utl.pt}, and
Vilson T.  Zanchin$^{1,}$\footnote{zanchin@ufabc.edu.br}}

\affiliation{\vspace{0.2cm}
$^{1}$Centro de Ci\^{e}ncias Naturais e Humanas,
Universidade Federal do ABC,
Avenida dos Estados
5001, 09210-580 - Santo Andr\'{e}, S\~{a}o Paulo, 
Brazil
\vskip .1cm
$^{2}$Centro de Astrof\'{i}sica e Gravita\c c\~ao - CENTRA,
Departamento de F\'{i}sica,
Instituto Superior T\'{e}cnico - IST,
Universidade de Lisboa - UL,
Avenida Rovisco Pais 1, 1049-001 Lisboa, Portugal}

\begin{abstract}
The stability of a class of electrically charged fluid spheres under
radial perturbations is studied. Among these spheres there are regular
stars, overcharged tension stars, regular black holes, quasiblack
holes, and quasinonblack holes, all of which have a
Reissner-Nordstr\"om exterior.  We formulate the dynamical
perturbed equations
by following the Chandrasekhar 
approach and investigate the stability
against radial perturbations through numerical methods.  It is found
that (i) under certain conditions that depend on the adiabatic index
of the radial perturbation, there are stable charged stars and stable
tension stars; (ii) also 
depending on the adiabatic index
there are stable regular black holes; (iii) quasiblack hole
configurations formed by, e.g., charging regular pressure stars or by
discharging regular tension stars, can be
stable against radial perturbations for reasonable values of the
adiabatic index; (iv) quasinonblack holes are unstable against radial
perturbations.
%For pressured configurations the adiabatic index of the perturbation
%is positive, for tensioned configurations the adiabatic index of the
%perturbation is negative.
\end{abstract}

\maketitle

\section{Introduction }

Solutions representing stars in general relativity are extremely
important as they can test general relativity itself in extreme
conditions.  Besides the gravitational field and matter one can put
some charge and electromagnetic fields into the solutions, which
allows the stars to be more compact.  An electrically charged
spherically symmetric solution was given by Guilfoyle
\cite{Guilfoyle1999} by first, giving a generalized ansatz of Weyl in
which a relation between the metric functions and the electric
potential is assumed, see \cite{LemosZanchin2009} for generalized
Weyl's ans\"atze, second, providing an electric version of the
constant density condition of the Schwarzschild interior solution,
and third, using the junction conditions,  performing 
a smooth matching to an electrovacuum Reissner-Nordstr\"om
spacetime.
Other
electric stars, like Bonnor stars where charged density equals energy
density have been found \cite{lemoszanchin2008}.  One of the main
aspects to seek in these solutions is to test for their compactness,
since then full general relativistic effects arise. 
There are bounds on the compactness of stars, in the case of
electrically charged stars these bounds were found in
\cite{AndreassonQ} yielding a generalization of the Buchdahl bound for
neutral general relativistic stars.  Interestingly, it has been shown
that the most compact
stars provided in Guilfoyle's solution saturate this bound
\cite{LemosZanchin2015}.  Now, the most extreme compactness
configuration is a quasiblack hole, a star on the verge of becoming a
black hole but never being one.  Quasiblack holes have been found in
\cite{LemosWeinberg2004} for stars with matter in which the charged
density equals the energy density,
which have been in turn compared with their
gravitational magnetic monopole analogues \cite{lzjmp},
have had their generic properties
studied \cite{lemoszaslavskii1,lemoszaslavskii2}, and have also been
discovered to exist in the most compact stars
of Guilfoyle's solution
\cite{LemosZanchin2010}, a review on these objects is in
\cite{lemoszaslavskii2020}.

Solutions representing black holes in general relativity are also
extremely important as they can also test general relativity itself in
extreme conditions. In general relativity, static vacuum solutions are
the Schwarzschild black hole which has an event horizon and a
singularity, and the electrically charged Reissner-Nordstr\"om black
hole which has Cauchy and event horizons and a singularity.  Regular
black holes, i.e., black holes without singularities, can be built in
general relativity in several ways and from several types of matter,
for instance regular black holes with phantom matter were found in
\cite{bron}, and
the matter energy conditions for regular black holes were
studied in \cite{zaslav}. Moreover,
a particular class of regular black
holes with a de Sitter core and a massless electric coat at the matter
boundary was found in \cite{LemosZanchin2011,Uchikata:2012zs},
a quasinormal mode analysis of regular black holes
was performed in \cite{flachilemos}, and a stability
analysis was done in \cite{masaoz}.
Now, the most extreme noncompact regular black hole is a
quasinonblack hole, a regular black hole on the verge of becoming a
star but never being one.  Quasinonblack holes have been found in
\cite{lemosluz2021}.

All these configurations, namely, stars, regular black holes,
quasiblack holes, and quasinonblack holes were discovered to exist
\cite{LemosZanchin2017} within Guilfoyle's solution.
By studying the full parameter space of this solution, which can be
put in the form $\frac{q^2}{R^2}\times \frac{r_0}{R}$, where $q$ is
the total electric charge, $r_0$ is
the radius of the object, and $R$ is a constant
with the dimension of length related to the effective energy density,
it was shown in \cite{LemosZanchin2017} that there are many different
types of compact objects such as Schwarzschild and
Reissner-Nordstr\"om black holes, Schwarzschild stars corresponding to
the Schwarzschild interior solution, electrically charged stars,
Bonnor stars, tension charged stars, regular charged black holes with
a phantom and a de Sitter core, quasiblack holes, quasinonblack holes,
among other singular compact objects.  Interesting to note that all
these configurations also exist in another exact solution of
electrically charged static thin shells \cite{lemosluz2021}.

The stability of a solution is always an important issue, and here it
is no exception, it is important to perform a stability analysis on
the whole set of solutions reveled in Guilfoyle's solution
\cite{LemosZanchin2017}.  To make the analysis one can use the method
developed by Chandrasekhar \cite{Chandre1964b} that can be extended to
electrically charged objects as has been done in some works.  Stettner
\cite{Stettner1973} considered the effect of a charged surface
distribution on the stability of a spherically symmetric fluid with
constant energy density and found that such a model is more stable
than the corresponding uncharged configuration.  Omote and Sato
\cite{Omote1974} developed the perturbation equations to arbitrary
charged fluid distributions and showed explicitly that Bonnor stars
are neutrally stable. Glazer \cite{Glazer1976,Glazer1979} also worked
with arbitrary charged fluid distributions, confirmed the stable
neutrality of Bonnor stars, and showed that stability of a homogeneous
configuration increases by adding electric charge.  De Felice and
collaborators \cite{Felice1999} stipulated a power law for
the electric charged function and Anninos and Rothman
\cite{Anninos2001} further considered a hyperbolic tangent function to
give a stability analysis of concrete examples.  Posada and Chirenti
\cite{PosadaChirenti2019} studied the radial stability of ultra
compact Schwarzschild stars beyond the Buchdahl limit.

The aim of this work is to do a stability analysis of the
Schwarzschild stars, electrically charged stars, Bonnor stars, tension
charged stars, regular charged black holes with a phantom core,
regular charged black holes with a de Sitter core, quasiblack holes,
and quasinonblack holes, contained in Guilfoyle's solution.  The
stability analysis is done against small radial adiabatic
perturbations, and since radial oscillations of the solutions do not
generate gravitational waves, the analysis is reduced to an eigenvalue
problem, where the oscillation frequencies are essentially the
eigenvalues of the perturbation equation. The methods employed here
stem and are adapted from all the works on perturbation analysis of
electrically charged stars that we mentioned.  A remark should perhaps
be made at this point. The stability analysis performed is only a
stability of the matter interior solution against radial perturbations
taking into account the boundary conditions at the junction to the
exterior. This means, that if for a certain interior solution
stability against radial perturbations follows, it is possible that
other types of perturbations, like nonspherical perturbations, scalar,
vector, and tensorial linear perturbations, and also generic nonlinear
perturbations, might give rise to instabilities. On the other hand, if
for a given interior solution instability against radial perturbations
follows, then the solution is certainly unstable.  In addition, some
of the solutions displayed by us have a Reissner-Nordstr\"om exterior
which is outside its own gravitational radius, other solutions also
displayed have a Reissner-Nordstr\"om exterior which is outside its
own Cauchy horizon radius. A stability analysis for the electrovacuum
exterior region is not performed, but it is known that a
Reissner-Nordstr\"om exterior region outside its own gravitational
radius is stable against any type of perturbation, which include
radial perturbations, while a Reissner-Nordstr\"om exterior region
containing a Cauchy horizon might be unstable to all sorts of
perturbations. So, for full stability one has take into account all
possible sources of perturbations that might arise in the full
solution, namely, in the interior and in the exterior regions.  In
brief, the upshot is that stability of the solution against radial
perturbations is a necessary but not a sufficient condition for the
solution to be stable. Our stability analysis is concerned with
radial perturbations of the interior solution alone. When we refer to
a solution being stable or unstable, although it might not be
explicitly stated, it is to mean specifically that the solution is
stable or unstable against this type of radial perturbations studied.
In summary, we perform a radial stability perturbation analysis
to a great variety of different objects that span a range
going from different sorts of star solutions to
different sorts of black hole solutions.

The present work is organized as follows. In Sec.~\ref{sec:basic}, the
basic equations describing a spherically symmetric electrically
charged fluid are presented, a perturbation analysis due to radial
oscillations of the configurations is thoroughly given with the
displaying of the master perturbation equation, and in addition the
numerical methods used to analyze this master perturbation equation
are stated.  In Sec.~\ref{sec:solutions} we describe all the
electrically charged solutions, namely, Schwarzschild and
Reissner-Nordstr\"om black holes, Schwarzschild stars, electrically
charged stars, Bonnor stars, tension charged stars, regular charged
black holes with a phantom core, regular charged black holes with a de
Sitter core, quasiblack holes, quasinonblack holes, among other
singular compact objects, which are contained in Guilfoyle's solution.
In Sec.~\ref{sec:results} we study carefully and thoroughly the
stability of all the interesting solutions against adiabatic radial
perturbations, in particular the stability of quasiblack hole and
quasinonblack hole configurations.  In Sec.~\ref{sec:final} we
conclude. In the Appendices \ref{sec:c}-\ref{appendixB} we perform some
calculations and give some results that are used in the main text.
%In the Appendices \ref{sec:c}, \ref{manipulation}, \ref{sec:SLP},
%\ref{sec:numeric}, and \ref{appendixB} we perform some calculations
%and give some results that are used in the main text.

\vskip 5cm

\section{Charged fluid spacetimes,
perturbation equations in static
spherical geometries, and numerical schemes}
\label{sec:basic}

\subsection{Basic equations}
\label{basicbasic}

The spacetimes and the  matter
we consider are described by the
Einstein-Maxwell equations with electrically charged matter,
namely,
\begin{equation}\label{eq:Einstein}
G_{\mu\nu}=8\pi T_{\mu\nu},
\end{equation}
\begin{equation}\label{eq:Maxwell}
\nabla_{\nu}F^{\mu\nu}=4\pi J^{\mu},
\end{equation}
where Greek indices range from $0$ to $3$, $0$ corresponding
to a timelike coordinate $t$, and $1,2,3$ to spatial coordinates,
$G_{\mu\nu}$ is the Einstein tensor, $T_{\mu\nu}$ is the
energy-momentum tensor, $\nabla_{\mu}$ represents the covariant
derivative, $F_{\mu\nu}$ is the Faraday-Maxwell
electromagnetic tensor, and $J^{\mu}$
is the charge current density.
The Einstein tensor
$G_{\mu\nu}$ is a function of the metric
$g_{\mu\nu}$ and its first two derivatives, and since
it is a long expression we
do not write it explicitly. 
The energy-momentum tensor
$T_{\mu\nu}$ has two
contributions, one contribution from
the matter distribution denoted by $M_{\mu\nu}$
and the other contribution from the
electromagnetic field denoted by $E_{\mu\nu}$,
so that
\begin{equation}\label{eq:tensortot}
T_{\mu\nu}=M_{\mu\nu}+E_{\mu\nu}\,.
\end{equation}
The contribution from the
matter is 
\begin{equation}\label{eq:tensormatter}
M_{\mu\nu}=\left(\rho+p\right)u_{\mu}u_{\nu}+pg_{\mu\nu},
\end{equation}
i.e., it is a perfect fluid contribution,
with $\rho$ being the fluid matter energy density,
$p$ being the isotropic
fluid pressure, and $u_{\mu}$ being
the fluid's four-velocity. The
contribution from the electromagnetic fluid $E_{\mu\nu}$ is
\begin{equation}\label{eq:tensorcharge}
E_{\mu\nu}=\frac{1}{4\pi}\left(F_{\mu}^{\gamma}F_{\nu\gamma} 
-\frac{1}{4}g_{\mu\nu}F_{\gamma\beta}F^{\gamma\beta}\right)\,.
\end{equation}
The Faraday-Maxwell tensor
$F_{\mu\nu}$ is defined in terms of a
vector potential $\mathcal{A}_{\mu}$ by 
\begin{equation}\label{eq:tensormax}
F_{\mu\nu}=\nabla_{\mu}\mathcal{A}_{\nu}-
\nabla_{\nu}\mathcal{A}_{\mu}\,.
\end{equation}
In turn this implies that
$F_{\mu\nu}$ obeys 
the internal Maxwell
equations $F_{[\mu\nu;\rho]}=0$, with all
the three indices being antisymmetrized.
For a charged fluid, the
current density is expressed as
\begin{equation}\label{eq:tensorj}
J^{\mu}=\rho_{e}u^{\mu}, 
\end{equation}
with $\rho_{e}$ standing for the electric charge density.
The constant of gravitation and the speed of
light are set to one.
Note that the system of equations given in
Eqs.~(\ref{eq:Einstein})-(\ref{eq:tensorj}) is consistent,
see Appendix~\ref{sec:c}.

\subsection{General spherical equations}
\label{generalsphericalequations}

We consider a static and spherically symmetric
spacetime with line element
in Schwarzschild coordinates 
$(t,r,\theta,\varphi)$ given by
\begin{equation}\label{eq:ds2}
ds^{2}=-B\left(r\right)dt^{2}+A\left(r\right)dr^{2}+
r^{2}d\Omega^2\,,
\end{equation}
where the
metric potentials $B\left(r\right)$ and $A\left(r\right)$
depend only
upon the radial coordinate $r$, and 
$d\Omega^2= d\theta^{2} +\sin^{2}
\theta d\varphi^{2}$
is the line element over the unit sphere.
The matter is composed of an
isotropic electrically charged perfect
fluid with energy density
$\rho(r)$, pressure $p(r)$, electric charge density
$\rho_e(r)$, and velocity
flow $u^{\mu}(r)$, with 
\begin{equation}
u^{\mu}=-B^{-\frac12}\left(r\right)\delta_{t}^{\mu}\,,
\end{equation}
where $\delta_{\mu}^{\nu}$
stands for the Kronecker delta.
The electromagnetic field is described by the
vector potential $\mathcal{A}_{\mu}(r)$ written as
\begin{equation}
\mathcal{A}_{\mu}=-\phi\left(r\right)\delta_{\mu}^{t}\,,
\end{equation}
where $\phi\left(r\right)$ is the scalar
electric potential.

The Einstein-Maxwell equations
given by Eqs.~(\ref{eq:Einstein})
and (\ref{eq:Maxwell})
together with 
Eqs.~(\ref{eq:tensortot})-(\ref{eq:tensorj})
and the corresponding definitions, 
yield a set of
three differential
equations. A combination of
the components $tt$ and $rr$ of these
equations provides two equations, namely,
\begin{equation}\label{eq:CombinacaoEin}
\frac{A^{\prime}(r)}{A(r)}+\frac{B^{\prime}(r)}{B(r)}=8\pi 
rA(r)\Big(\rho\left(r\right)+p\left(r\right)\Big),
\end{equation}
\begin{equation}\label{eq:CombinacaoEin2}
\left(\frac{r}{A(r)}\right)^{\prime}=1-8\pi 
r^{2}\left(\rho\left(r\right)+\frac{Q^{2}(r)}{8\pi r^{4}}\right),
\end{equation}
where a prime denotes derivative with respect to the
radial coordinate $r$.
In  analogy to the
Reissner-Nordstr\"om spacetime metric 
one often
writes the metric function $A(r)$ as 
$\frac1{A(r)}=1-
\frac{2{M}(r)}{r}+\frac{Q^2(r)}{r^2}$,
where ${M}$ is the mass function, i.e., 
the mass inside a surface of radius $r$,
and $Q(r)$ is the electric charge
function, i.e.,
the electric charge inside a surface of radius $r$.
In this case, instead of Eq.~\eqref{eq:CombinacaoEin2}
one has ${M}^\prime(r)=
4\pi 
r^{2}\left(\rho\left(r\right)+\frac{Q^{2}(r)}{8\pi r^{4}}\right)
+\frac12\left(\frac{Q^2(r)}{r}\right)^\prime$,
which integrates to 
$M(r) = \int_0 ^r 4\pi\, r^2 \left( \rho(r)
+\frac{Q^2(r)}{8\pi\,r^4}\right)dr + \frac{Q^2(r)}{2\,r}$.
The  electric charge function $Q(r)$
obeys $Q^\prime(r)=
4\pi\rho_{e}{r}^{2}
\sqrt{A\left({r}\right)
}$
which can be integrated to
$Q\left(r\right)=
4\pi\int_{0}^{r}\rho_{e}{r}^{2}
\sqrt{A\left({r}\right)
}dr$.  One can thus trade $\rho_e(r)$
with $Q(r)$ and vice versa,
noting that here and throughout we prefer
to use $Q(r)$.
The third Einstein-Maxwell equation
could be taken as its $\theta\theta$ component,
but it is more useful
to take it from the
contracted Bianchi identities, or equivalently
the energy-momentum
conservation equation, $\nabla_{\mu}T^{\mu\nu}=0$,
which gives
\begin{equation}\label{eq:equiHidro}
2p^{\prime}(r)+\frac{B^{\prime}(r)}{B(r)}\Big(\rho(r) + p(r)
\Big)-\frac{Q(r) Q^{\prime}(r)}{2\pi r^4}=0\,.
\end{equation}
The electric charge inside a surface within radius $r$,
$Q(r)$,
is given by the only nontrivial Maxwell equation, i.e.,
\begin{equation}
\label{eq:carga}
Q\left(r\right)=\frac{r^{2}\phi^{\prime}\left(r\right)}{\sqrt{
A\left(r\right)B\left(r\right)}}\,.
\end{equation}

The present problem is then formulated in terms of four equations,
Eqs.~\eqref{eq:CombinacaoEin}-\eqref{eq:carga}, for six functions,
namely, $A(r)$, $B(r)$, $\rho\left(r\right)$, $p\left(r\right)$,
$\phi(r)$, and $Q(r)$.  Thus, to solve the system two further
relations for the functions must be given.

\subsection{Radial perturbations of the fluid configurations}

\label{radialperturbationsoffluid}

\subsubsection{Lagrangian and Eulerian perturbations}

We now derive the equations governing small perturbations
in a static spherically symmetric general relativistic
spacetime coupled to an electrically charged perfect
fluid.
The equations of motion for
the perturbations are important
as they allow to calculate the normal modes of oscillation and their
frequencies, and thus permit to find  stability
criteria for the equilibrium static configuration.
We use the method developed by Chandrasekhar
for radial perturbations in 
stellar equilibrium configurations
in general relativity
and adapt it
to electrically charged fluid
spacetimes.

One can consider
that at the radius $r$ the physical quantity $f(r,t)$
suffers an Eulerian change, denoted
by $\delta f(r,t)$, so that
\begin{equation}\label{eq:EulPert}
f(r,t)=f_i(r)+\delta f(r,t)\,,
\end{equation}
where $f_{i}(r)$ is the initial value of $f$, i.e., the value of
$f$ in the static unperturbed equilibrium configuration.
There is another possible description for the perturbations.  
Any fluid element at $r$ is displaced to $r+\xi(r,t)$ in the perturbed
state with $\xi$ being the Lagrangian displacement of the fluid
element. This Lagrangian displacement quantity $\xi$ connects the
fluid element in the unperturbed configuration to the corresponding
element in the perturbed configuration, and in order
for the displacement $\xi$ to be a perturbation, it has to be small,
so one imposes $\big|\xi\big|\ll r$.  Due
to this displacement, any physical quantity $f(r,t)$ has a Lagrangian
change $\Delta f(r,t)$ when measured by an observer that moves with
the perturbation, so that
\begin{equation}\label{eq:LangPert}
f(r+\xi,t)=f_i(r)+\Delta f(r,t).
\end{equation}
This Lagrangian change is called Lie dragging
of the quantity $f$ 
in general relativity.
Comparing Eqs.~\eqref{eq:EulPert} and~\eqref{eq:LangPert},
the Eulerian perturbations $\delta f(r,t)$ and
the Lagrangian perturbations $\Delta f(r,t)$
are related in first
order by 
\begin{equation}\label{eq:LagEul}
\Delta f(r,t)
= f_{i}^{\prime}(r)\xi(r,t)+\delta f(r,t)\,,
\end{equation}
since in
Eq.~\eqref{eq:LangPert} one can write $f(r+\xi,t)$ as a first order
expansion in $\xi(r,t)$, namely,
$f(r+\xi,t)=f(r,t)+f_{i}^{\prime}(r)\xi(r,t)$,
where again a prime means derivative with respect to $r$.

\subsubsection{The perturbation equations}
\label{theperturbationequations}

Now we consider general radial perturbations in equilibrium
charged fluid spacetimes, i.e., we consider perturbations
in the quantities
$B(r)$,
$A(r)$,
$\rho(r)$,
$p(r)$,
$\phi(r)$, and
$Q(r)$, 
which were presented in the preceding section
and are by assumption  solutions of the Einstein-Maxwell equations.
Since we are considering radial displacements alone, there
is a small non-zero radial fluid flow that causes the spacetime
metric and fluid variables to depend on time maintaining its spherical
symmetry. 
Thus, the Eulerian perturbations can be written as 
\begin{equation}\label{eq:EulreianDis}
\begin{array}{lcl}
B(r,t)& = & B_{i}(r)+\delta B(r,t),\\
A(r,t)& = & A_{i}(r)+\delta A(r,t),\\
\rho(r,t)& = & \rho_{i}(r)+\delta \rho(r,t),\\
p(r,t)& = & p_{i}(r)+\delta p(r,t),\\
\phi(r,t)& = & \phi_{i}(r)+\delta \phi(r,t),\\
Q(r,t)& = & Q_{i}(r)+\delta Q(r,t),
\end{array} 
\end{equation}
where again
the subscript $i$
denotes the initial value of
the corresponding quantity.
Due to the
perturbation,
the fluid's four-velocity acquires a radial
component and can be expressed in the form $u^\mu = (u^t, u^r, 0,
0)$,
where 
$u^{t}=\dfrac{dt}{d\tau}$ and $u^{r}=\dfrac{dr}{d\tau}$,
$\tau$ being the proper time of the fluid element.
Thus, the components 
$u^{t}$ and $u^{r}$ are given up to first order by
\begin{equation}\label{eq:cv-1}
u^{t}  =  B_{i}^{-\frac12}\left(1-
\dfrac{\delta B}{2B_{i}}\right)\,,\quad
    u^{r} = 
    \dot{\xi} B_{i}^{-\frac12},
\end{equation}
where a dot  indicates partial
derivative with respect to the coordinates $t$, and so
the radial velocity $\dot{\xi}\equiv\frac{\partial \xi}{\partial t}$
is the time
variation of the displacement of a fluid element relative to its
equilibrium position.

The combination of
the $tt$ and $rr$ components 
of the Einstein equations given in Eqs.~(\ref{eq:CombinacaoEin})
and~(\ref{eq:CombinacaoEin2})
when perturbed
yield the following
two equations which link  the
perturbations 
$\delta A$, $\delta B$, $\delta\rho$, $\delta p$, and $\delta Q$,
\begin{equation}\label{eq:eint00}
8\pi r^{2} \delta\rho+\dfrac{2Q_{i}\delta Q}{ r^2}-
\left(\dfrac{r\delta A}{A_i^2}\right)^{\prime}= 0,
\end{equation}
\begin{equation}\label{eq:eint11}
\begin{split}
  8\pi r^2\delta p & -\dfrac{2Q_{i}\delta Q}{r^{2}}  +
  \dfrac{\delta A }{A_{i}^{2}} \\ & - \frac{r}{A_i}\left[
  \left(\dfrac{\delta B}{B_i} \right)^\prime - 
  \dfrac{B_i^\prime}{B_i} \frac{\delta A}{A_i} \right]  =0\,.
\end{split}
\end{equation}
The other Einstein equation, 
i.e., $\nabla_{\mu}T^{\mu\nu}=0$
given in Eq.~(\ref{eq:equiHidro}), when
perturbed gives the following equation
\begin{equation}\label{eq:motion}
\begin{split}
\dfrac{A_{i}}{B_{i}} & \left(\rho_i+p_i\right) \ddot{\xi} + \delta
p^{\prime}
+(\rho_i+p_i)\left(\dfrac{\delta B}{2B_i}\right)^{\prime}
\\ & + (\delta\rho+\delta p)\dfrac{B_{i}^{\prime}}{2B_{i}} -
\dfrac{Q_{i}^{\prime}\delta Q}{4\pi r^{4}} -\dfrac{Q_{i}\delta
Q^{\prime}}{4\pi r^{4}} =0\,.
\end{split}
\end{equation}
The $rt$ component
of the Einstein equation, which in the static case is
devoid of content, when perturbed
yields 
\begin{equation}\label{eq:eint01}
8\pi(\rho_i+p_i)A_{i}\dot{\xi}+\dfrac{(\delta A)\,\dot{}}{rA_{i}}=0\,.
\end{equation}
We still need to deal with
the perturbed quantities introduced into the
Maxwell equations.
The
perturbed Maxwell equations furnish now two differential
equations for the perturbed
electromagnetic potential $\delta \phi$ and for the perturbed
electric charge $\delta Q$. One of the perturbed equations
is found by perturbing the static equation
given in Eq.~\eqref{eq:carga},
the other equation is the $r$ component in the Maxwell
equations.
After some manipulation, the two equations
imply in
\begin{equation} \label{eq:deltaQL}
\delta Q+Q_{i}^{\prime}\xi=0\,,
\end{equation}
We have six unknowns, namely, $\delta A$,
$\delta B$,
$\delta \rho$,
$\delta p$,
$\delta Q$, and
$\xi$, and five equations, Eqs.~(\ref{eq:eint00})-(\ref{eq:deltaQL}).
Thus, we
still need a relation, which is going to be
a relation between the perturbed pressure and the
perturbed density.  The
new natural equation is to 
impose the condition  that the matter is perturbed 
adiabatically, and so 
one has 
\begin{equation}\label{eq:adindex}
\gamma=\dfrac{\rho_i+p_i}{p_i}\dfrac{\Delta p}{\Delta\rho},
\end{equation}
where $\gamma$ is the adiabatic index,
and $\Delta \rho$ and $\Delta p$ are the Lagrangian
perturbations of the energy density and the pressure, respectively.
So, Eqs.~(\ref{eq:eint00})-(\ref{eq:adindex})
form the set of equations that will give a
differential equation for $\xi$.

We have now to manipulate Eqs.~(\ref{eq:eint00})-(\ref{eq:adindex})
to obtain a set of equations in useful form.
The result is the following
set of six equations, see Appendix~\ref{manipulation},
\begin{equation}\label{eq:deltaA}
\delta A=-A_{i}
\left(\dfrac{A_{i}^{\prime}}{A_{i}}+
\dfrac{B_{i}^{\prime}}{B_{i}}\right)\xi\,,
\end{equation}
\begin{equation}\label{eq:deltaB}
\begin{split}
\left(\dfrac{\delta B}{B_{i}}\right)^{\prime}=8\pi
A_{i}&\Big(2rp_{i}^{\prime}-\left(\rho_{i}+p_{i}\right)\Big)\xi\\
&+8\pi A_{i}r\delta p-\dfrac{2A_{i}Q_{i}Q_{i}^{\prime}\xi}{r^{3}}\,,
\end{split}
\end{equation}
\begin{equation} \label{eq:deltarho}
\delta\rho=-\rho_{i}^{\prime}\xi-
\left(\rho_i+p_i\right)\dfrac{B_{i}^{{\frac12}}}{r^{2}} 
\left(r^{2}B_{i}^{-{\frac12}}\xi\right)^{\prime}\,,
\end{equation}
\begin{equation} \label{eq:deltaP}
\delta p=-p_{i}^{\prime}\xi-
\gamma\dfrac{p_{i}B_{i}^{{\frac12}}}{r^{2}} 
\left(r^{2}B_{i}^{-{\frac12}}\xi\right)^{\prime}\,,
\end{equation}
\begin{equation}\label{eq:motion2}
\begin{split}
\dfrac{A_{i}}{B_{i}} &
\left(\rho_i+p_i\right) \ddot{\xi} + \delta
p^{\prime}
+(\rho_i+p_i)
\left(\dfrac{\delta B}{2B_i}\right)^{\prime}
\\ & + (\delta\rho+\delta p)\dfrac{B_{i}^{\prime}}{2B_{i}} -
\dfrac{Q_{i}^{\prime}\delta Q}{4\pi r^{4}} -
\dfrac{Q_{i}\delta
Q^{\prime}}{4\pi r^{4}} =0,
\end{split}
\end{equation}
\begin{equation}
\delta Q=-Q_{i}^{\prime}\xi\,.
\label{eq:deltaQ}
\end{equation}
So there are six equations for six unknowns.

\subsubsection{Pulsation equation, boundary conditions, 
and  stability criteria for the equilibrium configuration} 

For the analysis of the stability or instability of the equilibrium
state of the fluid configurations, the equation of motion given in
Eq.~\eqref{eq:motion2} governing the perturbations
in $\xi$
can be rewritten in
a more useful form taking into account
$\delta A$ in Eq.~\eqref{eq:deltaA}, 
$\delta B$ in Eq.~\eqref{eq:deltaB}, 
$\delta \rho$ in Eq.~\eqref{eq:deltarho}, 
$\delta p$ in Eq.~\eqref{eq:deltaP}, and
$\delta Q$ in Eq.~\eqref{eq:deltaQ},
where all quantities are expressed in terms of $\xi$ and the
unperturbed variables. Considering that all perturbations have a
harmonic time dependence of the form $e^{i\omega t}$, where $\omega$
is the oscillation frequency, then Eq.~\eqref{eq:motion2}
together with all other equations becomes

\begin{eqnarray}
&
\left[\gamma\dfrac{p}{r^{2}} B^{{\frac32}}A^{{\frac12}}
\left(r^{2}B^{-{\frac12}}\xi\right)^{\prime}\right]^{\prime}
\nonumber\\
&-\Bigg[8\pi A(\rho+p)\left(p+
\dfrac{Q^{2}}{8\pi r^{4 }}\right)  
-\dfrac{1}{(\rho+p)}\left(\dfrac{QQ^{\prime}}
{4\pi r^{4}}-p^{\prime} \right)^{2}\nonumber\\
&+ \dfrac{4}{r} p^{\prime}-\omega^{2}(\rho+p)\dfrac{A}{B}
\Bigg]
BA^{{\frac12}}\xi=0\,,
\label{eq:eigenvalue}
\end{eqnarray} 
where we have dropped the subscript $i$ which is irrelevant from now
onward.
This is the modified Chandrasekhar radial pulsation equation
\cite{Chandre1964b} with the
inclusion of electric charge. It
serves to study the radial stability of the system. 
This pulsation equation, Eq.~\eqref{eq:eigenvalue},
has also been found
in~\cite{Omote1974,Glazer1976,Glazer1979,
Felice1999,Anninos2001}, although
\cite{Felice1999}
has a term in $Q^{\prime\prime}$ incorrect.

One still needs to provide boundary conditions for
Eq.~\eqref{eq:eigenvalue}.  One boundary condition is given at
the origin $r=0$, namely,
\begin{equation}\label{eq:ic1}
\xi(r=0)=0\,,
\end{equation} 
which means that the fluid does
not have radial motion at the center. 
In fact $\xi(r=0)$ only needs to be finite but we adopt
without loss of generality 
Eq.~(\ref{eq:ic1}).  The other boundary condition is
given at the surface of the star $r_0$, namely, $ \Delta
p(r=r_{0})=0$, i.e., the Lagrangian perturbation of the pressure is
zero, the pressure does not change when the boundary is moved, it
continues to be zero. In other words,
this condition expresses the fact that a fluid
element located at surface of the unperturbed configuration is
displaced to the perturbed surface.  Now, $\Delta p(r)$ can be taken
directly from its Eulerian perturbation $\delta p(r)$ using $\Delta p=
\delta p+p_{i}^{\prime}\xi$, which together with Eq.~\eqref{eq:deltaP}
yields $\Delta p=-\gamma\dfrac{pB^{\frac12}}{r^{2}}
\left(r^{2}B^{-\frac12}\xi\right)^{\prime}$ and this can then be
evaluated at $r_0$ so that $ \Delta p(r=r_{0})=0$.  This then means
that
\begin{equation}\label{eq:ic2}
\left(r^{2}B^{-\frac12}\xi\right)^{\prime}(r=r_0)=0\,,
\end{equation}
which is the second boundary condition.

The criteria for stability can now be established.
Equation~\eqref{eq:eigenvalue}, together with the boundary conditions
Eqs.~\eqref{eq:ic1} and~\eqref{eq:ic2}, is an $\omega^2$ eigenvalue
problem. Then, if $\omega^2>0$ the system oscillates and it is stable,
if $\omega^2=0$ then the system stays static and there is neutral
stability, and if $\omega^2<0$ the system expands or collapses
exponentially and is unstable.
We now turn to the formal
implementation of these stability criteria.

\subsubsection{The pulsation equation in a convenient Sturm-Liouville
form}
\label{convenientSturm-Liouvilleform}

To implement the stability analysis
and understand the various possibilities
related to stability or instability
it is important to rewrite
Eq.~\eqref{eq:eigenvalue}
in a convenient Sturm-Liouville (SL) form.
Thus, appropriate manipulation of Eq.~\eqref{eq:eigenvalue}
leads to the following second order ordinary homogeneous
differential equation,
\begin{equation}\label{eq:SLP}
F(r)\zeta^{\prime\prime}(r)+G(r)\zeta^{\prime}(r)+
\left[H(r)+\omega^2 W(r)\right]\zeta(r)=0,
\end{equation}
where 
\begin{equation}\label{eq:SLPzeta}
\zeta(r)=r^2 B^{-\frac12}\xi(r),
\end{equation}
and the coefficients $F(r)$, $G(r)$, $H(r)$, and $W(r)$ are given by
\begin{equation}\label{eq:SLPF}
F(r)=\dfrac{\gamma p B^{\frac32}A^{\frac12}}{r^2},
\end{equation}
\begin{equation}\label{eq:SLPG}
G(r)=\dfrac{dF(r)}{dr},
\end{equation}
\begin{equation}\label{eq:SLPH}
\begin{split}
H(r)=& \dfrac{B^{\frac32}A^{\frac12}}{r^{2}}
\left[\dfrac{1}{(\rho+p)}
\left(\dfrac{QQ^{\prime}}
{4\pi r^{4}}-p^{\prime} \right)^{2}\right.\\
 &\left.- 
\dfrac{4p^{\prime}}{r} -8\pi A(\rho+
p)\left( p+
\dfrac{Q^{2}}{8\pi r^{4}}\right)
\right], 
\end{split}
\end{equation}
\begin{equation}\label{eq:SLPW}
W(r)=\dfrac{(\rho+p) B^{\frac12}A^{\frac32}}{r^2}.
\end{equation}

The boundary conditions given in Eqs.~\eqref{eq:ic1}
and~\eqref{eq:ic2} are now
\begin{equation}\label{eq:ic1new}
\zeta(r=0)=0\,,
\end{equation} 
and 
\begin{equation}\label{eq:2c1new}
\zeta^\prime(r=r_0)=0\,,
\end{equation} 
respectively.
Depending on whether
$F$ is positive or negative and
whether $W$ is positive or negative
one can state various theorems
that indicate the stability
character of the solution,
see Appendix~\ref{sec:SLP}
for the details concerning the
theorems.

\subsubsection{Importance of the adiabatic index}
\label{Subec:IAI}

The coefficient $\gamma$ is defined in Eq.~\eqref{eq:adindex} and
is
an important quantity
for the stability analysis of compact
objects undergoing adiabatic perturbations,
see indeed 
the pulsation equation given by
Eq.~\eqref{eq:eigenvalue} or Eq.~\eqref{eq:SLP}.

For a classical ideal gas the adiabatic index
$\gamma$ is of the order of unity.
It may
assume very large values in the case of liquids,
and in the
case of noncompressible fluids $\gamma$ can be taken as
equal to infinity.  The adiabatic index can be a function of the
energy density and pressure so that when these change within the
fluid, the adiabatic index can also change.
For the study of radial
perturbations on
static and spherically symmetric configurations,
such as stars, positive constant values for $\gamma$, $\gamma>0$, are
assumed throughout the configurations, a procedure we follow here.
For a fluid supported by tension, i.e., negative pressure, one has
that $\frac{\Delta p}{\Delta \rho}$ is negative and so from the
definition of $\gamma$ one has that it is negative,
$\gamma<0$. Situations with negative $\gamma$ will appear in our
analysis.

\subsubsection{Numerical methods}
\label{Subec:Numericalmethods}
In order to solve the perturbation equation, being an eigenvalue SL
problem, we use numerical methods, namely, the shooting method,
borrowing analysis and results from 
\cite{PressBook1992,KongZettl1996,Moller1999,ZettlBook},
and the
Chebyshev finite difference method
borrwing analysis and results
from~\cite{Elgendi1969,Boyd19892013,Elbar2003,TMM2013,jansen17}.
For a detailed analysis of these methods see
Appendix~\ref{sec:numeric}.

\section{Electrically charged spheres: Guilfoyle's solution}
\label{sec:solutions}

\subsection{The analytical solutions}

Now we turn to the specific electrically
charged spacetimes containing charged fluids
that we will analyze.

The interior region solution, for which the radius
is interior to the boundary radius $r_0$,
$r\leq r_0$,
is composed of an electrically charged fluid.
We have seen that
in order to find solutions
for an electrically charged fluid
there are 
 four equations,
Eqs.~\eqref{eq:CombinacaoEin}-\eqref{eq:carga}, for six functions,
namely, $A(r)$, $B(r)$, $\rho\left(r\right)$, $p\left(r\right)$,
$\phi(r)$, and $Q(r)$, and
so to solve the system two further
relations for the functions must be given.
Guilfoyle \cite{Guilfoyle1999}
gave two further
relations with physical content and
mathematical motivation that make
the whole set of six equations
self contained. 
The first relation is
an assumption with respect to the effective energy
density defined as 
$\rho\left(r\right)+\dfrac{Q^{2}(r)}{8\pi 
r^{4}}$, and
one assumes that
a generalized Schwarzschild condition is obeyed,
namely
\begin{equation}\label{eq:energiatotal}
8\pi\rho\left(r\right)+\frac{Q^{2}\left(r\right)}{r^{4}}
=\frac{3}{R^{2}}\,,
\end{equation}
where $R$ is a new constant
parameter. The additional relation
is the assumption
that the metric potential $B(r)$ and the electric potential
$\phi(r)$
are related through a generalized Weyl condition,
namely
\begin{equation}
\label{eq:energiatotalweilrel}
B(r)=a\phi^2(r)\,,
\end{equation}
where $a$ is an arbitrary constant
that we call the Guilfoyle parameter.
With Eqs.~\eqref{eq:CombinacaoEin}-\eqref{eq:carga},
and these two new equations,
Eqs.~\eqref{eq:energiatotal}
and~\eqref{eq:energiatotalweilrel},
there is a closed system of equations
for the six unknowns
 $A(r)$, $B(r)$, $\rho\left(r\right)$, $p\left(r\right)$,
$\phi(r)$, and $Q(r)$ that can be solved exactly.
The interior region solution
 is then given by
explicit forms for the functions
$A(r)$,
$B(r)$,
$\rho(r)$,
$p(r)$,
$\phi(r)$,
and $Q(r)$.
The metric function
$A(r)$ is given by
\begin{equation}\label{eq:pmAex}
A(r)=\left(1-\frac{r^{2}}{R^{2}}\right)^{-1}.
\end{equation}
The metric function
$B(r)$ is given by
\begin{equation}\label{eq:pmBex}
B(r)=\left[\dfrac{(2-a)^{2}}{a^{2}}F^{2}(r)
\right]^{\frac{a}{a-2}}\,,
\end{equation}
where $F(r)$ is defined as
$F(r)=k_{0}\sqrt{1-\dfrac{r^{2}}{R^{2}}}-k_{1}$,
and the integration constants $k_{0}$ and $k_{1}$,
are found using the junction
conditions for a  smooth
matching to an exterior
Reissner-Nordstr\"om spacetime.
They 
are given
by
$k_{0}=\dfrac{R^{2}}{r_{0}^{2}}\left(\dfrac{m}{r_{0}}-
\dfrac{q^{2}}{r_0^{2}}
\right)\left(1-
\dfrac{r_{0}^{2}}{R^{2}}\right)^{\hskip -0.15cm-\frac1a}$,
and
$k_{1}=k_{0}\sqrt{1-\dfrac{r_{0}^{2}}{R^{2}}}
\left[1-\dfrac{a}{2-a}\dfrac{r_{0}^{
2}}{R^{2}}\left(\dfrac{m}{r_{0}}-
\dfrac{q^{2}}{r_0^{2}}\right)^{\!\!
-1}\right]$,
with $m$ and $q$ being
the  spacetime
mass and
electric charge
of
the exterior Reissner-Nordstr\"om spacetime, respectively.
The perfect
fluid quantities, namely
the energy density and the pressure, are 
\begin{equation}\label{eq:denEnerIn}
8\pi\rho\left(r\right)=\frac{3}{R^{2}}-
\dfrac{a}{(2-a)^{2}}\dfrac{k_{0}^{2}}{R^
{4}}\dfrac{r^{2}}{F^{2}(r)},
\end{equation}
\begin{equation}\label{eq:pressure}
\begin{array}{ccl}
8\pi p\left(r\right)&=&-\dfrac{1}{R^{2}}+\dfrac{a}{(2-a)^{2}} 
\dfrac{k_{0}^{2}}{R^{4}}\dfrac{r^{2}}{F^{2}(r)}+\\
\\                  & & 
+\dfrac{2a}{2-a}\dfrac{k_{0}}{R^{2}}
\dfrac{\sqrt{1-
\frac{r^2}{R^2}
}}{F(r)}\,,
\end{array}                    
\end{equation}
respectively.
The electric potential can be obtained from the relation
$\phi(r)=\epsilon\sqrt{\dfrac{B(r)}{a}}$,
see Eq.~\eqref{eq:energiatotalweilrel}, where
$\epsilon=\pm1$,
and so is given by
\begin{equation}\label{eq:phi}
\phi(r)=\dfrac{\epsilon}{\sqrt a}
\left[\dfrac{(2-a)^{2}}{a^{2}}F^{2}(r)\right]^{\frac{a}{2(a-2)}}
\,.
\end{equation}
The electric charge density $\rho_{e}(r)$ can be
written as
$
4\pi\rho_{e}(r)=\dfrac{k_{0}}{R^{2}} \dfrac{Q(r)}{r F(r)} \left(1+
\dfrac{3 R^2 F(r) }{k_{0}r^{2}}\sqrt{1-\frac{r^{2}}{R^{2}}}\right)
$
where $Q(r)$ is given by
\begin{equation}\label{eq:cargain}
Q(r)=\dfrac{\epsilon\sqrt{a}}{2-a}
\dfrac{k_{0}}{R^{2}}\dfrac{r^{3}}{F(r)}. 
\end{equation}
If one prefers to work with the mass ${M}(r)$
already defined and given by 
 ${M}(r)= \int_0 ^r 4\pi\, r^2 \left( \rho(r)
+\frac{Q^2(r)}{8\pi\,r^4}\right)dr + \frac{Q^2(r)}{2\,r}$, 
then one obtains 
$
{M}(r)=\dfrac{r^{3}}{2R^{2}}+
\dfrac{a}{2(2-a)^{2}}\dfrac{k_{0}^{2}}{R^{4
}}\dfrac{r^{5}}{F^{2}(r)}$. We stick to $A(r)$
given in Eq.~\eqref{eq:pmAex}
instead of ${M}(r)$.

The exterior region solution,
i.e.,  the region outside the distribution
of the electrically charged
fluid, with $r\geq r_0$,
is empty of matter, 
it is a vacuum solution, and so 
the solution of the Einstein-Maxwell
equations is given by the
Reissner-Nordstr\"om solution, 
\begin{equation}
A(r)=\left(1-\frac{2m}{r}+\frac{q^{2}}{r^{2}}\right)^{-1},
\label{eq:A(r)ext}
\end{equation}
\begin{equation}
B(r)=\frac{1}{A(r)}=1-\frac{2m}{r}+\frac{q^{2}}{r^{2}},
\label{eq:B(r)ext}
\end{equation}
with $\rho(r)=0$,
$p(r)=0$, such that $m$ is a constant defining the 
mass of the exterior spacetime, and
\begin{equation} \label{eq:phiext}
\phi(r)=\frac{q}{r}-\frac{q}{r_0}+\frac{\epsilon}{\sqrt{a}}
\sqrt{1-\dfrac{r_0^2
}{R^2}} \,,
\end{equation}
with $\rho_e(r)=0$, such that $q$ is a constant defining the total
electric charge, and the constant
of integration was adjusted such that
the electric potential is a continuous function through the boundary
$r=r_0$.
This exterior
spacetime has two important
intrinsic radii, namely,
the gravitational 
and the Cauchy radii, which are given in terms
of $m$ and $q$ through the relations
$r_+ = m +\sqrt{m^2 - q^2}$ and
$r_- = m - \sqrt{m^2 - q^2}$,
respectively.
These radii are real, and so physically relevant,
when $q\leq m$, i.e., for undercharged and
extremally charged
exterior spacetimes, and are imaginary, and so of no interest
when $q>m$, i.e., for overcharged
exterior spacetimes.
Moreover, in $q\leq m$ cases,
when $r_0\geq r_+$  one has that 
$r_+$ is simply the gravitational radius,
whereas when $r_0<r_+$  one has that 
$r_+$ is also an event horizon
radius. In the same manner 
in $q\leq m$ cases,
when $r_0\geq r_-$  one has that 
$r_-$ is simply the Cauchy radius,
whereas when $r_0<r_-$  one has that 
$r_-$ is also a Cauchy horizon
radius. 

At the interface, in between the interior and the exterior
regions, there is a smooth boundary.
By
imposing smooth boundary conditions of metric functions $A(r)$
and $B(r)$ at the surface $r=r_0$, one obtains
a relation between $m$, $q$, $r_0$, and $R$,
and
another relation
between $a$, $q$, $r_0$, and $R$. These relations are given
by
\begin{equation}\label{eq:m}
m=\dfrac{r_{0}}{2}\left(\dfrac{r_{0}^{2}}{R^{2}}+
\dfrac{q^{2}}{r_{0}^{2}}\right),
\end{equation}  
\begin{equation} \label{eq:a}
a=\dfrac{r_{0}^{2}}{4q^{2}}\left(\dfrac{r_{0}^{2}}{R^{2}}-
\dfrac{q^{2}}{r_{0}^{2}}\right)^2
\left(1-\dfrac{r_{0}^{2}}{R^{2}}\right)^{-1}.
\end{equation}
Thus,
there are only three free parameters in the model.
These are chosen to be $r_0$, $q^2$, and
$R$.
The other important parameters of the model are then written in
terms of these three, see~\cite{LemosZanchin2017}.

\subsection{Plethora of the solutions: Stars, regular black
holes, quasiblack holes, and quasinonblack holes}

\begin{figure}[h]
\centering 
\includegraphics[width=.43\textwidth]{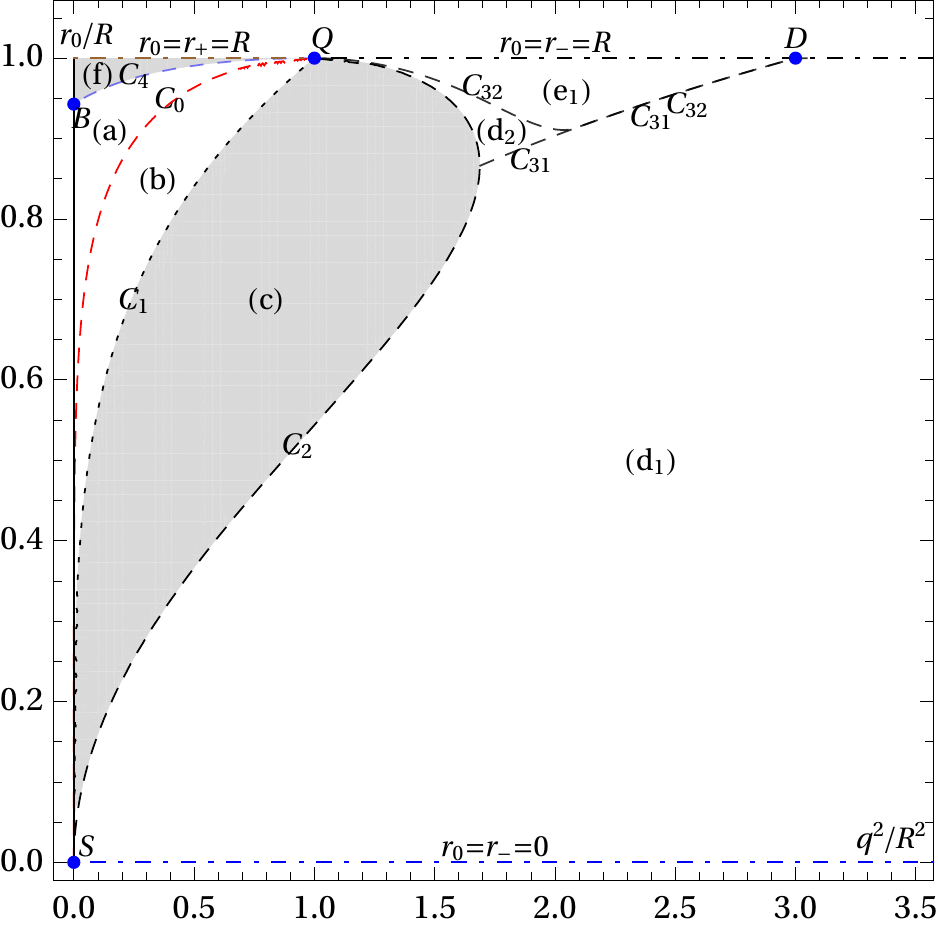}
\caption{
All electrically charged solutions in a $\frac{q^2}{R^2}\times
\frac{r_0}{R}$ space.  The regions, lines, and points shown in the
plot are explained in detail in the text.  White regions consist of
regular solutions and gray regions consist of singular solutions.}
\label{fig:regions} 
\vskip 1.5cm
\centering 
\includegraphics[width=.43\textwidth]{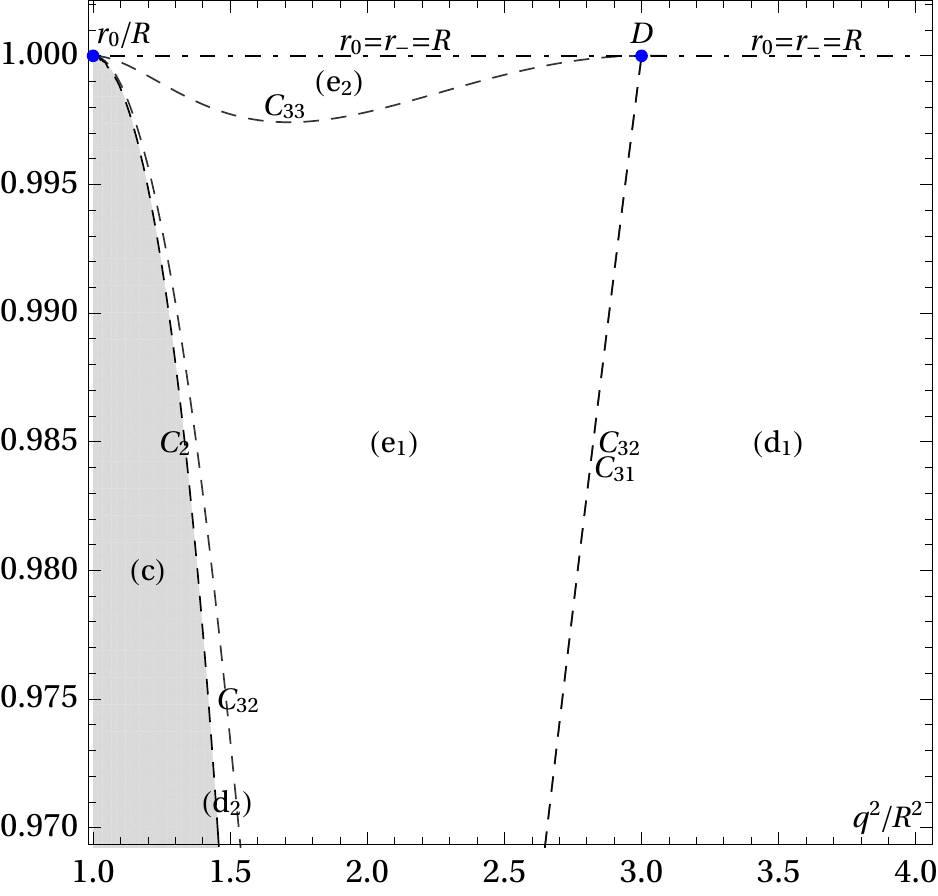}
\caption{A blow up of the previous figure
to show region (e2).
}
\label{fig:regionszoom} 
\end{figure}

The full spectrum of  Guilfoyle's solution was found
in~\cite{LemosZanchin2017}. Drawing on that work, we show in
Fig.~\ref{fig:regions} the relevant regions in the space of the
solutions defined by the parameters $\frac{q^2}{R^2}\times
\frac{r_0}{R}$, i.e., the space defined by the electric charge $q$ and
the radius of the configuration $r_0$, both quantities in units of the
radius $R$. The abscissa $\frac{q^2}{R^2}$ runs from zero to infinity,
and the ordinate $\frac{r_0}{R}$ possesses the remarkable feature that
it has a finite range, $0\leq\frac{r_0}{R}\leq1$, and so all the
possible configurations are displayed within this range.  Recall that
$R$ is an intrinsic radius, defined as the square
root of the inverse of the effective energy density.  This means that
giving $R$ as the unit of measure, a move along the configurations in
the space of the solutions in the figure can be seen as a change of
the parameters $q$ and $r_0$ in relation to $R$, and so in relation to
the defined constant effective energy density.
To emphasize this point we
refer to the figure, and 
note
that moving vertically in it
along $\frac{r_0}{R}$,
can be interpreted
as increasing the
radius of the configurations for the given fixed effective energy
density, and in doing so, the mass also increases, up to the point
where either
a singular configuration appears or
an event horizon and consequently a black hole appears in the
space of solutions.
This way of seeing stars, namely, constant
density and with the radius of the configuration increasing, was the
way envisaged by Michell and Laplace when they discussed dark stars
two hundred and fifty years ago.  Nowadays, the discussion hinges
often instead on the quotient $\frac{r_0}{r_+}$, and so by decreasing
$r_0$ maintaining the gravitational radius $r_+$ constant, one gets a
sequence of ever more compact objects. But here, in our context,
$\frac{r_0}{r_+}$, has drawbacks.
One is that there are cases in which $r_+$ does not exist, and
        another is that there are cases where,
	although $r_+$
	exists, 
$r_0$ is less than the Cauchy horizon radius
$r_-$, so clearly
outside the scope of $\frac{r_0}{r_+}$.
Definitely, $R$ is a
universal gauge for the full spectrum of the solutions and
so the quotient $\frac{r_0}{R}$ that we use is the perfect parameter
to deal with.

The regions, lines, and points described below are
referred to Fig.~\ref{fig:regions}, and when required
one refers to Fig.~\ref{fig:regionszoom},
which is a blow up of
a specific region of Fig.~\ref{fig:regions}.  The figures
will be important in the understanding
of the stability analysis. We will start the description
with the vertical axis
$\frac{q^2}{R^2}=0$ and then move
counterclockwise, as faithfully as possible, along the regions,
lines, and points, up to the very starting vertical axis.

Line $\dfrac{q^2}{R^2}=0$ is the vertical axis and corresponds to the
interior Schwarzschild solutions, i.e., Schwarzschild stars.  These
are solutions for a zero electrically charged incompressible perfect
fluid in static spherically symmetric spacetimes. The lower endpoint
in the limit, $\frac{r_0}{R}=0$, gives the Minkowski spacetime. This
is a line of interest for the stability problem.

Point $B$ represents the Buchdahl bound, for which
$r_0=\frac{9}{4}m$, i.e., $r_0=\frac{9}{8}r_+$, and also obeys
$\frac{q^2}{R^2}=0$, and for which the uncharged stars present an
infinite central pressure.  This is a point of interest for the
stability problem, as a limiting point.

Region (a) contains normal stars, i.e., regular undercharged stars, so
$m^2>q^2$, with positive energy density $\rho(r) > 0$,
positive pressure $p(r)> 0$, and positive enthalpy
$h(r)$, $h(r)=\rho(r)+p(r)>0$. This is a region of interest for the
stability problem.

Line $C_0$ obeys the
equation $a\left(\frac{r_0}{R}, \frac{q^2}{R^2}\right)=1$. The
pressure of all objects on this line is zero, and they are all
extremally charged objects
with $\rho(r)=\rho_e(r)$ and with $m^2=q^2$,
i.e., 
$r_-= r_+$, and such that $r_0 < r_- = r_+$.
On this line, the solutions are regular and are called  Bonnor
stars.  This is a line of interest for the stability problem.

Region (b) contains regular overcharged stars, so $m^2< q^2$. These
are all tension stars for which $\rho(r)> 0$
and $p(r) <0$, and that also satisfy the
positive enthalpy condition $h(r)=\rho(r)+p(r)>0$.  This is a region
of interest for the stability problem.

Point $Q$ from the left is extraordinarily interesting. It represents
quasiblack holes, which we abbreviate as
QBHs from now onward, and these obey $\frac{r_0}{R}=1$ and
$\frac{q^2}{R^2}=1$.  It is a degenerated point since at $Q$
there exist
many solutions with different physical and geometrical properties
which depend on the path followed to approach $Q$.  The solution may
be a pressure quasiblack hole if the point is reached from region (a),
a pressureless quasiblack hole if the point is reached by following
the line $C_0$, a tension quasiblack hole, if the point $Q$ is reached
from region (b).  We study below in detail these quasiblack hole
limits.  This is a point of high interest for the stability problem,
as a limiting point, indeed, to find the stability character of this
point from the left is one of the main motivations of the whole work.

Line $C_1$ obeys the equation $a\left(\frac{r_0}{R},
\frac{q^2}{R^2}\right)=0$.  It contains singular objects.  This is a
line of no interest for the stability problem.

Region (c) contains singular overcharged objects, so $m^2< q^2$. These
are weird objects having
the curvature scalars and the fluid quantities diverging at some
radius inside the matter distribution.  This is a region of no
interest for the stability problem.

Line $C_2$ obeys the equation $a\left(\frac{r_0}{R},
\frac{q^2}{R^2}\right)=1$. The pressure of all objects on this line is
zero, and they are all extremely charged singular black holes with
$m^2=q^2$, i.e., $r_-=r_+$, and such that $r_0<r_-=r_+$.  On this line,
the solutions are singular as the energy density and the charge
density, which obey $\rho(r)=\rho_e(r)$, diverge at $r=0$.  This line
$C_2$ has an elbow at $\frac{q^2}{R^2}=\frac{27}{16}=1.6875$.  Line
$C_0$ together with line $C_2$ form a closed curved in the parameter
with equation $a\left(\frac{r_0}{R}, \frac{q^2}{R^2}\right)=1$.  This
is a line of interest for the stability problem as it is a division
line to regular black holes.

Line $\dfrac{r_0}{R}=\dfrac{r_-}{R}=0$ is the horizontal axis. It is a
limiting line on which some of the quantities such as the mass
diverge. The solutions belonging to this line are not compact objects,
they correspond to Kasner spacetimes.  This is a line of no special
interest for the stability problem.

Point $S$ is the origin of the two axis, it obeys $\frac{q^2}{R^2}=0$
and $\frac{r_0}{R}=0$. It represents different spacetimes depending on
the path followed to get there.  For instance, it gives the
Schwarzschild black hole if the limit is taken by choosing the ratio
$\frac{q^2}{r_0}$ as a fixed finite number. This point is of no
special interest for the stability analysis.

Region (d1) contains regular black holes with a central core of
charged phantom matter for which $h(r)=\rho(r) + p(r) <0$ up to the
boundary radius $r_0$ with the radius of the object $r_0$ being inside
the Cauchy horizon, i.e., $r_0< r_-=m - \sqrt{m^2 -q^2}$.  The energy
density $\rho(r)$ is negative for a range of the radial coordinate $r$
inside the matter core, while the pressure $p(r)$ is negative
everywhere in the matter region. Regular black holes are always
interesting, so despite this negativity of the energy density, they
count as interesting solutions.  This is a region of interest for the
stability problem.

Line $C_{31}$ 
is drawn from two
conditions, the first is that $h(r_0)=\rho(r_0)=0$, which in turn
implies $\frac{q^2}{R^2}=\frac{3r_0^4}{R^4}$, and the second is that
it verifies that $h(r)$ has finite negative values for all $r$ inside
the region of matter distribution. A segment of this line separates
region (d1) from region (d2), the other segment of this line separates
region (d1) from region (e1).
This latter segment coincides with a segment of the
line $C_{32}$.
This conjoint segment
will then be called $C_{31}C_{32}$.  This is a line of interest for
the stability problem.

Region (d2) contains regular black holes with a central core
of charged phantom matter for which the enthalpy $h(r)=\rho(r) + p(r)
<0$ close to the center and changes sign toward the surface
at radius $r_0$,
with $r_0$ being inside the Cauchy horizon,
i.e., $r_0< r_-=m - \sqrt{m^2 -q^2}$, so that all the matter is
fully inside the Cauchy horizon.  The energy density is positive and
finite at the center, changes to negative values at some $r< r_0$ and
changes back to positive values close to the surface.  This kind of
configurations was not separated
in~\cite{LemosZanchin2017} where
region (d) is now  the region (d1) plus the region (d2).
It turns out that the sign change of
the enthalpy inside the matter core turns the region (d2)
different from region
(d1) regarding the stability analysis.  
This region is a region of interest for the
stability problem.

Line $C_{32}$ is drawn by the condition that the solution has
$\rho(r)=0$ for some $r$ inside the matter distribution region and
$\rho(r)\geq0$ for all $r$. A segment of this line separates region
(d2) from region (e1), the other segment of this line separates region
(d1) from region (e1). This is a line of interest for the stability
problem.

Region (e1) contains regular black holes with 
charged phantom matter for which $h(r)=\rho(r) + p(r) >0$
from some radius $r$ up to the
boundary radius $r_0$ with the radius of the object $r_0$ being inside
the Cauchy horizon, i.e., $r_0< r_-=m - \sqrt{m^2 -q^2}$, so that 
all the matter is fully inside the Cauchy horizon.  In this region
the energy
density is positive, $\rho(r)>0$, for all $0\leq r\leq r_0$.  Regular
black holes are always interesting, and these having positive energy
density are certainly
interesting solutions. This is a region of interest for the
stability problem.

Line $C_{33}$ is drawn by using the condition that the
solution has vanishing
central enthalpy density, i.e., $h(r=0)=\rho(r=0)+p(r=0)=0$,
and it also happens that the
configurations on this curve
have an enthalpy density
$h(r)=\rho(r) +p(r)$ which is positive  for all $r$
in the interval $0< r
\leq r_0$. This line is not explicitly shown
in~\cite{LemosZanchin2017}.  Line $C_{33}$  separates region
(e1) from region (e2), and is shown in Fig.~\ref{fig:regionszoom}
which is a blow of this zone of Fig.~\ref{fig:regions}.  This is
a line of interest for the stability problem.

Region (e2) contains regular black holes with a central core of
charged matter for which $h(r)=\rho(r) + p(r) >0$ up to the
boundary radius $r_0$, with  $r_0$ being inside
the Cauchy horizon, i.e., $r_0< r_-=m - \sqrt{m^2 -q^2}$, so that the
all the matter is fully inside the Cauchy horizon.  This region has
 positive energy $\rho(r)>0$
 and negative pressure $p(r)<0$.  This kind
of configurations was not shown  in~\cite{LemosZanchin2017} where
region (e) is the region (e1) plus the region (e2).  It turns out that
regarding the stability analysis the kind of configurations in (e1)
needs to be treated separately from objects of region (e2). The region
(e2) is shown in Fig.~\ref{fig:regionszoom}
which is a blow of this zone of Fig.~\ref{fig:regions}.
This is a region of interest for the
stability problem.

Line $\dfrac{r_0}{R}=\dfrac{r_-}{R}=1$ is the semi-infinite line with
$\frac{r_0}{R}=1$ and $1< \frac{q^2}{R^2}< \infty$ in
Fig.~\ref{fig:regions}. On this line, the
object has a 
boundary surface $r_0$ of the
matter that coincides with the de Sitter horizon of the inner metric and
with the inner horizon of the Reissner-Nordstr\"om exterior metric,
the matching being on a lightlike surface.  There are two
distinguished points on this line, the point $Q$ at
$\frac{q^2}{R^2}=1$, and the point $D$ at $\frac{q^2}{R^2}=3$, both
containing configurations with special properties.  The segment of
this line in the interval $1< \frac{q^2}{R^2}< 3$ is the top boundary
of region (e2), and the $\frac{q^2}{R^2}> 3$ line is part of the top
boundary of region (d1).  On this line, the metric coefficient $B(r)$
takes the simple form $B(r)=\frac{1}{4}\left(\frac{q^2}{R^2}-
1\right)^2\left(1-\frac{r^2}{R^2}\right)$.
After a time
reparameterization of the form $\frac{1}{4}\left(\frac{q^2}{R^2}-
1\right)^2 dt^2\to dt^2$ the metric potentials turn into a de Sitter
metric, i.e., $B(r)=A^{-1}(r) = \left(1-\frac{r^2}{R^2}\right)$.
In the
interior region, i.e.,  for $0\leq \frac{r}{R} <\frac{r_0}{R}$, 
the
energy density and pressure for the configurations in this line
are given by $8\pi\,\rho(r)
= \frac{3}{R^2}$ and $8\pi\,p(r)= -\frac{3}{R^2}$, so that the
equation of state is a de Sitter one, $\rho=-p$, and with the charge
density tending to a Dirac delta function centered at the boundary
surface $r=r_0$.  At the boundary $r_0$, where here $r_0=r_-=R$,
the energy density jumps from the value
$8\pi\,\rho(r) = \frac{3}{R^2}$ to the value $8\pi\,\rho(r_0) =
\frac{3}{R^2} -\frac{q^2}{R^4}$, and the pressure jumps from the value
$8\pi\,p(r) = - \frac{3}{R^2}$ to zero value, $8\pi\,p(r_0) = 0$.
Therefore, the enthalpy is zero throughout the matter region,
$\rho(r)+p(r)=0$, except at the boundary surface, since there
generically the energy density is nonzero, $\rho(r_0)\neq0$, and the
pressure is zero, $p(r_0) = 0$. There is an exception: at the point
$\frac{q^2}{R^2}=3$, point $D$, the energy density is zero,
$\rho(r_0)=0$, and since the pressure is also zero, the enthalpy is
zero, so that the enthalpy is zero throughout the matter region up to
and including the boundary $r_0$, making point $D$ a special point.
This is a line of interest for the stability problem.

Point $D$ is a special
and very interesting point. It represents a regular de Sitter
black hole with an electric charge coat at the boundary. This boundary
is a lightlike surface at $r_0=r_-$. The interior solution is pure de
Sitter. For $0\leq \frac{r}{R}< \frac{r_0}{R}$, the energy density is
given by $8\pi\,\rho(r) = \frac{3}{R^2}$ and the pressure by $8\pi\,
p(r) = -\frac{3}{R^2}$, so that $\rho+p=0$, i.e., a cosmological
constant equation of state is verified in this region.  For
$\frac{r}{R}= \frac{r_0}{R}$, i.e., $r=r_0$, the energy density is
given by $8\pi\,\rho(r_0) = 0$ and the pressure by $8\pi\, p(r) =0$,
so that obviously $\rho+p=0$ in this surface, a feature that does not
happen for the other black holes on the line $r_0=r_-=R$.  Thus, from
the interior up to the boundary itself, the enthalpy is zero,
$\rho+p=0$.  This solution is one of the regular black
holes found in \cite{LemosZanchin2011}.  Point $D$
has physical interest in itself
and surely is of interest for
the stability problem.

Point $Q$ from the right is also
extraordinarily interesting. It represents
quasinonblack holes, which we abbreviate as
QNBHs from now onward.  It is degenerated since there exist many solutions
with different physical and geometrical properties for the same
parameters, $\frac{r_0}{R}=1$ and $\frac{q^2}{R^2}=1$, depending on
the path followed from the right
to approach that point.  The solutions
have different characteristics
depending if point $Q$
is reached from regions (d2), (e1), or (e2).  All these differences
will be reflected in the stability analysis. We study in detail
below the QNBH limits.  Point $Q$ from the right
is a point
of high interest for the stability problem, as a limiting point.

Line $\dfrac{r_0}{R}=\dfrac{r_+}{R}=1$ is the segment of the line
$\frac{r_0}{R}=1$ with the electric charge in the interval
$0<\frac{q^2}{R^2}<1$. It contains singular objects.  This is a line
of no special interest for the stability problem.

Region (f) contains singular undercharged solutions, so $m^2> q^2$.
These are all objects for which the energy density and the
    pressure diverge at some $r$ inside the matter distribution.
This is a region of no
interest for the stability problem.

Line $C_4$ is the Buchdahl-Andr\'easson bound line characterized by
the central pressure of any object lying on this line being infinite.
One of the endpoints of this line is point $B$, the Buchdahl
bound point. 
This is a line of interest for the stability problem, as a limiting
line.

The description ends here, as the next line would be the vertical axis
that we have already described.

\subsection{The quasiblack hole 
and quasinonblack hole limits}
\label{sec:qbh}

\subsubsection{The five distinct quasiblack holes 
and quasinonblack holes}

As pointed out in~\cite{LemosZanchin2010}, the solutions
we are treating
admit QBHs.
This happens when the mass $m$ approaches the
electric charge $q$, $m^2\rightarrow q^2$,
and the boundary radius $r_0$ approaches
the gravitational radius $r_+$, $r_0\to r_+$,
or equivalently
$\frac{q^2}{R^2} \to 1$ and $\frac{r_0}{R} \to 1$, so that one reaches
the point $Q$ of Fig.~\ref{fig:regions}.
As also
pointed out in~\cite{LemosZanchin2017}, the point $Q$ is a
degenerate point which represents several different kinds of
objects. However, one has to distinguish when
$Q$ is approached from the left, which can give
rise to QBHs, from when 
$Q$ is approached from the right, which can give
rise to QNBHS.  QNBHs have
their own properties
distinct from the properties of QBHs,
as found in \cite{lemosluz2021}. 

QBHs are obtained by compressing star-like
configurations with radius $r_0$
quasistatically to the gravitational radius $r_+$,
$r_0\to r_+$.
In this limiting process one
arrives at point $Q$ from the
left, i.e., from the region of the parameter space for which
$\frac{q^2}{R^2} <1$,
and also leads to $r_0\to r_+ \to r_-$. The result is a starlike
configuration on the verge of being an extremely charged
Reissner-Nordstr\"om black hole, but instead becoming an
 extremely charged
Reissner-Nordstr\"om QBH.

QNBHs are obtained by decompressing regular black hole configurations
for which the radius $r_0$ is smaller than the inner radius $r_-$,
$r_0<r_-$, up to the radius $r_-$, $r_0\to r_-$.  In this limiting
process one arrives at point $Q$ from the right, i.e., from the region
of the parameter space for which $\frac{q^2}{R^2} >1$, and leads to
$r_0\to r_- \to r_+$. The result is a regular black hole on the verge
of not being an extremely charged regular Reissner-Nordstr\"om black
hole, but instead becoming an extremely charged Reissner-Nordstr\"om
QNBH.

From the analysis of the liming procedure on the several different
classes of regular objects just performed,
one finds conclusively that
QBH and QNBH configurations may be obtained.  This can also be
confirmed by direct inspection of Fig.~\ref{fig:regions}. We now
enumerate and describe
the five types that are obtained
in the limiting procedure,
with three types being 
QBHs and two types being QNBHs.

\vskip 0.2cm
\noindent
(i) QBHs from regular undercharged pressure stars:
These arise from region (a),
with
the parameter $a$ in the range
$1< a < 8$ approximately, below the line $C_4$ and above
the line $C_0$ in Fig.~\ref{fig:regions}.
These
QBHs form
from distributions of charged matter for which
the electric repulsion is less than the gravitational attraction and
there is matter pressure, $p>0$.  The resulting objects are pressure
QBHs, with $m^2=q^2$.
They are nonsingular, no curvature
invariant diverges for the whole spacetime. 
In the figure
this case corresponds
to taking the limit to the point $Q$ from region (a), which means
$a>1$ and $\frac{q^2}{R^2}< \frac{r_0^4}{R^4}\leq 1$ with the equality
holding just at $Q$.  These configurations satisfy all the energy
conditions and the causality condition as long as
$1< a< \frac43$.  This
type of QBHs has been
investigated in detail in~\cite{LemosZanchin2010}.
 
\vskip 0.2cm
\noindent
(ii) QBHs from extremal charged dust stars:
These arise from line $C_0$
in Fig.~\ref{fig:regions}, have $a=1$, which means 
$\frac{q^2}{R^2} = \frac{r_0^2}{R^2}\left(2 -
\frac{r_0^2}{R^2} - 2 \sqrt{1-\frac{r_0^2}{R^2} }\right)$.
These configurations
follow from distributions of extremal charged dust, for which the
electric repulsion counterpoises the gravitational attraction and
there is no matter pressure, $p=0$.  The resulting objects are
extremal QBHs, with with $m^2=q^2$.
They are nonsingular, no curvature
invariant diverges for the whole spacetime.  In
the figure, this case corresponds to taking the limit to
the point $Q$ along the curve $C_0$.
This type of QBHs has been investigated
in~\cite{LemosWeinberg2004}, see also
\cite{lemoszanchin2008}.

\vskip 0.2cm
\noindent
(iii) QBHs from overcharged tension stars:
These arise from
region (b), with $0< a < 1$
and $\frac{q^2}{R^2}< \frac{r_0^4}{R^4}$,
between lines $C_0$ and $C_1$
in Fig.~\ref{fig:regions}.
These configurations follow from distributions of charged matter, for
which the electric repulsion is greater than the gravitational
attraction and there is matter tension, $p<0$.  The resulting objects
are tension QBHs, with $m^2=q^2$.
They are nonsingular, no curvature
invariant diverges for the whole spacetime. 
This type of
QBHs has been investigated in~\cite{LemosZanchin2010}.

\vskip 0.2cm
\noindent
(iv)
QNBHs from regular phantom black holes:
These arise from
regions (d2) and (e1), with $1< a < 4$, and $\frac{q^2}{R^2}\geq 1$,
in Fig.~\ref{fig:regions}.
These configurations follow from regular electrically
charged black holes, for
which the matter is phantom, i.e., $\rho + p<0$ everywhere.  The
resulting objects are regular phantom QNBHs, with $m^2=q^2$.  In
the figure  this case corresponds to taking the limit to
the point $Q$ from regions (d2) and (e1).
This type of QNBHs has not been investigated,
see \cite{lemosluz2021} for an example of QNBHs.

\vskip 0.2cm
\noindent
(v)
QNBHs from regular normal black holes: These arise from region (e2),
with $a> 4$ approximately in Fig.~\ref{fig:regionszoom} which is a
zoom of Fig.~\ref{fig:regions} to see this region.  These
configurations follow from regular electrically charged regular black
holes, for which the matter is normal, i.e., $\rho + p>0$ everywhere,
with the pressure being negative.
The resulting objects are regular tension QNBHs, with $m^2=q^2$.  In
the Fig.~\ref{fig:regionszoom} this case corresponds to taking the
limit to the point $Q$ from region (e2). This type of QNBHs has not
been investigated, see \cite{lemosluz2021} for an example of QNBHs.

\subsubsection{Taking the limits to obtains  quasiblack holes 
and quasinonblack holes}

The Guilfoyle parameter $a$,
which by  Eq.~\eqref{eq:a}
is a function of $\frac{q^2}{R^2}$
and $\frac{r_0}{R}$ by Eq.~\eqref{eq:a}, is not well defined in the
limit to the point $Q$. In fact, the parameter $a$ may assume any 
value there, depending on the path followed in the
parameter space to reach that point.
To see this we write $\frac{q^2}{R^2} =
\left(1\pm2\sqrt{\varepsilon}\right)\frac{r_0^4}{R^4}$ and
$\frac{r_0^2}{R^2} = 1 -\delta$, where $\varepsilon$ and $\delta$ are
small nonnegative parameters. Upon substituting these expansions into
Eq.~\eqref{eq:a} one gets $ a= \frac{\varepsilon}{\delta}$ up to the
correct order.  Thus, clearly, in 
the limits $\varepsilon\to 0$ and
$\delta\to 0$ the parameter
$a(\frac{q^2}{R^2},\frac{r_0}{R})$ is
not a well defined function.
It follows that,
by parameterizing the problem in
terms of $\frac{q^2}{R^2}$
and $\frac{r_0}{R}$, it is
difficult to keep control of the values of $a$ during numerical
calculations when
approaching the point $Q$. This
control 
is necessary to analyze the stability
conditions of the QBH and QNBH limits within
each region of Fig.~\ref{fig:regions}
near the point $Q$.

In order to avoid such a lack of control, we should choose a
particular relation between $\varepsilon$ and $\delta$, $\varepsilon =
\varepsilon(\delta)$, and
in doing so a specific path has been chosen in
the parameter space. This is equivalent to choose a specific relation
between the two independent parameters $\frac{q^2}{R^2}$ and
$\frac{r_0}{R}$ and letting $a$ as a free parameter, as it was done in
\cite{LemosZanchin2010}.  To follow this rationale, we need to write
$\frac{q^2}{R^2}$ as a function of $a$ and $\frac{r_0}{R}$, which can
be done by means of Eq.~\eqref{eq:a}.
To proceed, we write
\begin{equation}
\frac{r_0^2}{R^2}= 1- \delta, \label{eq:delta}
\end{equation} with $\delta$ being a small positive number and
with Eq.~\eqref{eq:delta}
being valid in first order in $\delta$, i.e., in a region
close to the point $Q$.
With this assumption, the leading terms in the expression for
$\frac{q^2}{R^2}$ obtained from Eq.~\eqref{eq:a}
may be written as $\frac{q^2}{R^2}=  1 \pm
2\sqrt{a\,\delta} + 2\left(a-1\right) \delta$, or equivalently
$\frac{q^2}{R^2}=\left[1 \pm
2\sqrt{a\,\delta} + \left(2a-1\right)\delta\right]\frac{r_0^2}{R^2}$.
Since $\delta$ is small
and arbitrary, for finite $a$ we may write $a\,\delta \equiv
\varepsilon$, i.e., Eq.~\eqref{eq:a} together with 
Eq.~\eqref{eq:delta} leads to
\begin{equation}\label{eq:epsilon}
  \frac{q^2}{R^2}= \left[1
  \pm 2\sqrt{\varepsilon}+
  \left(2a-1\right)\frac{\varepsilon}{a}\right]\frac{r_0^2}{R^2} ,
 \end{equation}
 where $\varepsilon$ is a small nonnegative parameter given in terms
 of $a$ and $\delta$ by
 \begin{equation}
  \varepsilon = a \delta. \label{eq:epsilon-delta}
 \end{equation}
This relation means that the point $Q$ is approached by following
straight lines in the parameter space, with the $\pm$ signs indicating
if one reaches that point from the right side
or from the left side.  Since each
constant $a$ defines a curve in the parameter space, and all the
curves for different values of $a$ reach the point $Q$, the parameter
$a$ is the appropriate
parameter
to be used as a free parameter for the present
analysis.  Hence, from now on we
choose as free parameters $a$ and $\frac{r_0}{R}$, instead of
$\frac{q^2}{R^2}$ and $\frac{r_0}{R}$, with intervals $0< a< \infty$
and $0\leq \frac{r_0}{R}\leq 1$.
Moreover, on one hand,
the minus sign in Eq.~\eqref{eq:epsilon} indicates
paths approaching the point $Q$ from the left, i.e, with
$\frac{q^2}{R^2} <1$, which contains
star configurations in
regions (a) and (b), and other singular objects in regions (c) and
(f).  In this case, the limits $\varepsilon\to 0$ and $\delta\to 0$
take the radius
of the object under consideration, be it a star or a singular
configuration,
to the limit of the gravitational radius, i.e.,
almost to a black hole which is 
the QBH limit \cite{LemosWeinberg2004}.
On the other hand, the plus sign in Eq.~\eqref{eq:epsilon} indicates paths
approaching the point $Q$ from the right, i.e., with $\frac{q^2}{R^2} >1$,
which corresponds to other singular objects in the region (c), and
black hole configurations in the regions (d2), (e1), and (e2).
In this case, the limits $\varepsilon \to 0$ and $\delta \to 0$ take
the boundary matter in region (c) to the limit of the gravitational
radius, i.e., almost to a black hole which is the QBH limit [6], and
take the boundary matter in the regions (d2), (e1), and (e2) to the
Cauchy horizon radius which is equal to the event horizon radius,
i.e., to the QNBH limit \cite{lemosluz2021}.
Then, we rewrite the relevant equations of the model in
terms of $a$ and $\frac{r_0}{R}$ up to first order in $\varepsilon$
and, at the end, take the limit $\frac{r_0}{R}\to 1 -\frac\delta2$, with
$\delta$ related to $\varepsilon$ through Eq.~\eqref{eq:epsilon-delta}.
For instance, one finds that, at the lowest orders in $ \varepsilon$,
Eq.~\eqref{eq:m} implies in $\frac{m}{r_0}= 1
\pm \sqrt\varepsilon+\left(a-1\right)\frac{\varepsilon}{a} $ which,
together with Eq.~\eqref{eq:epsilon}, gives $\frac{m^2}{q^2}=
1 + \frac{a-1}{a}\varepsilon$.
Notice then that one gets $\frac{m^2}{q^2}<1$ for $0<a<1$ as expected,
and $\frac{m^2}{q^2}\geq1$ for $a\geq1$ as also expected.
Using the same procedure one also finds that
the constants $k_0$ and $k_1$
that appear in the expressions for the metric functions,
matter functions, and electric functions are
$k_0 = \mp a^{\frac1a}\varepsilon^{\frac{a-2}{2a}}$, and
$k_1=\pm \frac{\sqrt a\, k_0}{2-a}=-\frac{a^{\frac{a+2}{2a}} }{2-a}
\varepsilon^{\frac{a-2}{2a}}$. All equalities
are approximate equalities, correct up to the first order
in the expansion.

We now find the expressions for the metric potentials,
the matter functions, the electric
potential, and the electric charge,
when the configurations approach
the QBH or the QNBH limits.
Taking the expansions given in
Eqs.~(\ref{eq:delta})-\eqref{eq:epsilon-delta} 
and the approximations for $k_0$ and $k_1$ presented in the last paragraph 
into the corresponding equations for the metric potentials,
Eqs.~\eqref{eq:pmAex}-\eqref{eq:pmBex}, we find  
\begin{eqnarray}
& A(r)  = \left(1-\dfrac{r^2}{r_0^2}\left[1-
\dfrac{\varepsilon}{a}\right]\right)^{-1}, \label{A(r)-qbh}  \\
& B(r) = \left(1 \mp\dfrac{2-a}{\sqrt{a\,}}
\sqrt{1-\dfrac{r^2}{r_0^2}}
\right)^{\!\!\frac{2a}{a-2}}\!\!\!
\dfrac{\varepsilon}{a}\,, \label{B(r)-qbh}
\end{eqnarray}
where we have written $R$, $m$, and $q$ in terms of $a$, $r_0$,
and $\varepsilon$. 
All equalities are valid up to first order.
Now, taking the expansions given in
Eqs.~(\ref{eq:delta})-\eqref{eq:epsilon-delta} into the corresponding
equations for the fluid quantities, namely, the energy density and
the pressure, Eqs.~\eqref{eq:denEnerIn}-\eqref{eq:pressure}, one
finds
\begin{eqnarray}\!\!\! 8\pi\rho(r)& = & \dfrac{3}{r_0^2} -
\dfrac{r^2}{r_0^4\,} \left(\frac{2-a}{\sqrt{a\,}}\,
\sqrt{1-\dfrac{r^2}{r_0^2}}\mp 1\right)^{\hskip-0.15cm-2} ,
\label{rhom-qbh} \\
8\pi p(r) &\!=\!& -\frac{1}{r_0^2} + \frac{r^2}{r_0^4\,}
\left(\frac{2-a}{\sqrt{a\,}}\, \sqrt{1-\dfrac{r^2}{r_0^2}} \mp 1
\right)^{\hskip-0.15cm-2}
\nonumber\\ & &
\hspace{-1.2cm} + \dfrac{2\,\sqrt{a\,} }{r_0^2}\sqrt{1-\dfrac{r^2}{r_0^2}}\,
\left(\frac{2-a}{\sqrt{a}}\sqrt{1-\dfrac{r^2}{r_0^2}}
\mp 1\right)^{\hskip-0.15cm -1}.\label{p-qbh}
\end{eqnarray}
These are the zeroth order approximations for $\rho(r)$
and $p(r)$ in which
$\frac{r_0^2}{R^2} =1$ and,
as a consequence, $m^2=q^2=r_0^2$.

Taking the expansions into the corresponding equation for the
electric potential of the interior region, Eq.~\eqref{eq:phi}, one finds
\begin{eqnarray}
\phi(r) = \epsilon\left(1 \mp \dfrac{2-a}{\sqrt{a\,}}
\sqrt{1-\dfrac{r^2}{r_0^2}}\right)^{\!\!\frac{a}{a-2}}\!\!\!
\dfrac{\sqrt{\varepsilon}}{a}\,.   \label{phi(r)-qbh}
\end{eqnarray}
Taking the expansions given in
Eqs.~(\ref{eq:delta})-\eqref{eq:epsilon-delta} into 
the corresponding equation for the electric charge of the interior
region, Eq.~\eqref{eq:cargain},
one finds
\begin{eqnarray}
Q(r) =  \dfrac{\epsilon  r^3}{r_0^2} \left(\frac{2-a}{\sqrt{a\,}}\, \sqrt{1-
\dfrac{r^2}{r_0^2}}\mp 1\right)^{\hskip-0.15cm -1} .
\label{charge-qbh}
\end{eqnarray} 
that is also a zeroth order approximation such that $r_0=R$,

Similarly, the approximated expressions of all quantities related to
the exterior solution are obtained. The corresponding limits for the
metric potentials of the exterior region solution, with $r\geq r_0$,
can be obtained.
The potential $A(r)$ in 
Eq.~\eqref{eq:A(r)ext} is
then 
\begin{eqnarray}
\hskip -0.2cm A(r) \hskip -0.15cm = \hskip -0.15cm \left[
\left(1
\hskip -0.05cm 
-
\hskip -0.05cm 
\dfrac{r_0}{r}
\left[1\pm\sqrt{\varepsilon}\right]\right)^{\!\! 2}
\hskip -0.05cm 
-
\hskip -0.05cm 
\dfrac{r_0}{r}\left(2
\hskip -0.05cm 
-
\hskip -0.05cm 
\dfrac{r_0}{r}\right)\dfrac{a-1}{a}\varepsilon
\right]^{-1}\hskip -0.3cm,
\label{Aext-qbh}
\end{eqnarray}
and 
the potential $B(r)= \dfrac{1}{A(r)}$ in 
Eq.~\eqref{eq:B(r)ext} is
then 
\begin{eqnarray}
\hskip -0.3cm  B(r)=
\left(1 - \dfrac{r_0}{r}
\left[1\pm\sqrt{\varepsilon}\right]\right)^{\!\! 2}
 -\dfrac{r_0}{r}\left(2-\dfrac{r_0}{r}\right)
 \dfrac{a-1}{a}\varepsilon.
\label{ABext-qbh}
\end{eqnarray}
Taking the expansions into the corresponding equation for the
electric potential of the exterior region, Eq.~\eqref{eq:phiext}, one finds
\begin{eqnarray}
\phi(r) = \epsilon \left(\dfrac{r_0}{r}-1\right)\left(1\pm
\sqrt{\varepsilon}\right)+ \epsilon\dfrac{\sqrt{\varepsilon}}{a},
\label{phiext-qbh}
\end{eqnarray}
where the integration constant has been adjusted so that the function
in  \eqref{phiext-qbh} equals the function for the interior electric
potential given in \eqref{phi(r)-qbh} at the boundary.

The approximate relations given in this section hold for both
QBH and QNBH cases, with the lower sign in $\pm$ or $\mp$ holding
for QBH configurations while the upper sign holds for QNBH configurations.
QBHs occur for $a>1$ with $\frac{q^2}{R^2}=1-2\sqrt{\varepsilon}$ and
for $0< a\leq 1$ with $\frac{q^2}{R^2} = 1-
2\sqrt{\varepsilon\,}$.
QNBHs occur for $a>1$ with $\frac{q^2}{R^2} =
1 +2\sqrt{\varepsilon\,}$.
The relations given in Eqs.~\eqref{rhom-qbh}, \eqref{p-qbh}, and
\eqref{charge-qbh} are the zeroth order approximations in
$\sqrt{\varepsilon\,}$. The other relations are first order
approximations in $\sqrt{\varepsilon\,}$.

\subsection{Summary of the plethora of solutions}

Within the electrically charged spherically symmetric solutions
presented here, there are many solutions of interest, either because
they may represent actual objects within the physical universe, or
they have in themselves interesting physical features, like a rich
causal behavior, relevant matter characteristics, or some other
important aspect. Almost all these solutions have a core of
electrically charged matter and a Reissner-Nordstr\"om exterior,
excluding some degenerate cases that we have mentioned.

A sketch of all solutions that naturally appeared within the class
studied is given in Table \ref{summarytable}.
A concise description of these solutions is now given.
With respect to objects that can be classified as stars, i.e., star
solutions, there is a list that we should refer to.  There are the
interior Schwarzschild solutions, i.e., Schwarzschild stars, the first
member of this family being a black hole, and the last member of the
family being the Schwarzschild star saturating the Buchdahl
bound. There are also uncharged singular star solutions.  There are
undercharged stars, the last members of this family are stars
saturating the Buchdahl-Andr\'easson bound.  There are also
undercharged singular star solutions.  There are extremally charged
objects, i.e., the Bonnor stars, the exterior being an extremely
Reissner-Nordstr\"om spacetime. There are tension overcharged stars.
There are QBHs that appear in distinct forms, namely, pressure QBHs,
pressureles QBHs, and tension QBHs.  There are also singular
overcharged objects and singular extremely charged objects.  There are
Kasner like objects, which are highly singular.
With
respect to objects that can be classified as regular black holes there
is the following list. There are regular black holes with negative
energy densities, regular black holes with a central core of charged
phantom matter, regular tension black holes with positive enthalpy
density, and there is a regular de Sitter black hole with an electric
charge coat at the boundary. There are QNBHs.  There is also a number
of nonregular black holes.  All these different solutions are found
within the class of the Guilfoyle solution presented above.

The stability of an object and of a solution is an important feature
that it must possess in order to be considered of relevance in the set
of natural objects. Thus, we now turn to the stability problem of
these objects.

\begin{widetext}

\vskip 0.8cm

\begin{table}[htbp]
\begin{ruledtabular}
\begin{tabular}{l l l}
\textrm{Configurations }& \textrm{ Features} &
\textrm{Location in the parameter space}\\
\colrule
Schwarzschild stars & uncharged, regular & Line $\frac{q^2}{R^2}=0$,
$0< \frac{r_0}{R}
< \frac{2\sqrt{2}}{3}$ \\
Buchdahl limit & singular Schwarzschild star & Point $B$: $\frac{q^2}{R^2}=0$,
$\frac{r_0}{R} =\frac{2\sqrt{2}}{3}$ \\
Undercharged stars & regular, $q^2 < m^2$ & Region ($a$) \\
Buchdahl-Andréasson limit & singular undercharged star & Line $C_4$\\
Undercharged stars & singular, $q^2< m^2$ & Region ($f$) \\
Extremely charged stars & regular, $q^2=m^2$ & Line $C_0$\\
Overcharged tension stars & regular, $q^2 > m^2$, $-1 < \frac{p}{\rho} <0$ &
Region ($b$)\\
Overcharged tension stars & singular, $q^2 > m^2 $ & Line $C_1$ and
region ($c$)\\
Extremely charged stars & singular, $q^2=m^2$ & Line $C_2$\\
Regular phantom black holes & $q^2< m^2 $, phantom matter: $\frac{p}{\rho}< -1$ &
Regions ($d_1$) and ($d_2$), and line $C_{31}$\\
Regular phantom black holes & $q^2< m^2 $, phantom matter: $-1<\frac{p}{\rho}< 0$,
$\rho<0$ & Region ($e_1$) and line $C_{32}$ \\
Regular tension black holes & $q^2< m^2 $, $-1 < \frac{p}{\rho} < 0$, $\rho>0$
& Region ($e_2$) \\
Regular de Sitter black hole & $q^2< m^2 $, $ \frac{p}{\rho} = -1$, $\rho>0$ &
Point $D$: $\frac{q^2}{R^2}=3$, $\frac{r_0}{R}=1$ \\
Regular pressure quasiblack holes & $q^2= m^2 $, regular pressure
core: $ \frac{p}{\rho} >0 $ & Point Q from region ($a$) \\
Singular quasiblack holes & $q^2= m^2 $, singular pressure core: $
\frac{p}{\rho}>0 $ &Point Q from region ($f$) and line $C_4$ \\
Regular quasiblack holes & $q^2= m^2 $, regular pressureless core: $
\frac{p}{\rho} =0 $ & Point Q from line $C_0$ \\
Regular tension quasiblack holes & $q^2= m^2 $, regular tension core: $
-1 <\frac{p}{\rho} < 0 $ & Point Q from region ($b$)\\
Singular quasiblack holes & $q^2= m^2 $, singular tension core: $ -1 <
\frac{p}{\rho} <0 $ & Point Q from line $C_1$ and region ($c$)\\
Singular quasiblack holes & $q^2= m^2 $, singular pressureless core: $
\frac{p}{\rho} =0 $ & Point Q from line $C_2$\\
Regular quasinonblack holes & $q^2= m^2 $, phantom matter:
$\frac{p}{\rho}< -1$ &Point Q from region ($d_2$) \\
Regular quasinonblack holes & $q^2< m^2 $, phantom matter: $-1<\frac{p}{\rho}<
0$, $\rho<0$ & Point Q from region ($e_1$) and line $C_{32}$\\
Regular tension quasinonblack holes & $q^2< m^2 $, $-1 < \frac{p}{\rho} < 0$,
$\rho>0$ & Point Q from region ($e_2$) \\
Kasner spacetimes & & Line $\frac{r_0}{R}=0$, $\frac{q^2}{R^2}>0$ \\
\end{tabular}
\end{ruledtabular}
\caption{The plethora of solutions.}
\label{summarytable}
\end{table}

\vskip 0.8cm

\end{widetext}

\centerline{}
\newpage

\centerline{}
\newpage

\section{Stability analysis of the electric charged spheres: Results
for regular stars, regular black holes,  quasiblack holes,
and quasinonblack holes}
\label{sec:results}

\subsection{The stability of regular stars}

\subsubsection{Zero electric charged stars:
$\frac{q^2}{R^2}=0$, i.e.,  Schwarzschild stars}
\label{Sec:USS}

The Schwarzschild star solutions, composed of a Schwarzschild interior
and a Schwarzschild exterior vacuum solution, are given by
$\frac{q^2}{R^2}=0$ with variable $\frac{r_0}{R}$.  The expressions
for the metric potentials, the fluid quantities, and the electric
quantities, are obtainable from Guilfoyle's solution,
see~\cite{LemosZanchin2017}.  In Fig.~\ref{fig:regions} these
Schwarzschild stars correspond to the vertical axis.

In this case, the energy density and the pressure are positive
functions everywhere inside the matter. Thus, one finds
that the enthalpy
$h(r)=\rho(r)+p(r)>0$ and, as a consequence,
assuming that the adiabatic index obeys
$\gamma> 0$, the
coefficients $F(r)$ and $W(r)$ that appear in Eq.~\eqref{eq:SLP} are
both positive functions, and so in the SL problem this leads to the
case (A) of the theorem given in Appendix \ref{sec:SLP}. Therefore,
stable solutions to radial perturbations
are found for positive adiabatic indices such that
$\gamma>\gamma_{\rm cr}$, where $\gamma_{\rm cr}$ is
the critical value, the
minimum value of $\gamma$ for which the solution is stable,
see~\cite{Chandre1964b}.
We note here, in passing, that for numerical analysis the frequency
$\omega$ is normalized as $\omega R$. From this section onward,
including all the figures dealing with stability, we drop $R$ to
simplify notation.

\begin{figure}[h] 
\centering
\includegraphics[width=0.235\textwidth]{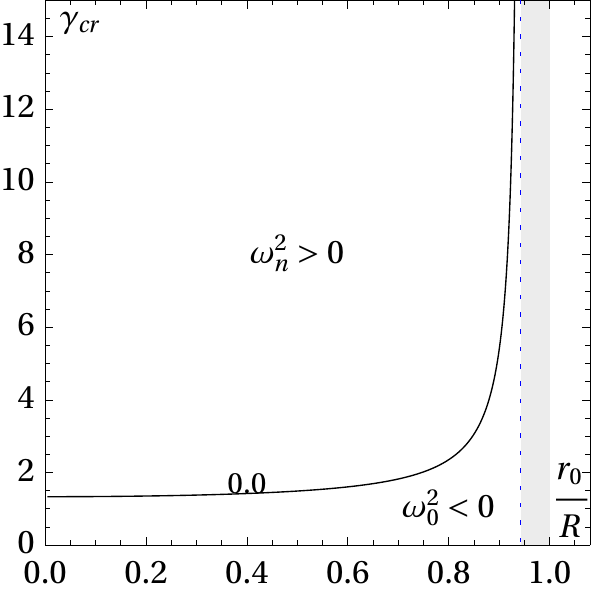}
\caption{Stability of zero charge stars, i.e.,
Schwarzschild stars. These
stars are on the vertical line $\frac{q^2}{R^2}=0$ of
Fig.~\ref{fig:regions}. The critical adiabatic index $\gamma_{\rm cr}$,
which gives the $\gamma$ for which $\omega_{0}^{2}=0$, is shown as a
function of the radius $\frac{r_0}{R}$. The region above the line is
stable to radial perturbations, the region below the line is unstable.
It is seen that $\gamma_{\rm cr}$ starts at $\frac43$, and
as the
star gets more compact the $\gamma_{\rm cr}$ gets higher and
higher.  The
light gray region corresponds to stars that are not regular and are
beyond the Buchdahl limit.}
\label{fig:unchargedNormalStars}
\end{figure}

In Fig.~\ref{fig:unchargedNormalStars} we show the numerical results
for the critical adiabatic index $\gamma_{\rm cr}$ as a function of
the normalized radius of the star $\frac{r_0}{R}$ for zero electric
charge, $\frac{q^2}{R^2}=0$.
The vertical axis  bounds the plot on the left.
The vertical dotted line on the right in
the plot is the Buchdahl bound \cite{AndreassonQ}, see also
\cite{LemosZanchin2015}, which is represented by point $B$ in
Fig.~\ref{fig:regions}. In the plot there is the white region that
represents the range of the parameter $\frac{r_0}{R}$ where regular
Schwarzschild stars are found, and the light gray region that contains
Schwarzschild stars that are singular.  The solid line drawn is for
the vanishing fundamental oscillation frequency squared, i.e., for
$\omega_{0}^{2}=0$, which means that $\omega_{0}^{2}$ changes sign
across this curve. All configurations represented by points located
above the $\omega_{0}^{2}=0$ line are stable stars, i.e., all
$\omega_{n}^{2}$ are positive, all configurations represented by
points located below the $\omega_{0}^{2}=0$ line are unstable stars.
The solid line starts at $\frac{r_0}{R}=0$ and extends to point
$B$, the Buchdahl limit, given by $\frac{r_0}{R}=\frac{2\sqrt{2}}{3}=
0.943$, where this last equality is approximate, and where
$\gamma_{\rm cr}$ diverges.
Let us comment in more detail on these configurations and their
stability.
The
limit $\frac{r_0}{R}=0$ for zero charge stars
means that there is no star.
Indeed, for $R$ fixed, taking the limit of
$r_0$ going to zero means that the mass of the stars goes
to zero sufficiently fast so that in the 
$r_0=0$ limit there is no mass and so no star. But since 
$R$ is fixed, and so the effective density is fixed,
although there is no mass, no star, and no gravity,
there is a fluid, and this means that the spacetime is that of
a fluid composed of test fluid elements
in Minkowski spacetime. 
In this case to be stable the lowest $\gamma_{\rm cr}$
is the $\gamma_{\rm cr}$ for a fluid in the laboratory,
with no gravity, and it is $\gamma_{\rm cr}=\frac43=1.33$,
where this last equality is approximate.
It is worth noting that such an interpretation
can be given only
after the stability analysis is made, 
because
only then it is possible to understand that
in this limit there is a test fluid
in a Minkowski spacetime rather than pure
empty Minkowski spacetime.
At the other end of the plot, at the point $B$
in
Fig.~\ref{fig:regions}, i.e., for
$\frac{r_0}{R}= \frac{2\sqrt{2}}{3}= 0.943$,
where this latter value is an approximate value, 
it can be taken to mean that for some $R$ fixed,
and since $R$ is the inverse of the effective
density,  for some fixed effective
density, there is a sufficiently high $r_0$
that makes the star relatively large but compact.
It is is indeed a 
Schwarzschild star at the Buchdahl limit.
In this case to be stable a very high $\gamma_{\rm cr}$
is necessary, in the limit $\gamma_{\rm cr}$
has to be infinite
to provide a stable star against radial perturbations.
Since in this picture
we are fixing $R$ and so the effective density
of the star,
it is the way of considering a compact star
as Michell and Laplace have done, namely,
the density of the star is given and fixed,
the star has relatively
large mass and  large radius, but is
in all measures compact.

\begin{figure}[h] 
\centering
\includegraphics[width=0.235\textwidth]{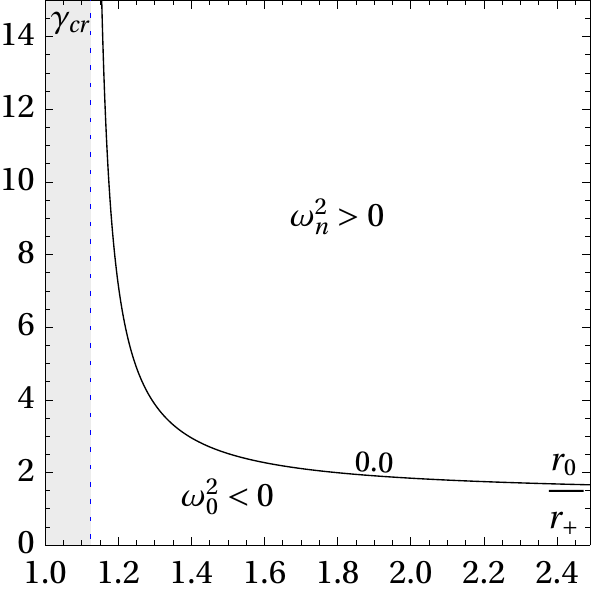}
\caption{Stability of zero charge stars, i.e., Schwarzschild
stars. These stars are on the vertical line $\frac{q^2}{R^2}=0$ of
Fig.~\ref{fig:regions}. The critical adiabatic index $\gamma_{\rm
cr}$, which gives the $\gamma$ for which $\omega_{0}^{2}=0$, is shown
as a function of the radius $\frac{r_0}{r_+}$. The region above the
line is stable
against
radial perturbations, the region below the line is unstable.  It is seen
that for a Schwarzschild star at the Buchdahl limit $\gamma_{\rm cr}$
is unlimited and then decreases up to $\frac43$.  The light gray
region corresponds to stars that are not regular and are beyond the
Buchdahl limit.}
\label{fig:unchargedNormalStarsr+}
\end{figure}

In Fig.~\ref{fig:unchargedNormalStarsr+} we show the numerical results
for the critical adiabatic index $\gamma_{\rm cr}$ as a function of
the normalized
radius $\frac{r_0}{r_+}$ for zero electric charge, namely,
$\frac{q^2}{R^2}=0$. It is interesting to show this new plot of
$\gamma_{\rm cr}$ as a function of $\frac{r_0}{r_+}$ as some features
are highlighted and complementary to the plot of
Fig.~\ref{fig:unchargedNormalStars}, noting that $\frac{r_0}{r_+}$ and
$\frac{r_0}{R}$ are convertible from one to the other.  
The vertical dotted line on the left in the plot is the Buchdahl bound
\cite{AndreassonQ}, see also \cite{LemosZanchin2015}, which is
represented by point $B$ in Fig.~\ref{fig:regions}.
On the right the plot extends to infinity.
In the plot there
is
 the
light gray region that contains singular Schwarzschild stars,
and the
white region that represents the range of parameter
$\frac{r_0}{r_+}$ where regular Schwarzschild stars are found.  The
solid line drawn is for the vanishing fundamental oscillation
frequency squared, i.e., for $\omega_{0}^{2}=0$, which means that
$\omega_{0}^{2}$ changes sign across such a curve. All configurations
represented by points located above the $\omega_{0}^{2}=0$ line are
stable stars, i.e., all $\omega_{n}^{2}$ are positive, all
configurations represented by points located below the
$\omega_{0}^{2}=0$ line are unstable stars.
The solid line starts at $\frac{r_0}{r_+}=\frac98$ that corresponds to
Buchdahl bound, and extends to $\frac{r_0}{r_+}$ infinitely large.
Let us comment in more
detail on these configurations and their stability.  The limit
$\frac{r_0}{r_+}=\frac98$ means that the radius of the star is very
compact, indeed it is a Schwarzschild star at the Buchdahl limit,
almost at the $r_0=r_+$ QBH limit.  In this case to be
stable a very high $\gamma_{\rm cr}$ is necessary, in the limit
$\gamma_{\rm cr}$  has
to be infinite to provide a stable star
against radial perturbations.
Since in this
picture we are fixing $r_+$, and so the spacetime mass, it is the way
of considering a compact star as it is nowadays usually done, as for
instance in the work of Chandrasekhar~\cite{Chandre1964b}.  With the
parameter $\frac{r_0}{r_+}$ one gets the compactness of the star
immediately.
At the other end, for
$\frac{r_0}{r_+}$ indefinitely large, one has that the radius of the
star is very large compared with $r_+$ and so the star is extremely
disperse.  In the limit that $r_0$ is infinite there is a fluid made
of test fluid elements in a Minkowski background.  In this case to be
stable the lowest $\gamma_{\rm cr}$ is the $\gamma_{\rm cr}$ for a
fluid in the laboratory, with no gravity, and it is $\gamma_{\rm
cr}=\frac43=1.33$, where this last equality is approximate.  Again,
this interpretation can be given only after the stability analysis is
made, because only then
it is possible to
understand that in this limit there is
a test fluid in a Minkowski spacetime rather than pure empty Minkowski
spacetime.

In Table~\ref{tab:zerocharge} we give details of the numerical results
for the stability of the Schwarzschild stars, i.e., zero charged
stars.  The behavior of $\gamma_{\rm cr}$ as a function of the radius
$\frac{r_0}{R}$ and $\frac{r_0}{r_+}$, for $\frac{q^2}{R^2}=0$, is
displayed.
\begin{table}[htbp]
\begin{ruledtabular}
\begin{tabular}{c c c c c}
\textrm{$\frac{r_0}{R}$}&
\textrm{$\frac{r_0}{r_+}$}&
\textrm{$\gamma_{\rm cr}$}&
\textrm{$\gamma_{\rm ch}(1)$~\cite{Chandre1964b}}&
\textrm{$\gamma_{\rm pc}$~\cite{PosadaChirenti2019}}
\\
\colrule
0.342 & 8.549 & 1.39406 &  1.3940& 1.394010 \\
0.500 & 4.000 & 1.48957 & 1.4890 & 1.489546  \\
0.707 & 2.000 & 1.84347 & 1.8375& 1.843456  \\
0.819 & 1.490 & 2.55434 & 2.5204 & 2.554324  \\
0.907 & 1.217 & 6.12566 & 5.5802& 6.125634  \\
\end{tabular}
\end{ruledtabular}
\caption{\label{tab:zerocharge}
The critical adiabatic index $\gamma_{\rm cr}$ for the radial
perturbations of zero
charged stars, $\frac{q^2}{R^2}=0$, i.e.,
Schwarzschild stars, with different radii $\frac{r_0}{R}$ and
$\frac{r_0}{r_+}$.  For not so large $\frac{r_0}{R}$,
our results are in good agreement with the values $\gamma_{\rm ch}$ found
by Chandrasekhar~\cite{Chandre1964b} for various values of the
parameter $0.00<\frac{r_0}{R}<0.819$, and for all $\frac{r_0}{R}$ our
results are in good agreement with the values $\gamma_{\rm pc}$ found
in~\cite{PosadaChirenti2019}.  These zero charged stars are in the
vertical axis $\frac{q^2}{R^2}=0$ of Fig.~\ref{fig:regions}.  }
\end{table}
The values of the critical adiabatic index $\gamma_{\rm
cr}$ are obtained from the shooting and the pseudospectral methods,
and are in agreement to each other up to six decimal places.  Our
results are in good agreement with the values of the critical
adiabatic index $\gamma_{\rm ch}$ calculated in~\cite{Chandre1964b},
see the fourth column of the table, and are in very good agreement
with the values of the critical adiabatic index $\gamma_{\rm pc}$
calculated in~\cite{PosadaChirenti2019}, see the fifth column of the
table.  Note, however, that there is a difference between the critical
$\gamma_{\rm ch}$ calculated by Chandrasekhar~\cite{Chandre1964b} and
the critical $\gamma_{\rm pc}$ calculated in~\cite{PosadaChirenti2019}
and by us
as $\frac{r_0}{R}$ 
approaches from below
$\frac{r_0}{R}= \frac{2\sqrt{2}}{3} = 0.943$, with the
latter number being approximate, 
and as $\frac{r_0}{r_+}$
approaches from above
$\frac{r_0}{r_+}=\frac98=1.125$,
i.e., the Buchdahl point $B$ in
Fig.~\ref{fig:regions}. 
This difference may be explained by the fact
that the trial functions used by Chandrasekhar do
not approximate the true eigenfunctions in the limit of large
$\frac{r_0}{R}$~\cite{PosadaChirenti2019}.

\subsubsection{Undercharged pressure stars: $0<q^2<m^2$}
\label{Sec:UPS}

Undercharged pressure stars are stars with $0<q^2<m^2$ and also obey
$0<\frac{q^2}{R^2}<1$.  These configurations belong to region (a)
between lines $C_0$ and $C_4$ in Fig.~\ref{fig:regions}.

In this case, the energy density and the pressure are positive
functions everywhere inside the matter. Thus, one finds
$h(r)=\rho(r)+p(r)>0$ and, as a consequence, assuming $\gamma> 0$ the
coefficients $F(r)$ and $W(r)$ that appear in Eq.~\eqref{eq:SLP} are
both positive functions, and so in the SL problem this leads to the
case (A) of the theorem given in Appendix \ref{sec:SLP}. Therefore,
similarly to the case of the zero charged Schwarzschild stars, stable
solutions to radial perturbations
are found for positive adiabatic indices such that
$\gamma>\gamma_{\rm cr}$.

\begin{figure}[htbp] 
\centering
\includegraphics[width=0.235\textwidth]{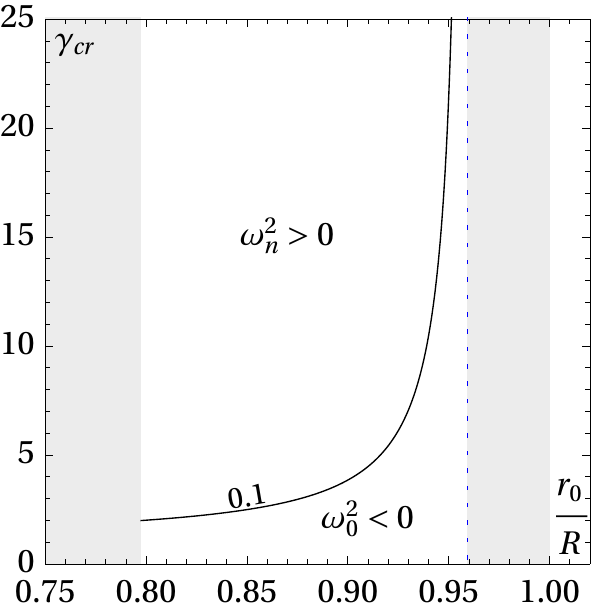}
\includegraphics[width=0.235\textwidth]{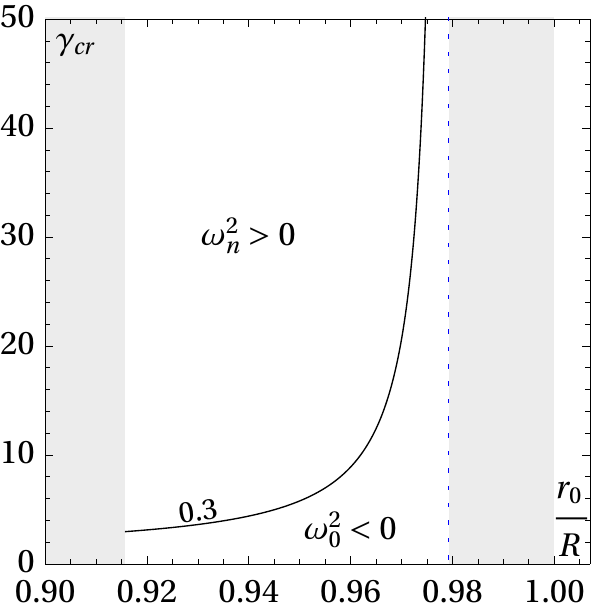}
\includegraphics[width=0.235\textwidth]{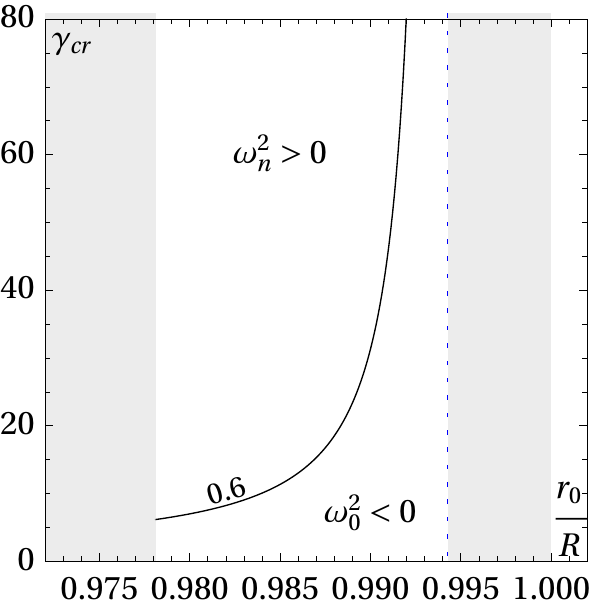}
\includegraphics[width=0.235\textwidth]{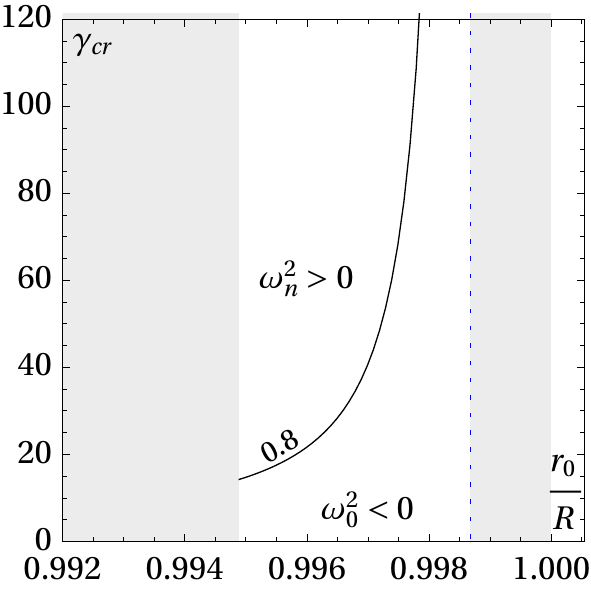}
\caption{Stability of regular undercharged pressure stars.  These
stars are stars with $0<q^2<m^2$, also obey $0<\frac{q^2}{R^2}<1$, and
belong to region (a) between lines $C_0$ and $C_4$ in
Fig.~\ref{fig:regions}.  The critical adiabatic index $\gamma_{\rm
cr}$ for four values of the electric charge parameter
$\frac{q^2}{R^2}=0.1$, $\frac{q^2}{R^2}=0.3$, $\frac{q^2}{R^2}=0.6$,
and $\frac{q^2}{R^2}=0.8$, is shown as a function of the radius
$\frac{r_0}{R}$.
In each of the four plots, the line starts at a
minimum radius $\frac{r_0}{R}$ which correspond to a value
$\gamma_{\rm cr}$ at some point on the curve $C_0$, and extends to
relatively large values as $\frac{r_0}{R}$ grows and approaches the
line $C_4$.
The light gray region on the left  side of each plot
corresponds to stars that are not undercharged, and the light gray
region on the right  side of each plot corresponds to stars that
are not regular and are beyond the Buchdahl-Andr\'easson limit.}
\label{fig:NormalStars}
\end{figure}

In Fig.~\ref{fig:NormalStars} we show the numerical results for the
critical adiabatic index $\gamma_{\rm cr}$ as a function of the radius
$\frac{r_0}{R}$ for four values of the electric charge, namely,
$\frac{q^2}{R^2}=0.1$, $\frac{q^2}{R^2}=0.3$, $\frac{q^2}{R^2}=0.6$,
and $\frac{q^2}{R^2}=0.8$, as indicated in the figure.
In each plot
the light gray region on the left side contains solutions that
are overcharged stars, so require a different analysis.
The white region represents the
range of the parameter $\frac{r_0}{R}$ where regular undercharged stars
are found.
The vertical
dotted line on the right side of each of the four plots indicates the
Buchdahl-Andr\'easson bound \cite{AndreassonQ}, see also
\cite{LemosZanchin2015}, which is represented by the curve $C_4$ in
Fig.~\ref{fig:regions}.
The light gray region on the right side contains
solutions for singular undercharged stars, i.e., undercharged
configurations with higher radii, namely, the ones whose values of
$\frac{r_0}{R}$ are on or above the curve $C_4$, i.e., in the region
(f) in Fig.~\ref{fig:regions}. Since they are singular undercharged star
solutions they are of little interest in general and in particular
for the stability analysis.
The solid
curved line in each of the four plots is for the vanishing fundamental
oscillation frequency squared, i.e., for $\omega_{0}^{2}=0$, which
means that $\omega_{0}^{2}$ changes sign across such a curve.  All
configurations represented by points located above the
$\omega_{0}^{2}=0$ line are stable stars, i.e., all $\omega_{n}^{2}$
are positive, all configurations represented by points located below
the $\omega_{0}^{2}=0$ line are unstable stars.  Each solid curved
line starts at some radius $\frac{r_0}{R}$ that corresponds to a point
just outside
the curve $C_0$ with a
relatively low $\gamma_{\rm cr}$
and extends to some point $\frac{r_0}{R}$ on the
curve $C_4$ at the Buchdahl-Andr\'easson bound where $\gamma_{\rm cr}$
diverges.
For instance, the range of radii $\frac{r_0}{R}$
corresponding to 
regular undercharged stars 
for the case $\frac{q^2}{R^2}=0.3$
is from
$\frac{r_0}{R} = 0.915703$ to
$\frac{r_0}{R} = 0.979269$,
where the numbers are approximate values, 
as can be confirmed from the 
top right panel of Fig.~\ref{fig:NormalStars}.
Note that, for a fixed finite
adiabatic index $\gamma$, the undercharged pressure
stars are stable configurations
against radial perturbations
for relatively small radius, i.e., small
$\frac{r_0}{R}$ which, since $R$ is a constant with the meaning
of inverse effective energy density, means a normal star far from
the Buchdahl-Andr\'easson bound and so far from forming a horizon.
At the Buchdahl-Andr\'easson bound,
these stars are unstable as they need an infinite
$\gamma_{\rm cr}$.

\begin{figure}[ht] 
\centering
\includegraphics[width=0.235\textwidth]{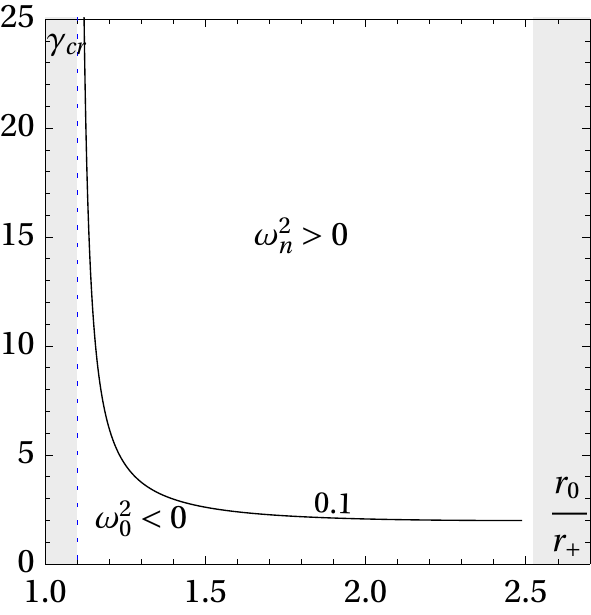}
\includegraphics[width=0.235\textwidth]{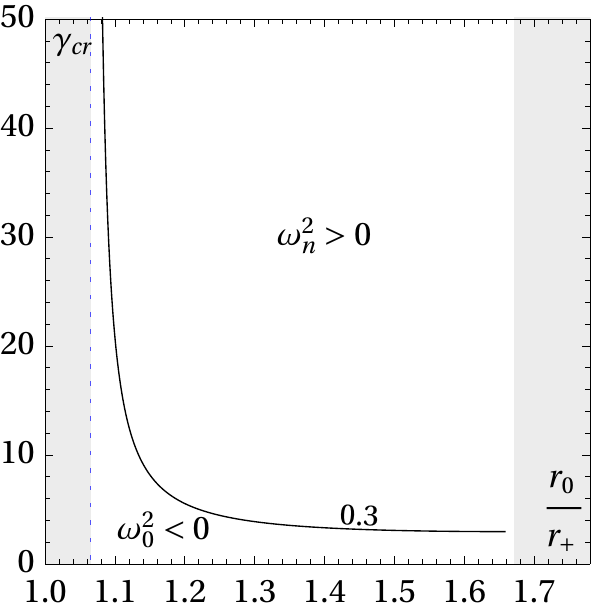}
\includegraphics[width=0.235\textwidth]{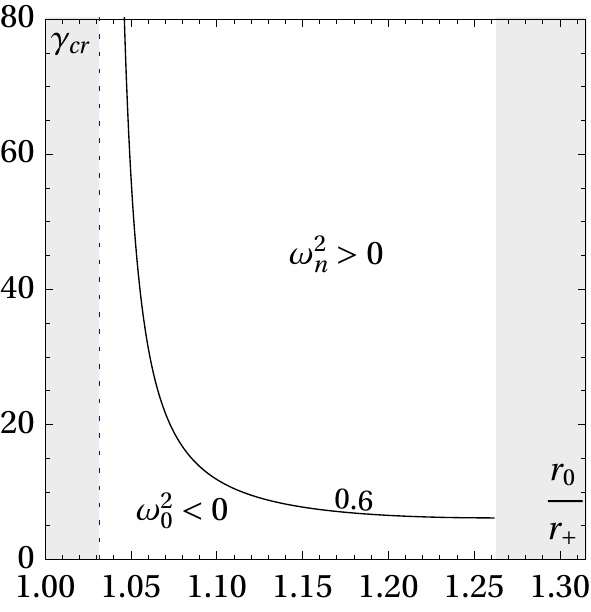}
\includegraphics[width=0.235\textwidth]{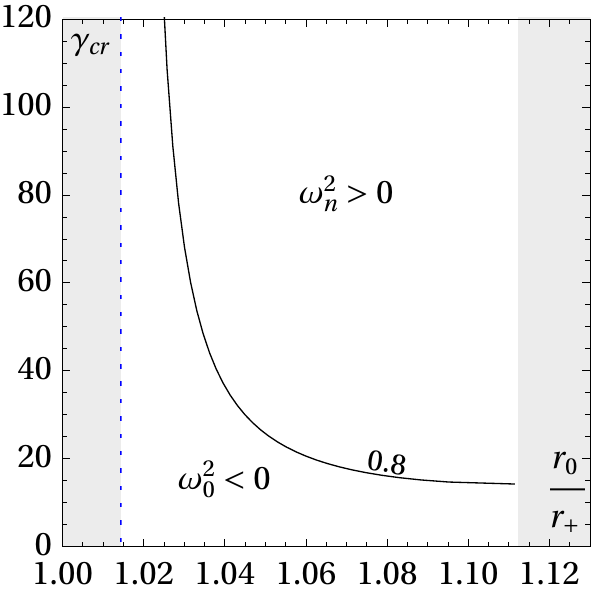}
\caption{ Stability of regular undercharged pressure stars.  These
stars are stars with $0<q^2<m^2$ and also obey $0<\frac{q^2}{R^2}<1$
and belong to region (a) between lines $C_0$ and $C_4$ in
Fig.~\ref{fig:regions}.  The critical adiabatic index $\gamma_{\rm
cr}$ for four values of the electric charge parameter
$\frac{q^2}{R^2}=0.1$, $\frac{q^2}{R^2}=0.3$, $\frac{q^2}{R^2}=0.6$,
and $\frac{q^2}{R^2}=0.8$, is shown as a function of the radius
$\frac{r_0}{r_+}$.
In each of the four plots, the line starts on the
right at some radius $\frac{r_0}{r_+}$ on the line $C_4$
which corresponds
to the Buchdahl-Andr\'easson bound and where $\gamma_{\rm cr}$
diverges, and extends to some $\frac{r_0}{r_+}$ on the line $C_0$
and where
$\gamma_{\rm cr}$
has some positive given value.
The light gray region on the left side
of each plot corresponds to stars that are not regular and are beyond
the Buchdahl-Andr\'easson limit, and the light gray region on the
right  side of each plot corresponds to stars that are not
undercharged.}
\label{fig:NormalStarsr+}
\end{figure}

In Fig.~\ref{fig:NormalStarsr+} we show the numerical results for the
critical adiabatic index $\gamma_{\rm cr}$ but now as a function of the
radius $\frac{r_0}{r_+}$, instead of $\frac{r_0}{R}$.
The radius
$\frac{r_0}{r_+}$ helps in a better
understanding of the compactness
of the star, i.e., in the
relation between the star radius $r_0$
and its gravitational radius
$r_+$, which is now the quantity kept constant,
rather than $R$.
The critical adiabatic index $\gamma_{\rm cr}$ is
shown for  the same four values of the electric charge,
namely, $\frac{q^2}{R^2}=0.1$, $\frac{q^2}{R^2}=0.3$,
$\frac{q^2}{R^2}=0.6$, and $\frac{q^2}{R^2}=0.8$, as indicated in the
figure.
In each plot, the light gray region on the left 
side contains solutions for singular undercharged stars, i.e.,
undercharged configurations with small radii, namely,
configurations for which the
values of $\frac{r_0}{r_+}$ are on or above the curve $C_4$ in the region
(f) of Fig.~\ref{fig:regions}, and since
they represent singular solutions, they are of
little interest.
The vertical dotted line in the left side of each of the four
plots indicates the Buchdahl-Andr\'easson bound \cite{AndreassonQ},
see also \cite{LemosZanchin2015}, which is represented by the curve
$C_4$ in Fig.~\ref{fig:regions}.  The white region
represents the range of the parameter $\frac{r_0}{r_+}$ where regular
undercharged stars are found.
The light gray region on the right  side
contains solutions that are overcharged stars, so require a different
analysis.
The solid curved line in each of the four plots is for the vanishing
fundamental oscillation frequency squared, i.e., for
$\omega_{0}^{2}=0$, which means that $\omega_{0}^{2}$ changes sign
across such a curve.  All configurations represented by points located
above the $\omega_{0}^{2}=0$ line are stable stars, i.e., all
$\omega_{n}^{2}$ are positive, all configurations represented by
points located below the $\omega_{0}^{2}=0$ line are unstable stars.
The solid curved line starts from the
left at the curve $C_4$ at the Buchdahl-Andr\'easson
bound where the stars are very compact and $\gamma_{\rm cr}$ diverges
and extends to the 
right at some minimum for
$\frac{r_0}{r_+}$ relatively large, that corresponds to a point on the curve
$C_0$ where the stars are not anymore undercharged.
Stability of stars with $\frac{r_0}{r_+}$
approaching the curve $C_4$, i.e., the Buchdahl-Andr\'easson bound
occurs just for arbitrarily large values of the adiabatic index.
For a fixed adiabatic index, the undercharged pressure
stars are stable configurations
against radial perturbations for relatively large radius, i.e., large
$\frac{r_0}{r_+}$.

In Table~\ref{tab:undercharge1} we give details of the numerical
results for the stability of an undercharged star.  The behavior of
$\gamma_{\rm cr}$ as a function of the radius $\frac{r_0}{R}$ and
$\frac{r_0}{r_+}$, for $\frac{q^2}{R^2}=0.3$, is displayed.  The
values of the critical adiabatic index $\gamma_{\rm cr}$ are obtained
from the
shooting and the
pseudospectral methods, and are
in agreement to each other up to six decimal places.
\begin{table}[h]
\begin{ruledtabular}
\begin{tabular}{c c c c}
\textrm{$\frac{r_0}{R}$}& 
\textrm{$\frac{r_0}{r_+}$}& 
\textrm{$\gamma_{\rm cr}$}& \\
\colrule
$0.915704$ & 1.66855 & $2.95794$\\
$0.924784$ & 1.39611 & $3.32947$\\
$0.933863$ & 1.30128 & $3.86695$\\
$0.942943$ & 1.23458 & $4.70936$\\
$0.952022$ & 1.18179 & $6.20193$\\
$0.961102$ & 1.13763 & $9.47855$\\
$0.970181$ & 1.09943 & $21.1295$\\
$0.979261$ & 1.06562 & $440359$\\
\end{tabular}
\end{ruledtabular}
\caption{\label{tab:undercharge1}
The critical adiabatic index $\gamma_{\rm cr}$ for the radial
perturbations of undercharged stars with $\frac{q^2}{R^2}=0.3$ and for
various values of the parameter $\frac{r_0}{R}$ and the compactness
ratio $\frac{r_0}{r_+}$.  These undercharged stars are in the region (a)
of Fig.~\ref{fig:regions}.
}
\end{table}
We have calculated the zero mode frequencies squared $\omega_{0}^{2}$
and the first mode frequencies squared $\omega_{1}^{2}$ for these
$\frac{q^2}{R^2}=0.3$ stars with $\gamma=4$.  We find that a star with
$\frac{r_0}{R}=0.933863$ and so $\frac{r_0}{r_+}=1.3012$
has
$\omega_{0}^{2}=2.34262\times10^{-3}$, $\omega_{1}^{2}=0.541388$, and
$\gamma_{\rm cr}=3.86695$, so $\gamma=4$ being above $\gamma_{\rm cr}$ this
star is stable
to radial perturbations, while a star with $\frac{r_0}{R}=0.942943$ and so
$\frac{r_0}{r_+}=1.23458$  has
$\omega_{0}^{2}=-0.01355012$,
$\omega_{1}^{2}=0.628574$, and $\gamma_{\rm cr}=4.70936$, so $\gamma=4$
being below $\gamma_{\rm cr}$ this star is unstable.
The solutions for these undercharged pressure stars having radii
extending from
approximately $\frac{r_0}{R} = 0.915703$ to
approximately
$\frac{r_0}{R}=
0.979269$, in the $\gamma=4$
adiabatic index case
have $\omega_{0}^2$ 
positive in the range $0.915703\leq \frac{r_0}{R} \leq
0.942943$, where the values given are
approximate values, and  $\omega_{0}^2$ negative in
the range $0.942943\leq
\frac{r_0}{R} \leq 0.979269$, where the values given are
approximate values, as it can be seen in more detail in
Appendix~\ref{appendixB}.

Undercharged stars that are singular, 
are stars with $q^2<m^2$ and also  are
above the Buchdahl-Andr\'easson curve $C_4$.
These configurations belong to region (f), the
region between the
horizontal line $r_0=r_+=R$ and
the line $C_4$ in Fig.~\ref{fig:regions}.
They are of no interest for the stability problem
since the curvature scalars and the fluid quantities
diverge at some radius inside the matter distribution.

\subsubsection{Extremally charged dust stars}

Extremely charged dust stars or Bonnor stars
\cite{lemoszanchin2008}
are configurations
that have charge density equal to mass density,
$\rho_e=\rho$, the pressure is zero, obey
$q^2=m^2$ and also obey $0<\frac{q^2}{R^2}<1$.
These configurations are on the
line $C_0$ in Fig.~\ref{fig:regions}.

In this case, the energy density is positive
and since the pressure is zero everywhere inside the matter
one has 
$h(r)=\rho(r)+p(r)>0$.
We can analyze the stability in this case directly,
without having to resort to the theorem in
Appendix~\ref{sec:SLP}.
Indeed, from Eqs.~(\ref{eq:SLP})-(\ref{eq:SLPW}) one finds that since
$\rho_e=\rho$ and $p=0$, one has
$F(r)=0$, $G(r)=0$, and $H(r)=0$, and so Eq.~(\ref{eq:SLP})
reduces to $\omega^2W(r)\zeta(r)=0$, i.e.,
\begin{equation}\label{stabilityextremalstars}
\omega^2 \rho(r) A^{\frac32}(r)\xi(r)=0\,,
\end{equation} 
where we have used Eqs.~(\ref{eq:SLP})
and~(\ref{eq:SLPzeta}).
One can find Eq.~(\ref{stabilityextremalstars})
directly from Eq.~(\ref{eq:eigenvalue}).
For generic 
$\rho(r)$,
$A(r)$, and
$\xi(r)$,
the solution is 
\begin{equation}\label{stabilityextremalstarszero}
\omega^2 =0\,.
\end{equation} 
Thus, extremely charged dust stars have a neutral stability
against radial perturbations. If
displaced in a spherically symmetric way they stay put or increase or
decrease their radius homothetically and uniformly.
An extremely charge dust star by itself neither expands nor collapses.
Note, however, that for a
nongeneric $A(r)$, namely, $A(r)=0$, for some $r$, than $\omega^2$ can
be anything, we return to this case later.

Numerically, the behavior of this type of solutions against small
radial perturbations can be displayed through the region (a) when the
parameters of the stars in that region are very close to the line
$C_0$.  With an adiabatic index $\gamma=4$ and
$\frac{q^2}{R^2}=0.3$ in the region (a), the frequencies are very close to
zero for $\frac{r_0}{R}=0.915704$, i.e., near the line $C_0$, see also
Appendix~\ref{appendixB}.  This implies that along the line $C_0$, the
square frequencies for the fundamental and the first excited modes are
very close to zero or vanish as
Eq.~(\ref{stabilityextremalstarszero}) implies.  This case has also
been worked out in~\cite{Omote1974,Glazer1976}.

\subsubsection{Overcharged tension stars}
\label{Sec:OTS}

Overcharged regular tension stars 
are stars with $m^2<q^2$ and also obey
$\frac{q^2}{R^2}<1$.  These configurations belong to region (b), the
region between lines $C_0$ and $C_1$ in Fig.~\ref{fig:regions}.

In this case, the energy density is positive and the pressure is
negative, it is a tension. The enthalpy density $h(r)=\rho(r)+p(r)$ is
always greater than zero and, as a consequence, the function $W(r)$
that appears in Eq.~\eqref{eq:SLP} is positive. However, the sign of
the function $F(r)$ depends on the product $\gamma p(r)$.  Therefore,
if $\gamma$ is assumed to be positive, the SL problem falls into the
case (B) of the theorem summarized in Appendix \ref{sec:SLP}. The
corresponding theorem implies that for a positive function $W(r)$ and
a negative function $F(r)$, the sequence of eigenvalues is bounded
from above, with fundamental frequency $\omega_0^2$ being the largest
among all of them, i.e., $\cdots<\omega_2^2<\omega_1^2<\omega_0^2<
\infty$. Hence, if the restriction $\gamma>\gamma_{\rm cr}>0$ is
fulfilled, $\omega_0^2$ will be positive but the largest excited modes
would have negative square frequencies and the configurations will be
unstable against radial perturbations, see Appendix~\ref{appendixB}
for more details.  Let us give physical arguments for the instability
of these configurations when one considers $\gamma$ positive.
Equation~\eqref{eq:adindex} can be cast as $ \Delta p=c_s^2\Delta\rho
$, where $c_s^2$ is the sound speed squared defined as $c_s^2=
\dfrac{\gamma p}{\rho+p}$. In the interior region of tension stars the
conditions $p<0$ and $\rho+p>0$ hold, implying that for $\gamma>0$ one
has $c_s^2=\frac{\Delta p}{\Delta\rho}<0$, which means that when the
density increases the tension increases and conversely when the
density decreases the tension decreases.  Then, when perturbing the
system, if the fluid is compressed, and so the density increases, so
also the tension grows, favoring the system to get even more
compressed in a runaway process. Conversely, when perturbing the
system, if the fluid is expanded, and so the density decreases, so
also the tension diminishes, favoring the system to get even more
expanded in a runaway process.  This implies that, once started, the
perturbed configuration never stops its process of compression or
expansion, indicating an instability of the system.  Another way of
seeing this is that for $\gamma>0$, the sound speed squared obeys
$c_s^2<0$, the sound speed is imaginary, and so there is no
propagation of the perturbation and no possibility for stability.
This leads to the conclusion that for tension stars, i.e., stars
supported by negative pressure, one should assume that the
radial perturbations are governed by a negative $\gamma$, and ask whether
there are stable configurations for overcharged tension stars when
$\gamma<0$ or not.  If the adiabatic index $\gamma$ is negative, the
coefficients $F(r)$ and $W(r)$ that appear in Eq.~\eqref{eq:SLP} are
both positive functions, and so in the SL problem this leads to the
case (A) of the theorem given in Appendix \ref{sec:SLP}.  In this case
the stable solutions are found for negative adiabatic index such that
$\gamma<\gamma_{\rm cr}$, where negative $\gamma_{\rm cr}$ is the
critical, i.e., maximum negative number, value of $\gamma$, or in
terms of absolute value which makes things clearer, one has
$|\gamma|>|\gamma_{\rm cr}|$ for stability.  Let us give physical
arguments for the possible stability of these configurations when one
considers $\gamma$ negative.  Equation~\eqref{eq:adindex}, as we have
already seen, can be cast as $ \Delta p=c_s^2\Delta\rho $, where
$c_s^2$ is the sound speed squared defined as $c_s^2= \dfrac{\gamma
p}{\rho+p}$. In the interior region of tension stars the conditions
$p<0$ and $\rho+p>0$ hold, implying that for $\gamma<0$ one has
$c_{s}^{2}>0$.  Moreover, now if the density increases the tension
decreases, and conversely if the density decreases the tension
increases.  Then, when perturbing the system radially,
if the fluid is
compressed, and so the density increases, so the tension diminishes
favoring the system to get less compressed in a possible stable
process. Conversely, when perturbing the system
radially, if the fluid is
expanded, and so the density decreases, so the tension grows, favoring
the system to get less expanded in a possible stable process.  This
implies that, once started, the perturbed configuration can return to
the original configuration, the process of compression and expansion
can be halted, indicating stability of the system.  Another way of
seeing this is that for $\gamma<0$, the sound speed squared obeys
$c_s^2>0$, the sound speed is real, and so there is propagation of the
perturbation and possibility for stability.

\begin{figure}[h] 
\centering
\includegraphics[width=0.235\textwidth]{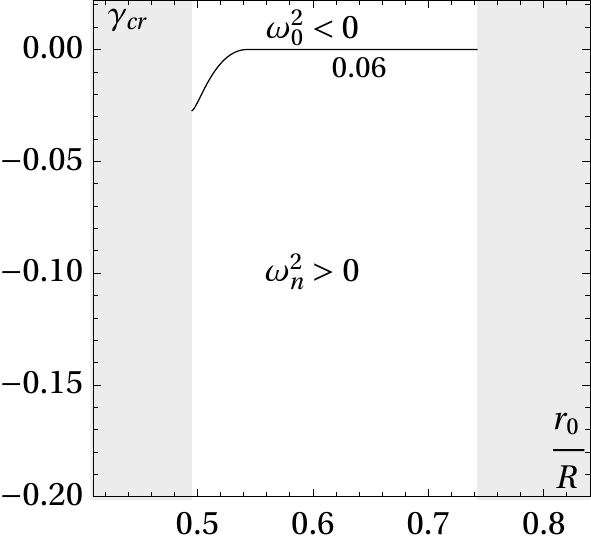}
\includegraphics[width=0.235\textwidth]{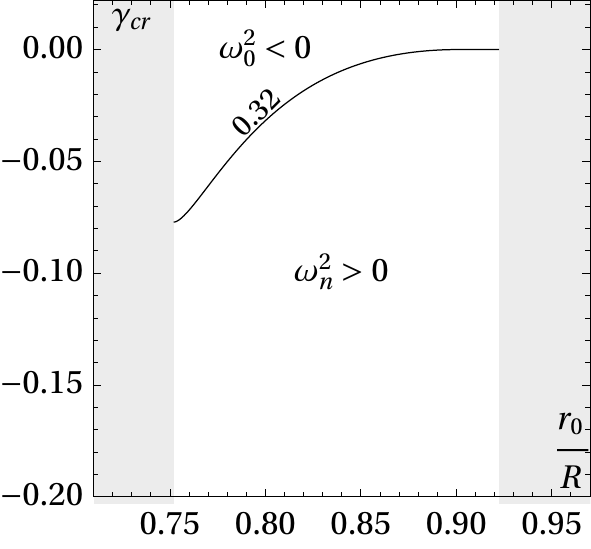}
\includegraphics[width=0.235\textwidth]{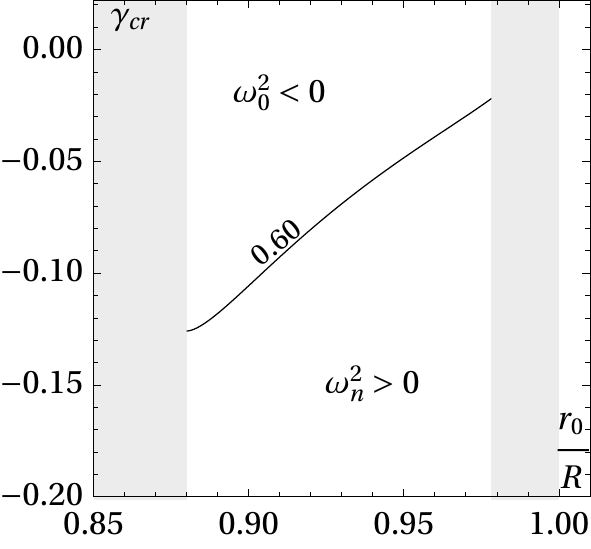}
\centering
\includegraphics[width=0.235\textwidth]{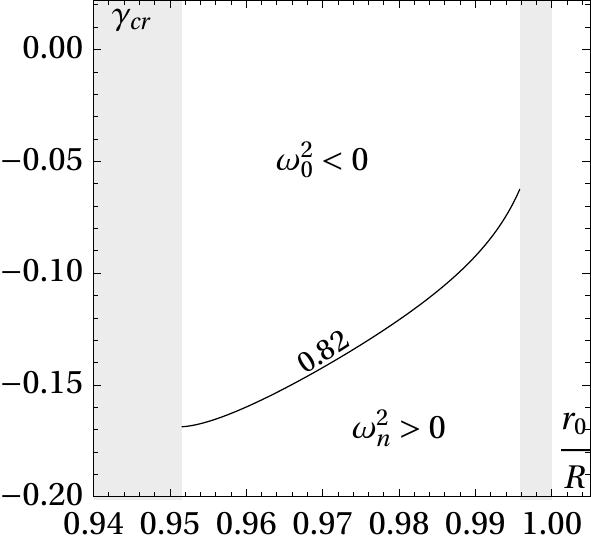}
\caption{
Stability of regular overcharged tension stars.  These stars are stars
with $m^2<q^2$ and also obey $0<\frac{q^2}{R^2}<1$ and belong to
region (b) between lines $C_0$ and $C_1$ in Fig.~\ref{fig:regions}.
The critical adiabatic index $\gamma_{\rm cr}$ for four values of the
electric charge parameter $\frac{q^2}{R^2}=0.06$,
$\frac{q^2}{R^2}=0.32$, $\frac{q^2}{R^2}=0.60$, and
$\frac{q^2}{R^2}=0.82$, is shown as a function of the radius
$\frac{r_0}{R}$.  In each of the four plots, the line starts at a
minimum radius $\frac{r_0}{R}$ which correspond to a negative value
$\gamma_{\rm cr}$ at some point on the curve $C_1$, and extends to some
negative value as $\frac{r_0}{R}$ grows and approaches the line
$C_0$. The light gray region on the left  side of each plot
corresponds to stars that are overcharged and singular, beyond the
curve $C_1$, and the light gray region on the right  side of each
plot corresponds to stars that are not overcharged, beyond the curve
$C_0$.
}
\label{fig:TensionStars}
\end{figure}

In Fig.~\ref{fig:TensionStars}, we show the numerical results for the
critical adiabatic index $\gamma_{\rm cr}$, negative here, as a
function of the radius $\frac{r_0}{R}$ for four values of
$\frac{q^2}{R^2}$, namely, $\frac{q^2}{R^2}=0.06$,
$\frac{q^2}{R^2}=0.32$, $\frac{q^2}{R^2}=0.60$, and
$\frac{q^2}{R^2}=0.82$, i.e., for overcharged stars.  In each plot,
the light gray region on the left side is for solutions that are
singular overcharged stars, i.e., stars beyond the curve $C_1$
of Fig.~\ref{fig:regions}.
The
white region represents the range of the parameter $\frac{r_0}{R}$
where regular overcharged stars are found.  The light gray region on
the right side contains solutions that are not overcharged,
i.e., stars beyond the
curve $C_0$, and do not belong here.  The solid curved line in each of
the four plots is for the vanishing fundamental oscillation frequency
squared, i.e., for $\omega_{0}^{2}=0$, which means that
$\omega_{0}^{2}$ changes sign across such a curve.
All configurations represented by points located below the
$\omega_{0}^{2}= 0$ line are stable stars, i.e., all
$\omega_{n}^{2}$ are
positive, all configurations represented by points located above the
$\omega_{0}^{2}= 0$ line are unstable stars.
Each solid curved line
starts at some radius $\frac{r_0}{R}$ that corresponds to a point on
the curve $C_1$ and corresponds to the first nonsingular
overcharged stars on the curve,
and extends to some point $\frac{r_0}{R}$ on the curve
$C_0$ where the solutions have charge density equal to
mass density. Along the solid line, from left to
right as  $\frac{r_0}{R}$ increases, the stars get
more mass, and so need less tension to support the interior
against expansion. 
For overcharged stars there is no gravitational radius $r_+$
and it means there is no possibility of interchanging 
$\frac{r_0}{R}$ with $\frac{r_0}{r_+}$.
One could think in plotting the critical adiabatic index
$\gamma_{\rm cr}$
as a function of  $\frac{r_0}{m}$ instead, where $m$ is the spacetime
mass,
but there is no gain in it clearly,
the only difference would be a
reverse of the sign in the slope of the curve.

In Table~\ref{tab:overcharge2} we give details
of the numerical results
for the stability of overcharged stars.  The behavior of
$\gamma_{\rm cr}$ as a function of the radius $\frac{r_0}{R}$ for
$\frac{q^2}{R^2}=0.6$, is displayed.  The values of the critical
adiabatic index $\gamma_{\rm cr}$ are obtained
from the shooting and
pseudospectral methods, and are in agreement to each
other to six decimal places.
We have calculated the zero mode frequencies squared
$\omega_{0}^{2}$ and the first mode frequencies squared
$\omega_{1}^{2}$ for these $\frac{q^2}{R^2}=0.6$ stars with
$\gamma=-0.06$.  We find that
for a star with $\frac{r_0}{R}=0.936112$ one
has $\omega_{0}^{2}=-1.69590 \times 10^{-3}$, and
$|\gamma_{\rm cr}|=0.0623795$, so $|\gamma|=0.06$
being below $|\gamma_{\rm cr}|$
means that
this star is unstable
against radial perturbations, while
for a star with $\frac{r_0}{R}=0.950112$ one
has $\omega_{0}^{2}=5.03598 \times10^{-3}$,
$\omega_{1}^{2}=0.0529323$, and $|\gamma_{\rm cr}|=0.0483227$, so
$|\gamma|=0.06$ being above $|\gamma_{\rm cr}|$
means that this star is unstable.
\begin{table}[h]
\begin{ruledtabular}
\begin{tabular}{c c c c c c c c}
\textrm{$\frac{r_0}{R}$}& 
\textrm{$\gamma_{\rm cr}$}\\
\colrule
0.880113 & -0.125874 \\
0.894113 & -0.113132 \\
0.908113 & -0.0952036 \\
0.922113 & -0.0779790 \\
0.936112 & -0.0623795 \\
0.950112 & -0.0483227 \\
0.964112 & -0.0352687 \\
0.978111 & -0.0220528  \\
\end{tabular}
\end{ruledtabular}
\caption{The critical adiabatic index $\gamma_{\rm cr}$ for the radial
perturbations of overcharged stars with $\frac{q^2}{R^2}=0.6$ and for
various values of the parameter $\frac{r_0}{R}$.
These overcharged stars are in the region (b)
of Fig.~\ref{fig:regions}.}
\label{tab:overcharge2}
\end{table}
The solutions for these overcharged tension stars having radii
extending from approximately $\frac{r_0}{R} = 0.880113$ to
approximately $\frac{r_0}{R}= 0.978111$ in the $\gamma=-0.06$
adiabatic index case,
have $\omega_{0}^{2}$ negative approximately in the range
           $0.880113 \leq  \frac{r_0}{R} \leq 0.938387$,
	   and $\omega_{0}^{2}$
           positive approximately in the range
	   $0.938387 \leq  \frac{r_0}{R} \leq
           0.978111$.
Thus, stars with larger
$\frac{r_0}{R}$, i.e., overcharged stars with more mass and less
electric charge, and thus less tension, are stable
to radial perturbations.
Moreover,
for $\frac{r_0}{R}$
close to the line $C_0$ in the region (b) of Fig.~\ref{fig:regions},
one has that the
corresponding stars tend to electrically charged dust stars
with $m^2=q^2$, the tension
on these tension stars being very small. One finds numerically 
that $\omega_{0}^2$ as well as all other higher tones
tend to zero and so in the limit these stars are
neutrally stable, as we have discussed in the
undercharged case and have found the exact
stability solutions in the charge density equal
energy density case. Some
more detail is given in
Appendix~\ref{appendixB}.
An interesting case needing further investigation occurs when
$\frac{r_0}{R}\to 1$
and $\frac{q^2}{R^2}
\to 1$ for the solutions in the region (b), see below.

Overcharged stars that are singular 
are stars with $m^2<q^2$ and also obey
$\frac{q^2}{R^2}<\frac{27}{16}=1.6875$.
These configurations belong to region (c), the
region between lines $C_1$ and $C_2$ in Fig.~\ref{fig:regions}.
They are of no interest for the stability problem
since the curvature scalars and the fluid quantities
diverge at some radius inside the matter distribution.

\subsection{The stability of regular black holes}

\subsubsection{Regular black holes with
negative energy densities}
\label{rbhsned}

Regular black holes with negative energy densities are also phantom
regular black holes with no singularities, for which the electric
charge obeys $\frac{q^2}{R^2}>0$, and the radius $r_0$ is inside the
Cauchy horizon, $r_0<r_-$. These configurations belong to region (d1),
i.e., to the right of lines $C_2$ and below the line $C_{31}$ plus
$C_{31}C_{32}$ in Fig.~\ref{fig:regions}.

In this case, the energy density $\rho(r)$ is negative for a range of
the radial coordinate $r$ inside the matter and the pressure $p(r)$ is
always negative. The enthalpy density $h(r)=\rho(r)+p(r)$ is
everywhere less than zero, and as a consequence, assuming $\gamma> 0$
the coefficients $F(r)$ and $W(r)$ that appear in Eq.~\eqref{eq:SLP}
are both positive functions, and so in the SL problem this leads to
the case (A) of the theorem given in Appendix
\ref{sec:SLP}. Therefore
stable solutions to radial perturbations
of these regular black holes are found for
positive adiabatic indices such that $\gamma>\gamma_{\rm cr}$.

\begin{figure}[h] 
\centering
\includegraphics[width=0.235\textwidth]{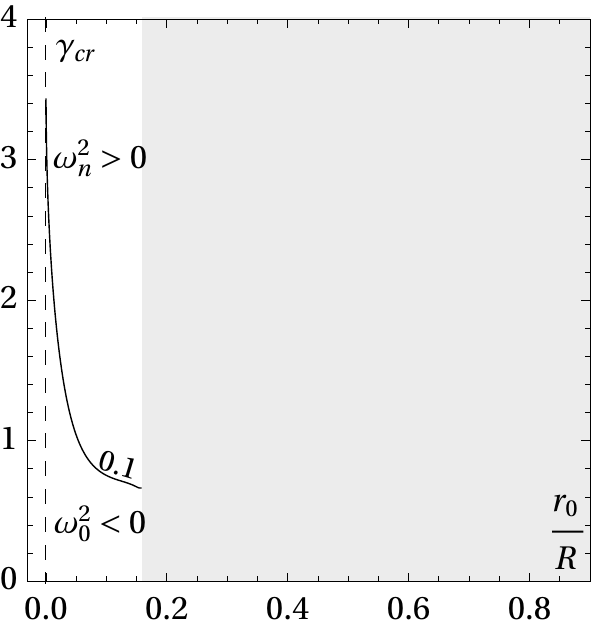}
\includegraphics[width=0.235\textwidth]{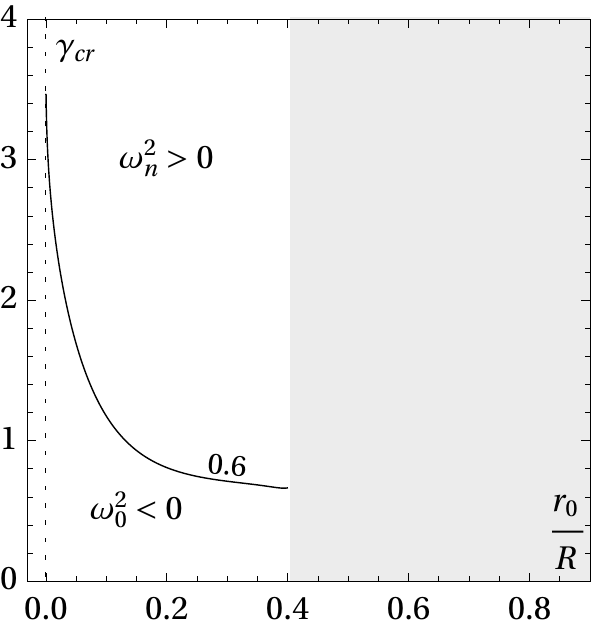}
\includegraphics[width=0.235\textwidth]{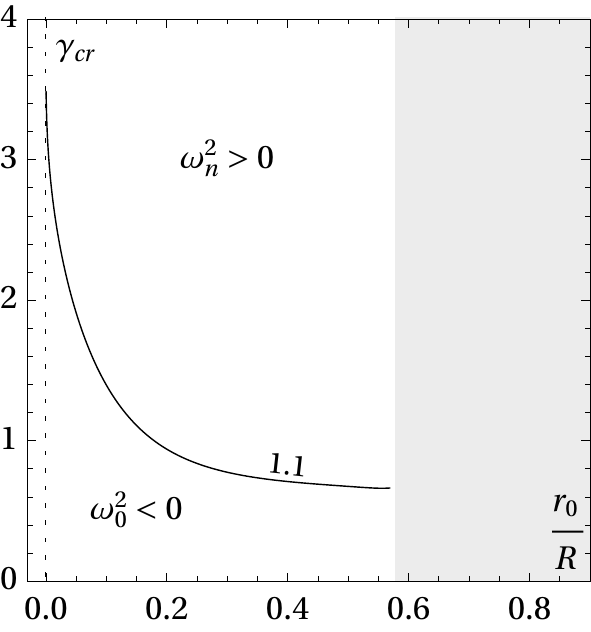}
\centering
\includegraphics[width=0.235\textwidth]{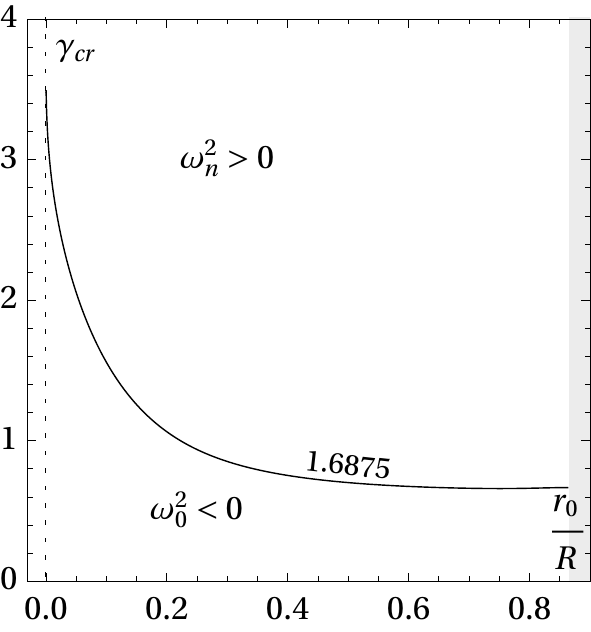}
\caption{
Stability of regular black holes with negative energy density.  These
regular black holes belong to region (d1) to the right of line $C_2$
and to the left of the vertical line
$\frac{q^2}{R^2}=\frac{27}{16}=1.6875$ in Fig.~\ref{fig:regions}.  The
critical adiabatic index $\gamma_{\rm cr}$ for four values of the
electric charge parameter $\frac{q^2}{R^2}=0.1$,
$\frac{q^2}{R^2}=0.6$, $\frac{q^2}{R^2}=1.1$, and
$\frac{q^2}{R^2}=\frac{27}{16}=1.6875$, is shown as a function of the
radius $\frac{r_0}{R}$.  In each of the four plots, the line starts
just above $\frac{r_0}{R}=0$ for some $\gamma_{\rm cr}$ and extends to
some value as $\frac{r_0}{R}$ grows and approaches the line $C_2$.
In each of the four plots,
the adiabatic index
$\gamma_{\rm cr}$ decreases up to some radius $\frac{r_0}{R}$ where,
although not discernible in the plots, it starts to grow again
slightly, the reason for this behavior being not clear.  The light
gray region on the right side of each plot corresponds to objects
beyond the curve $C_2$, that are overcharged and singular, and are not
black hole configurations.
}
\label{fig:RBHNE1}
\end{figure}

In Fig.~\ref{fig:RBHNE1}, we show the numerical results for the
critical adiabatic index $\gamma_{\rm cr}$ as a function of the radius
$\frac{r_0}{R}$ for
four values of the electric charge, namely, $\frac{q^2}{R^2}=0.1$,
$\frac{q^2}{R^2}=0.6$, $\frac{q^2}{R^2}=1.1$, and
$\frac{q^2}{R^2}=1.6875$, as indicated in the figure,
with
the value
$\frac{q^2}{R^2}=\frac{27}{16}=1.6875$
being the value of the elbow in
curve $C_2$ of Fig.~\ref{fig:regions}.  
In each plot the left part is the axis $\frac{r_0}{R}=0$.  The white
region represents the range of the parameter $\frac{r_0}{R}$ where
regular black holes are found. The light gray region on the right side
of each plot contains solutions of singular charged stars, i.e.,
overcharged configurations with higher radii, namely, the ones whose
values of $\frac{r_0}{R}$ are on or above the curve $C_2$ of mass
equal to charge, i.e., in the region (c) in Fig.~\ref{fig:regions},
and so are of no interest.
The solid curved line in each of the four plots is for the vanishing
fundamental oscillation frequency squared, i.e., for
$\omega_{0}^{2}=0$, which means that $\omega_{0}^{2}$ changes sign
across such a curve.  All configurations represented by points located
above the $\omega_{0}^{2}=0$ line are stable regular black holes,
i.e., all $\omega_{n}^{2}$ are positive, all configurations
represented by points located below the $\omega_{0}^{2}=0$ line are
unstable regular black holes.  Each solid curved line starts
just above $\frac{r_0}{R}=0$ and extends to some point
$\frac{r_0}{R}$ on the curve $C_2$.  One sees that the critical
adiabatic index $\gamma_{\rm cr}$ on each line of the four different
$\frac{q^2}{R^2}$, starts at approximately the same value for
$\frac{r_0}{R}$ very small, and then decreases for larger
$\frac{r_0}{R}$.
One can make some further remarks for all the four plots with the
numbers given meaning approximate rather than exact numbers.  In each
of the four plots the critical adiabatic index decreases to a minimum
value close to $\gamma_{\rm cr}=0.66$ for relatively large
$\frac{r_0}{R}$, and then grows again,
although only a little
not visible in the plots,
to approximately $\gamma_{\rm
cr}=0.6667$ when $\frac{r_0}{R}$ approaches the line $C_2$, i.e., the
line $m^2=q^2$.  
For a fixed adiabatic index $\gamma$ above about
$\gamma_{\rm cr}=3.5$, all regular black holes are stable
to radial perturbations.
In each of the four plots, the line starts immediately after
$\frac{r_0}{R}=0$ for some $\gamma_{\rm cr}$ and decreases up to some
radius $\frac{r_0}{R}$ where it starts to grow again slightly.
We have not been able to give a heuristic explanation for the reason
of this change of stability in each of the four plots.  For fixed
adiabatic index $\gamma$ in the range $\gamma<0.6$ all regular black
holes are unstable.  In the limit $\frac{r_0}{R}=0$, the mass
diverges, the metric turns into a Kasner metric, and the stability
problem set here does not apply.

In Table~\ref{tab:regularbhne}, we give details  of the numerical
results for the stability of regular black holes with
negative energy densities.  The behavior of
$\gamma_{\rm cr}$ as a function of the radius $\frac{r_0}{R}$,
for $\frac{q^2}{R^2}=\frac{27}{16}=1.6875$, is displayed.  The
values of the critical adiabatic index $\gamma_{\rm cr}$ are obtained
from the shooting
and the pseudospectral  methods, and are
in agreement to each other up to six decimal places.
\begin{table}[h]
\begin{ruledtabular}
\begin{tabular}{c c c c}
\textrm{$\frac{r_0}{R}$}& 
\textrm{$\gamma_{\rm cr}$}\\
\colrule
 0.0186989 & 2.59470 & \\
 0.139572 & 1.30905\\
 0.260445 & 0.916685\\
 0.381318 & 0.766649  \\
 0.502191 & 0.701179 \\
 0.623064 & 0.670727 \\
 0.743937 & 0.659659 \\
 0.864810 & 0.666663 \\
\end{tabular}
\end{ruledtabular}
\caption{The critical adiabatic index $\gamma_{\rm cr}$ for the radial
perturbations of regular black holes with negative energy densities
with $\frac{q^2}{R^2}=\frac{27}{16}=1.6875$ and for various values of
the parameter $\frac{r_0}{R}$.  These regular black holes are in
the left part of
region (d1) of Fig.~\ref{fig:regions}. }
\label{tab:regularbhne}
\end{table}
We have calculated the zero mode frequencies squared $\omega_{0}^{2}$
and the first mode frequencies squared $\omega_{1}^{2}$ for these
$\frac{q^2}{R^2}=\frac{27}{16}$ regular black holes with
negative energy densities for the adiabatic
index $\gamma=4$.  We find that all
these regular black holes are stable,
all eigenfrequencies squared are
positive. 
The solutions for these regular black holes have radii extending from
above $\frac{r_0}{R} = 0$ to approximately $\frac{r_0}{R}= 0.866025$.
Note that as $\frac{r_0}{R}$ increases the
value of $\gamma_{\rm cr}$
decreases up
to approximately $\frac{r_0}{R}=0.66$ where it increases again.  All
this can be seen in more detail in Appendix~\ref{appendixB}.

\vskip 1cm

In Fig.~\ref{fig:RBHNE2}, we show the numerical results for the
critical adiabatic index $\gamma_{\rm cr}$ as a function of the radius
\begin{figure}[h] 
\centering
\includegraphics[width=0.235\textwidth]{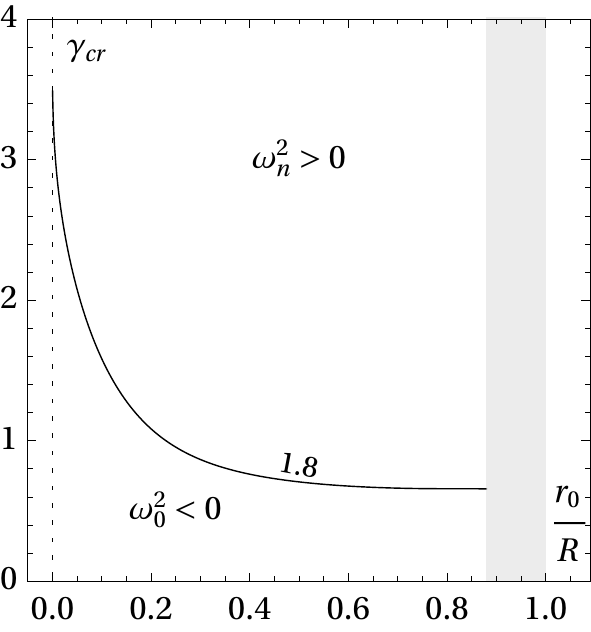}
\includegraphics[width=0.235\textwidth]{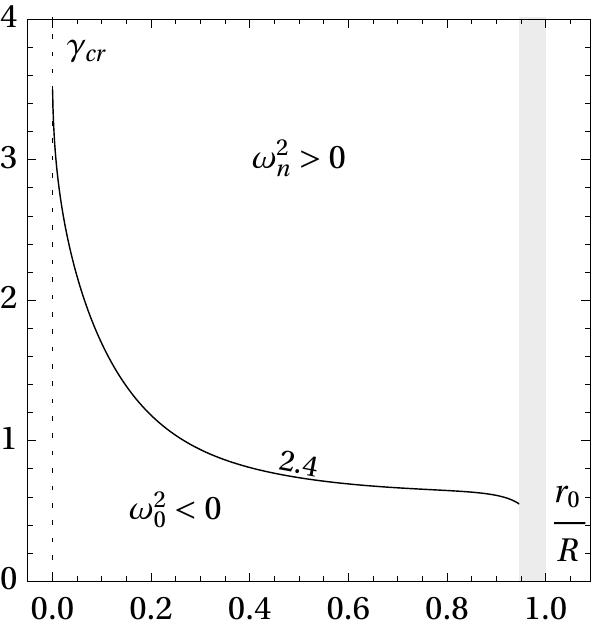}
\includegraphics[width=0.235\textwidth]{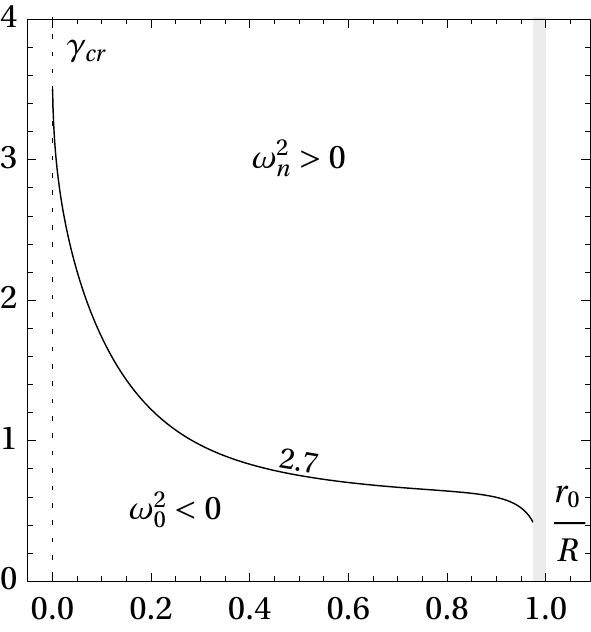}
\includegraphics[width=0.235\textwidth]{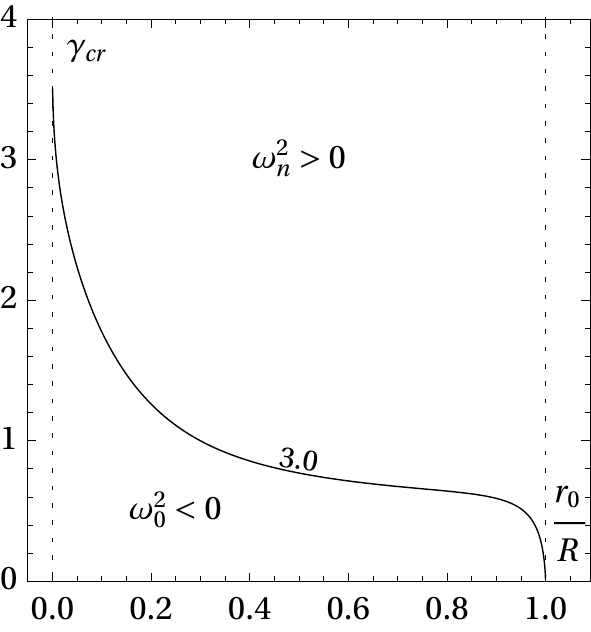}
\caption{
Stability of regular black holes with negative energy density,
continuation from the previous figure.  These regular black holes
belong to region (d1) to the right of the vertical line
$\frac{q^2}{R^2}=\frac{27}{16}=1.6875$, below the line $C_{31}$ plus
$C_{31}C_{32}$, and to the left of the vertical line
$\frac{q^2}{R^2}=3.0$ in Fig.~\ref{fig:regions}.  The critical
adiabatic index $\gamma_{\rm cr}$ for four values of the electric
charge parameter $\frac{q^2}{R^2}=1.8$, $\frac{q^2}{R^2}=2.4$,
$\frac{q^2}{R^2}=2.7$, and $\frac{q^2}{R^2}=3.0$, is shown as a
function of the radius $\frac{r_0}{R}$.  In each of the four plots,
the line starts just above $\frac{r_0}{R}=0$ for some $\gamma_{\rm
cr}$ and extends to some value as $\frac{r_0}{R}$ grows and approaches
the line $C_{31}$ plus $C_{31}C_{32}$.  The adiabatic index
$\gamma_{\rm cr}$ decreases up to some large $\frac{r_0}{R}$ and it
decays abruptly for large $\frac{r_0}{R}$ being even zero when
$\frac{q^2}{R^2}=3.0$, which is consistent since in this case the
regular black hole is made of a pure de Sitter interior, and the de
Sitter solution is stable
against radial perturbations.  The light gray region on the right side of
each plot corresponds to different regular black holes, beyond the
line $C_{31}$ plus $C_{31}C_{32}$.
}
\label{fig:RBHNE2}
\end{figure}
$\frac{r_0}{R}$ for four values of the electric charge, namely,
$\frac{q^2}{R^2}=1.8$, $\frac{q^2}{R^2}=2.4$, $\frac{q^2}{R^2}=2.7$,
and $\frac{q^2}{R^2}=3.0$, as indicated in the figure.
In each plot the left part is the axis $\frac{r_0}{R}=0$.
The white region represents the range of the parameter $\frac{r_0}{R}$
where regular black holes are found. The light gray region on the
right  side contains solutions that
do not belong here, the solutions correspond
to
different regular
black holes, namely, the ones whose values of $\frac{r_0}{R}$ are on
or above the curve $C_{31}$ plus $C_{31}C_{32}$, i.e., in the regions
(d2), (e1), and (e2) in Fig.~\ref{fig:regions},
with (e2) only appearing explicitly in 
Fig.~\ref{fig:regionszoom}.  The solid curved line in
each of the four plots is for the vanishing fundamental oscillation
frequency squared, i.e., for $\omega_{0}^{2}=0$, which means that
$\omega_{0}^{2}$ changes sign across such a curve.  All configurations
represented by points located above the $\omega_{0}^{2}=0$ line are
stable regular black holes, i.e., all $\omega_{n}^{2}$ are positive,
all configurations represented by points located below the
$\omega_{0}^{2}=0$ line are unstable regular black holes.
Each solid curved line starts just above
$\frac{r_0}{R}=0$ and extends to some point
$\frac{r_0}{R}$ on the curve $C_{31}$ plus $C_{31}C_{32}$.  The four
plots fall within the range $\frac{q^2}{R^2}>\frac{27}{16}$, so are
to the right of the elbow in curve $C_2$.
In
all the four cases, 
for a fixed adiabatic index $\gamma$ above
about $\gamma_{\rm cr}=3.5$,
all regular black holes are stable
against radial perturbations.
For a fixed adiabatic index $\gamma$ 
 below
about $\gamma_{\rm cr}=3.5$,
there are stable
regular black holes for $\frac{r_0}{R}$
greater than some value. 
For small $\frac{r_0}{R}$ the behavior of
$\gamma_{\rm cr}$ is practically the same for each of the four plots.
For
large $\frac{r_0}{R}$ the decrease in $\gamma_{\rm cr}$ is
very rapid, abrupt in some cases,
when the radius $\frac{r_0}{R}$ approaches the curve
$C_{31}$ plus
$C_{31}C_{32}$.
In the case $\frac{q^2}{R^2}=3.0$ and in the limit
$\frac{r_0}{R}=1$, the critical adiabatic index
indeed vanishes, and the regular black hole is
stable independently of the $\gamma$, it
is absolutely stable to these perturbations.
This is because this case is
of a regular black hole made of an
interior which is purely de Sitter up to the boundary
which is at the Cauchy horizon
radius, $r_0=r_-$, where, in turn,
there is a massless electric coat
\cite{LemosZanchin2011},
and as it is known the de Sitter
solution 
is stable.

In Table~\ref{tab:regularbhn3}, we give details
of the numerical
\begin{table}[h!]
\begin{ruledtabular}
\begin{tabular}{c c}
\textrm{$\frac{r_0}{R}$}&
\textrm{$\gamma_{\rm cr}$}\\
\colrule
0.0186989 & 2.71551 \\
0.158885 & 1.42949 \\
0.299070 & 1.00152 \\
0.439256 & 0.817217 \\
0.579442 & 0.723642 \\
0.719628 & 0.667765 \\
0.859813 & 0.616897 \\
0.999999 & 0.000367 \\
\end{tabular}
\end{ruledtabular}
\caption{The critical adiabatic index $\gamma_{\rm cr}$ for the radial
perturbations of regular black holes with negative energy densities
with $\frac{q^2}{R^2}=3$ and for various values of the parameter
$\frac{r_0}{R}$.  These regular black holes are in a part of region
(d1) of Fig.~\ref{fig:regions}. }
\label{tab:regularbhn3}
\end{table}
results for the stability of regular black holes with
negative energy densities.  The behavior of
$\gamma_{\rm cr}$ as a function of the radius $\frac{r_0}{R}$,
for $\frac{q^2}{R^2}=3$, is displayed.  The
values of the critical adiabatic index $\gamma_{\rm cr}$ are obtained
from the shooting
and the pseudospectral methods, and are
in agreement to each other up to six decimal places.
The solutions for these regular black holes have radii extending from
above $\frac{r_0}{R} = 0$ to  $\frac{r_0}{R}= 1$.
Note that as $\frac{r_0}{R}$ increases the $\gamma_{\rm cr}$
decreases up to zero when
$\frac{r_0}{R}$ is equal to one,
so that this regular black hole is stable for  this
type of perturbations.

\begin{figure}[h!] 
\centering
\includegraphics[width=0.235\textwidth]{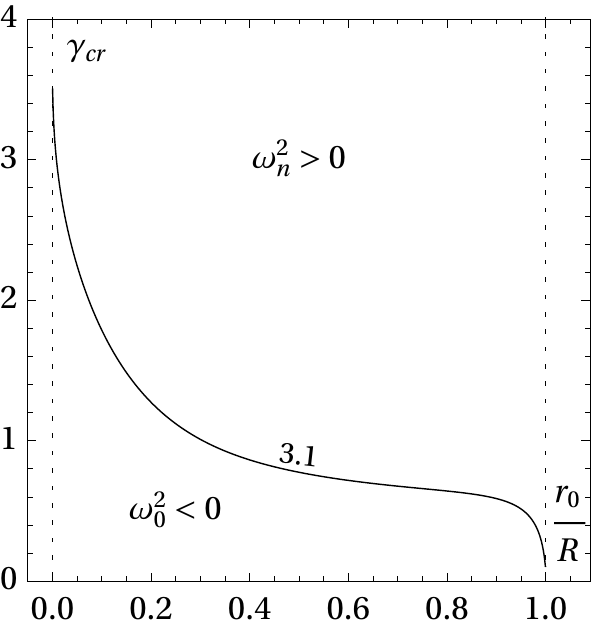}
\includegraphics[width=0.235\textwidth]{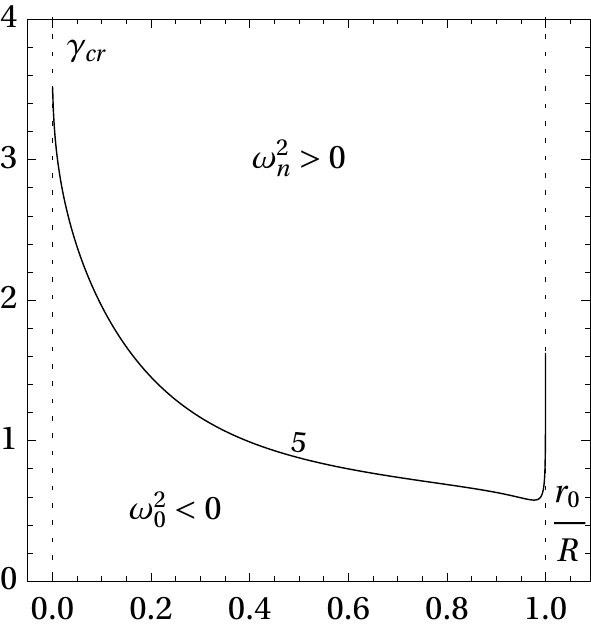}
\includegraphics[width=0.235\textwidth]{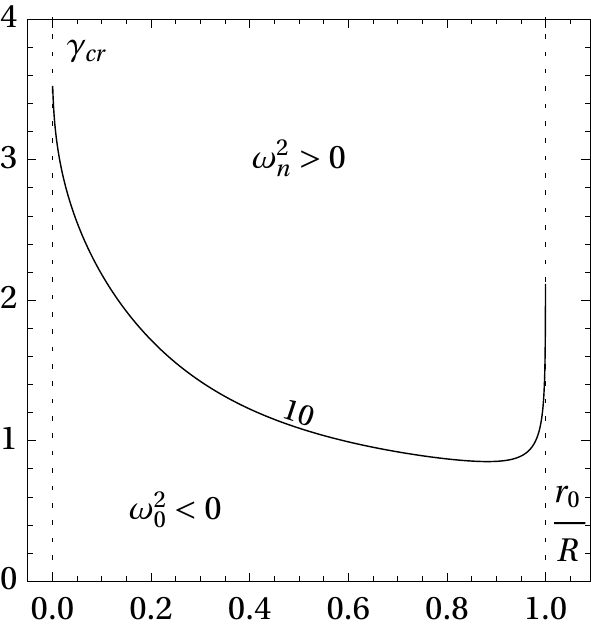}
\includegraphics[width=0.235\textwidth]{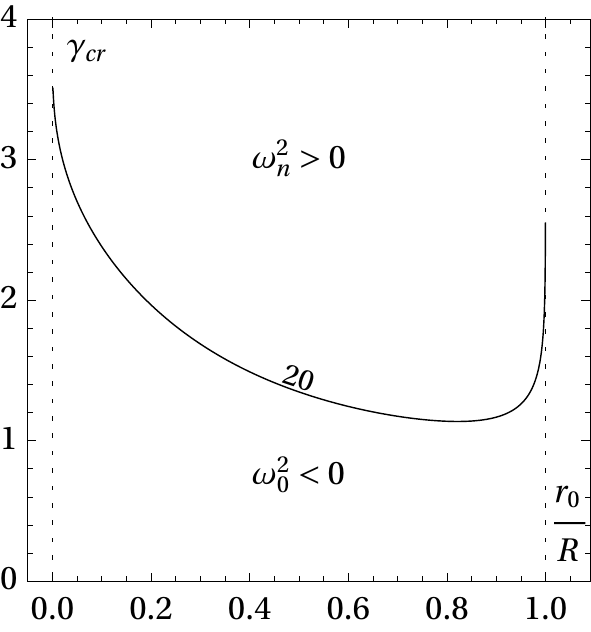}
\caption{
Stability of regular black holes with negative energy density,
continuation from the previous figures, i.e., Figs.~\ref{fig:RBHNE1}
and \ref{fig:RBHNE2}.  These regular black holes belong to region (d1)
to the right of the vertical line $\frac{q^2}{R^2}=3$ and have as
upper boundary the line $r_0=r_-=R$ in Fig.~\ref{fig:regions}.  The
critical adiabatic index $\gamma_{\rm cr}$ for four values of the
electric charge parameter $\frac{q^2}{R^2}=3.1$,
$\frac{q^2}{R^2}=5.0$, $\frac{q^2}{R^2}=10$, and $\frac{q^2}{R^2}=20$,
is shown as a function of the radius $\frac{r_0}{R}$.  In each of the
four plots, the line starts just above $\frac{r_0}{R}=0$ for some
$\gamma_{\rm cr}$ and extends to some value as $\frac{r_0}{R}$ grows
and approaches the line $r_0=r_-=R$. The adiabatic index $\gamma_{\rm
cr}$ decreases up to some large $\frac{r_0}{R}$ and in the last three
plots it starts to grow for large $\frac{r_0}{R}$ reaching a finite
value at the $\frac{r_0}{R}=1$.
}
\label{fig:RBHNE3}
\end{figure}

In Fig.~\ref{fig:RBHNE3}, we show the numerical results for the
\noindent
critical adiabatic index $\gamma_{\rm cr}$ as a function of the radius
$\frac{r_0}{R}$ for four values of the electric charge, namely,
$\frac{q^2}{R^2}=3.1$, $\frac{q^2}{R^2}=5$, $\frac{q^2}{R^2}=10$,
and $\frac{q^2}{R^2}=20$, as indicated in the figure.
In each plot the left part is the axis $\frac{r_0}{R}=0$.
The white region represents the range of the parameter $\frac{r_0}{R}$
where regular black holes are found,
all situated in the region (d1) 
in Fig.~\ref{fig:regions}. The
right part is the axis $\frac{r_0}{R}=1$.
The solid curved
line in each of the four plots is for the vanishing fundamental
oscillation frequency squared, i.e., for $\omega_{0}^{2}=0$, which
means that $\omega_{0}^{2}$ changes sign across such a curve.  All
configurations represented by points located above the
$\omega_{0}^{2}=0$ line are stable regular black holes, i.e., all
$\omega_{n}^{2}$ are positive, all configurations represented by
points located below the $\omega_{0}^{2}=0$ line are unstable regular
black holes.  Each solid curved line starts
just above $\frac{r_0}{R}=0$ and
extends to a point on the line $\frac{r_0}{R}=\frac{r_-} {R}=1$.
The limit $\frac{r_0}{R}\to 1$ with $\frac{q^2}{R^2} >3$ gives the top
boundary of region (d1) of Fig.~\ref{fig:regions}.
For small $\frac{r_0}{R}$ the behavior of
$\gamma_{\rm cr}$ is practically the same for each of the four plots,
it starts at about $\gamma_{\rm cr}=3.5$.
For
large $\frac{r_0}{R}$
and $\frac{q2}{R^2}$ a little larger 
than $3$, $\frac{q2}{R^2}=3.1$ in the
plot, $\gamma_{\rm cr}$ decreases
to some value greater than zero
when
$\frac{r_0}{R}\to1$.
For
large $\frac{r_0}{R}$
and $\frac{q2}{R^2}$ relatively large,
as is shown in the plots
for $\frac{q^2}{R^2}=5$, $\frac{q^2}{R^2}=10$,
and $\frac{q^2}{R^2}=20$,
the increase in $\gamma_{\rm cr}$ is
very rapid, even abrupt,
when the radius $\frac{r_0}{R}$ approaches 
$\frac{r_0}{R}=1$. 
In this limit, the regular black holes
are stable against
radial perturbations
for positive adiabatic indices larger than
some $\gamma_{\rm cr}$. This
$\gamma_{\rm cr}$ increases with the
electric charge, starting from $\gamma_{\rm cr}=0$ at
$\frac{q2}{R^2}=3$,
and diverges in the limit $\frac{q^2}{R^2} \to
\infty$.

In Table~\ref{tab:regularbhn4}, we give details of the numerical
\begin{table}[h]
\begin{ruledtabular}
\begin{tabular}{c c}
\textrm{$\frac{r_0}{R}$}&
\textrm{$\gamma_{\rm cr}$}\\
\colrule
0.0186989 & 2.81135 \\
0.158885 & 1.62442 \\
0.299070 & 1.16791 \\
0.439256 & 0.941473 \\
0.579442 & 0.813839 \\
0.719628 & 0.729194 \\
0.859813 & 0.656414 \\
0.999999 & 1.86624 \\
\end{tabular}
\end{ruledtabular}
\caption{The critical adiabatic index $\gamma_{\rm cr}$ for the radial
perturbations of regular black holes with negative energy densities
with $\frac{q^2}{R^2}=5$ and for various values of
the parameter $\frac{r_0}{R}$.  These regular black holes are in
right part of region (d1) of Fig.~\ref{fig:regions}. }
\label{tab:regularbhn4}
\end{table}
results for the stability of regular black holes with negative energy
densities.  The behavior of $\gamma_{\rm cr}$ as a function of
the radius
$\frac{r_0}{R}$, for $\frac{q^2}{R^2}=5$, is displayed.  The values of
the critical adiabatic index $\gamma_{\rm cr}$ are obtained from the
shooting and the pseudospectral methods, and are in agreement to each
other up to six decimal places.
The solutions for these regular black holes have
boundary radii extending from
above $\frac{r_0}{R} = 0$ to approximately $\frac{r_0}{R}= 1$.  Note
that as $\frac{r_0}{R}$ increases the index $\gamma_{\rm cr}$
decreases to
a minimum at some $\frac{r_0}{R}$ close to $\frac{r_0}{R}=1$, and then
increases with $\frac{r_0}{R}$ up to a finite value in the limit
$\frac{r_0}{R} \to 1$. The change of behavior of $\gamma_{\rm cr}$ in
comparison to the region for smaller electric charges occurs exactly
at $\frac{q^2}{R^2}=3$. Such a change is not visible in the case of
$\frac{q^2}{R^2}=3.1$, whose $\gamma_{\rm cr}$ curve is shown
in the top
left panel of Fig.~\ref{fig:RBHNE3}, because the turning point is very
close to the boundary line.

\subsubsection{Regular black holes with a phantom matter core} 
\label{regbhspmc}

Regular black holes with a phantom matter core have no singularities
and the radius $r_0$ is inside the Cauchy horizon, $r_0<r_-$. These
configurations belong to regions (d2) and (e1) above the curve $C_{31}$
plus $C_{31}C_{32}$ in Fig.~\ref{fig:regions} and below the line
$C_{33}$ of Fig.~\ref{fig:regionszoom}, with
Fig.~\ref{fig:regionszoom} being an enlargement of
Fig.~\ref{fig:regions} in that region of interest.

In the region (d2) the energy density is positive and finite
at the center of the distribution of matter, changes to
negative values at some $r < r_0$, and changes back to positive
values close to the surface, the pressure is negative and,
in modulus is larger than the energy density at the center
of the distribution, and it goes to zero at the surface
$r_0$.
In the region (e1) the energy density is positive
everywhere inside matter and the pressure is negative.  Thus, for a
finite region inside the matter one finds $\rho+p<0$. As a
consequence, the coefficient $F(r)$ is a negative function on the
whole interval $0\leq r\leq r_0$ if $\gamma$ is a positive number,
and is a
positive function on the whole interval $0\leq r\leq r_0$
if $\gamma$ is a
negative number.  The coefficient $W(r)$ is a
negative function in
$0\leq r\leq r_d$, for some $r_d$, and it is positive in
$r_d\leq r\leq r_0$. This case falls into case (D) of the
theorem in the
Appendix~\ref{sec:SLP}, and the behavior of the eigenvalues of the SL
problem is tortuous.  The upshot is that there are no stable
configurations
against radial perturbations
for regular black holes with a phantom matter core, as
it can be seen in more detail in Appendix~\ref{appendixB}.

\subsubsection{Regular tension black holes with positive enthalpy
density}
\label{rbhpositiveh}

Regular tension black holes with positive enthalpy density are not
phantom and obey
$r_0<r_-$.  These configurations belong to region (e2), above the
curve $C_{33}$ of Fig.~\ref{fig:regionszoom} which is an enlargement of
Fig.~\ref{fig:regions} to precisely see this region.

In this case, the energy density is positive everywhere inside the
matter and the pressure is negative, i.e.,
the matter is constrained from bursting by
tension.
The
enthalpy density is always positive $h(r)=\rho(r)+p(r)>0$.  Since
$h(r)>0$ one has that the coefficient $W(r)$
is a positive function. Thus, the
SL problem falls into the case (A) of the theorem of Appendix
\ref{sec:SLP} only if the adiabatic index is a negative number.
In
this case the stable solutions to radial perturbations
are found for $\gamma<\gamma_{\rm cr}$
where $\gamma_{\rm cr}$ here is a negative number,
so in absolute values
$|\gamma|>|\gamma_{\rm cr}|$.

\begin{figure}[h] 
\centering
\includegraphics[width=0.235\textwidth]{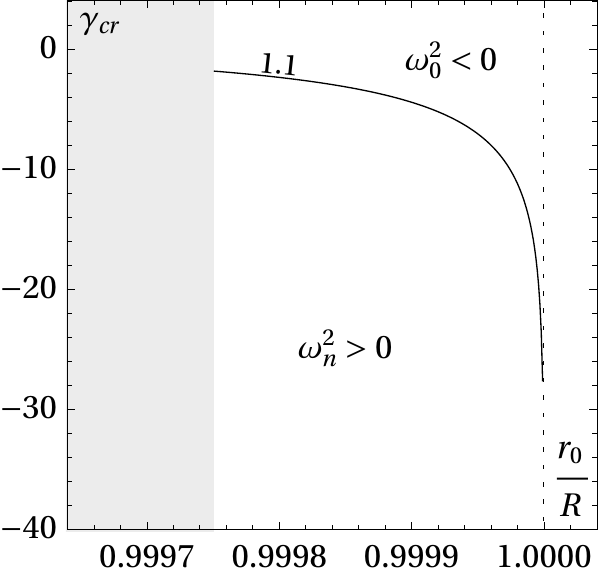}
\includegraphics[width=0.235\textwidth]{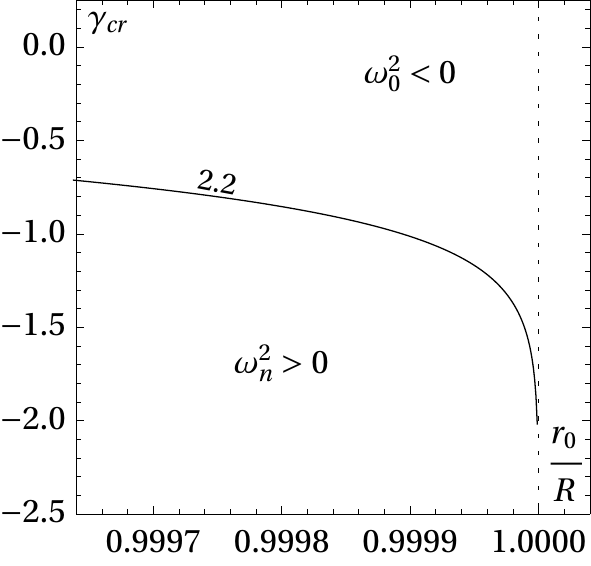}
\centering
\includegraphics[width=0.235\textwidth]{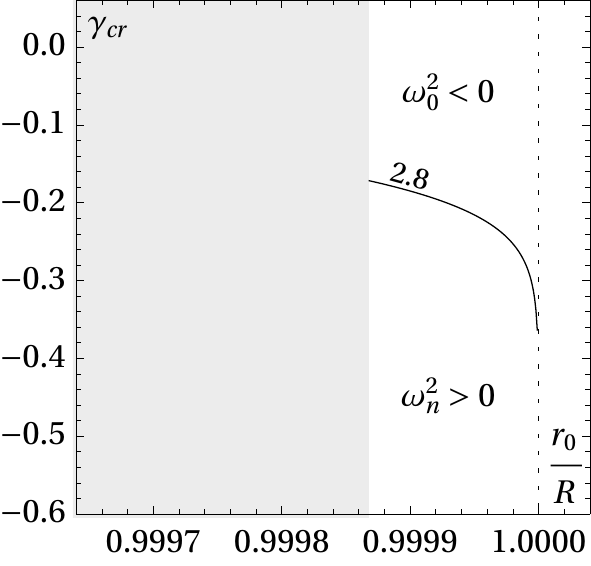}
\includegraphics[width=0.235\textwidth]{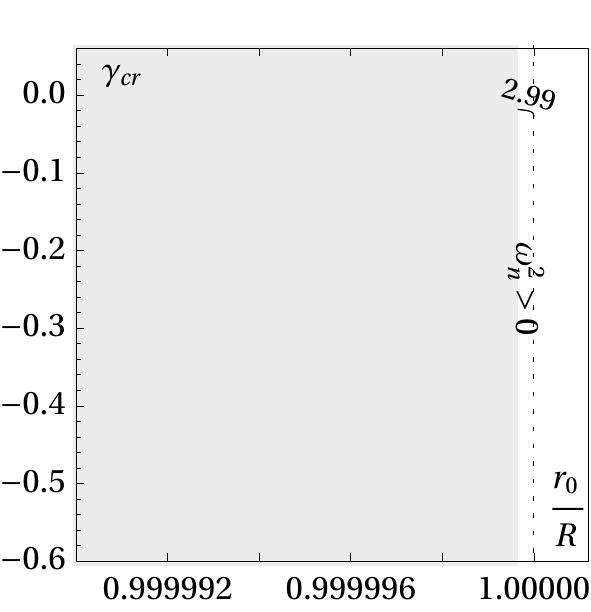}
\caption{
Stability of regular tension black holes with positive enthalpy
density. These regular black holes belong to region (e2), above the
curve $C_{33}$ of Fig.~\ref{fig:regionszoom} which is an enlargement
of Fig.~\ref{fig:regions}.  The critical adiabatic index $\gamma_{\rm
cr}$ for four values of the electric charge parameter
$\frac{q^2}{R^2}=1.1$, $\frac{q^2}{R^2}=2.2$, $\frac{q^2}{R^2}=2.8$,
and $\frac{q^2}{R^2}=2.99$, is shown as a function of the radius
$\frac{r_0}{R}$.  In each of the four plots, the line starts at a
minimum radius $\frac{r_0}{R}$ on the curve $C_{33}$ for which
$\gamma_{\rm cr}$ is negative and for larger $\frac{r_0}{R}$,
$\gamma_{\rm cr}$ becomes more negative up to the line
$\frac{r_0}{R}=1$.  The light gray region on the left side of each
plot corresponds to other different regular tension black holes.}
\label{fig:RTBH}
\end{figure}

In Fig.~\ref{fig:RTBH} we show the numerical results for the critical
adiabatic index $\gamma_{\rm cr}$ as a function of the radius for four
values of the electric charge, namely, $\frac{q^2}{R^2}=1.1$,
$\frac{q^2}{R^2}=2.2$, $\frac{q^2}{R^2}=2.8$, and
$\frac{q^2}{R^2}=2.99$, as indicated in the figure.
In each plot
the light gray region on the left side 
contains solutions that are regular black holes but not of this kind,
they are solutions below line $C_{33}$.
The
white region represents the range of the parameter $\frac{r_0}{R}$
where regular tension black holes with positive enthalpy density are
found. The vertical line
$\frac{r_0}{R}=1$ on the right side marks the end of the plots.
The solid curved line in each of the
four plots is for the vanishing fundamental oscillation frequency
squared, i.e., for $\omega_{0}^{2}=0$, which means that
$\omega_{0}^{2}$ changes sign across such a curve.  All configurations
represented by points located below the $\omega_{0}^{2}=0$ line are
stable regular black holes
against radial perturbations, i.e., all $\omega_{n}^{2}$ are positive,
all configurations represented by points located above the
$\omega_{0}^{2}=0$ line are unstable regular black holes.  Each solid
curved line starts at some $\frac{r_0}{R}$ and extends to
$\frac{r_0}{R}=1$.  The behavior of the critical adiabatic index
$\gamma_{\rm cr}$ is such that it decreases slowly for relatively small
radius, but then it decreases very fast when $\frac{r_0}{R}$ is near
1.  So, in modulus, $|\gamma_{\rm cr}|$ is small for relatively low
$\frac{r_0}{R}$ close to the curve $C_{33}$, and the maximum values of
$|\gamma_{\rm cr}|$ in modulus are obtained for $\frac{r_0}{R}=1$, when
$r_0$ is also equal to $r_-$.  Note also that, for a fixed negative
adiabatic index, the regular black hole configurations are stable for
relatively small radius, but are unstable for large radius.  In the
$\frac{q^2}{R^2}$ near 1 case, of which $\frac{q^2}{R^2}=1.1$ shown is
an example, for $\frac{r_0}{R}$ close to 1 stability is only achieved
for large values of the adiabatic index $\gamma_{\rm cr}$, near
$\gamma_{\rm cr}=-29$.  In the $\frac{q^2}{R^2}$ far from 1 case, of which
$\frac{q^2}{R^2}=2.8$ shown is an example, for $\frac{r_0}{R}$ close
to 1 stability is now achieved for relatively small values of the
adiabatic index $\gamma_{\rm cr}$, near $\gamma_{\rm cr}=-0.34$.
In the limit $\frac{q^2}{R^2}\to 3$, i.e., approaching point $D$ of
Figs.~\ref{fig:regions} and \ref{fig:regionszoom}, of which
$\frac{q^2}{R^2}=2.99$ is the closest value the numerical methods
furnish good results, stability is achieved for all values of the
adiabatic index $\gamma_{\rm cr}<0$, since $\gamma_{\rm cr}\to 0$.
The limit $\frac{r_0}{R}= 1$ with the electric charge in the
interval $1<\frac{q^2}{R^2} <3$ gives the top boundary of region (e2)
of Fig.~\ref{fig:regions}.  As seen in Fig.~\ref{fig:RTBH}, in this
limit, the regular tension black holes with positive enthalpy are
stable against
radial perturbations for negative adiabatic indices smaller
than $\gamma_{\rm cr}$. The
critical adiabatic index decreases in modulus with the electric
charge, starting from arbitrarily large negative values close to
$\frac{q^2}{R^2}=1$, i.e., at the QNBH configuration,
and reaching $\gamma_{\rm cr}=0$ at
$\frac{q2}{R^2}=3$, i.e., at the de Sitter regular black hole
mentioned already.

In Table~\ref{tab:regularbhn7},
\begin{table}[htbp]
\begin{ruledtabular}
\begin{tabular}{c c}
\textrm{$\frac{r_0}{R}$}&
\textrm{$\gamma_{\rm cr}$}\\
\colrule
0.998379 & -0.327587 \\
0.998610 & -0.368522 \\
0.998842 & -0.416910 \\
0.999073 & -0.475289 \\
0.999305 & -0.549655 \\
0.999536 & -0.651395 \\
0.999768 & -0.819164 \\
0.999999 & -2.02034 \\
\end{tabular}
\end{ruledtabular}
\caption{
The critical adiabatic index $\gamma_{\rm cr}$ for the radial
perturbations of regular black holes with  positive enthalpy density
with $\frac{q^2}{R^2}=2.2$ and for various values of
the parameter $\frac{r_0}{R}$.  These regular black holes are in
region (e2) of Fig.~\ref{fig:regionszoom} which is an enlargement of
Fig.~\ref{fig:regions}. 
}
\label{tab:regularbhn7}
\end{table}
we give details of the numerical results for the stability of regular
black holes with positive enthalpy density.  The behavior of
$\gamma_{\rm cr}$ as a function of the radius $\frac{r_0}{R}$, for
$\frac{q^2}{R^2}=2.2$, is displayed.  The values of the critical
adiabatic index $\gamma_{\rm cr}$ are obtained from the shooting and
the pseudospectral methods, and are in agreement to each other up to
six decimal places.
The
solutions for these regular black holes are in the region (e2)
        of Fig.~\ref{fig:regions}  and have boundary
	radii extending from relatively
        high $\frac{r_0}{R}$
	up to $\frac{r_0}{R}= 1$.
Note that
as $\frac{r_0}{R}$ increases the index $\gamma_{\rm cr}$ decreases
negatively, i.e, its modulus $|\gamma_{\rm cr}|$ increases up to a
maximum finite value.  So the regular black holes in this region can
be stable to radial perturbations.

\subsubsection{Regular de Sitter black hole:
$r_0=r_-$ and $\frac{q^2}{R^2}=3$}

Point $D$
in Fig.~\ref{fig:regions}
is a special configuration, a pure de
Sitter interior solution that obeys the equation of state $p(r) =
-\rho(r)$ up to the lightlike surface boundary
$r_0=r_-$, where there is a coat
of electric charge, and where both
the energy density and the pressure
drop to zero. The solution is a regular black hole with a de
Sitter interior
and a lightlike boundary,
and is a particular case of the
regular black holes studied in~\cite{LemosZanchin2011}.

This regular
de Sitter black hole is stable, it has $\gamma_{\rm cr}=0$. The
vanishing of $\gamma_{\rm cr}$ when approaching the point $D$ can be
seen in the bottom right panel of Fig.~\ref{fig:RBHNE2}, where
the curve for $\gamma_{\rm cr}$ was drawn by taking
$\frac{q^2}{R^2}=3.0$. The same behavior is verified in the top left
panel of Fig.~\ref{fig:RBHNE3}, where the curve for $\gamma_{\rm cr}$
was drawn by taking $\frac{q^2}{R^2}=3.1$, as well as in the bottom
right panel of Fig.~\ref{fig:RTBH}, where the curve for $\gamma_{\rm
cr}$ was drawn by taking $\frac{q^2}{R^2}=2.99$.  Our analysis
confirms the previous studies on the stability of regular black holes
with a de Sitter core inferred in \cite{Uchikata:2012zs}. Moreover,
we give a definite answer, that this de Sitter regular black hole is
stable against radial perturbations.

\subsection{The stability of quasiblack holes
and quasinonblack holes}

%\vskip -0.2cm

\subsubsection{Quasiblack holes
from regular undercharged pressure stars}

%\vskip -0.3cm

QBHs from regular undercharged pressure stars are obtained by
approaching the point $Q$ from region (a) of Fig.~\ref{fig:regions},
obey $q^2=m^2$ and also obey $\frac{q^2}{R^2}=1$.

The resulting objects are pressure QBHs that satisfy all
the energy conditions and, as long as the
 parameter $a$ obeys $1< a< \frac43$,
they also satisfy
the causality condition \cite{LemosZanchin2010}.

\begin{figure}[h] 
\centering
\includegraphics[width=0.235\textwidth]{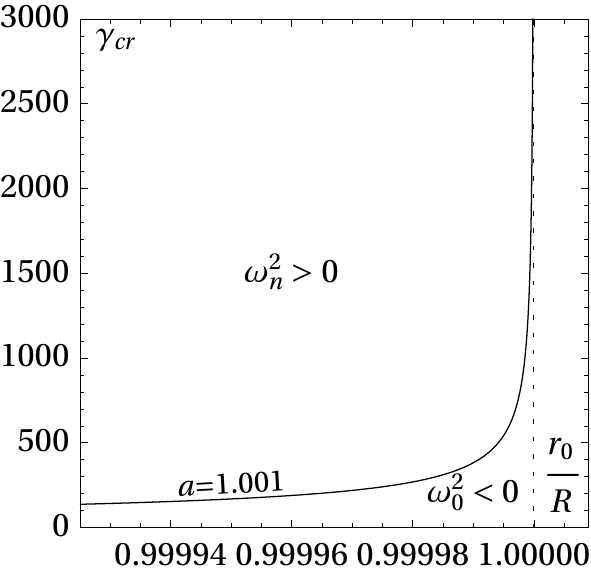}
\includegraphics[width=0.235\textwidth]{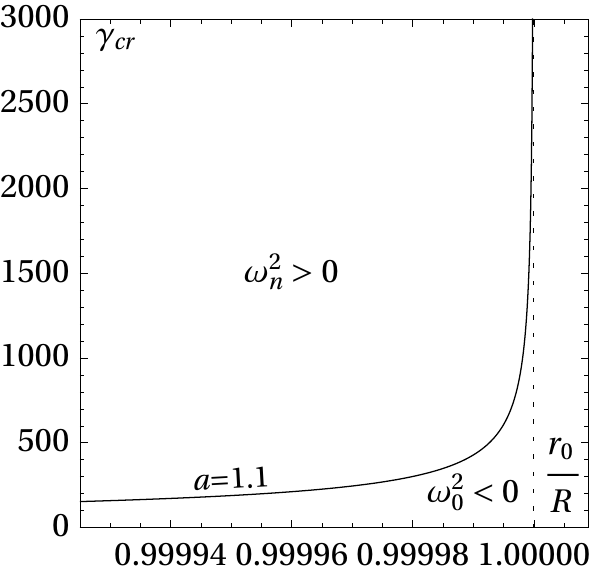}
\includegraphics[width=0.235\textwidth]{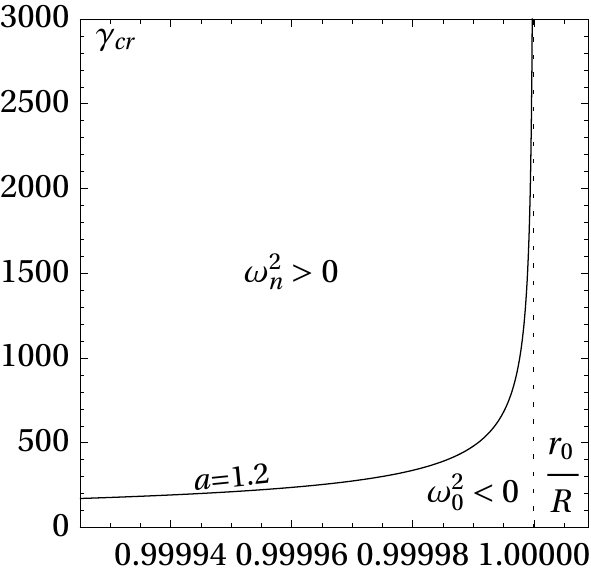}
\centering
\includegraphics[width=0.235\textwidth]{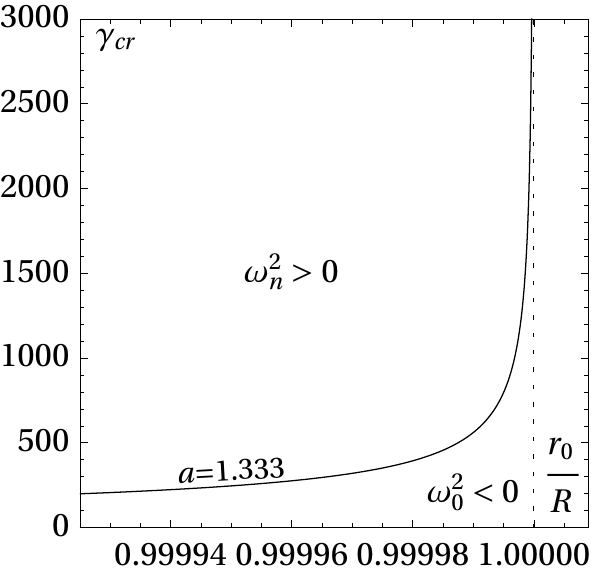}
\caption{
Stability of QBHs from regular undercharged pressure stars.
These QBHs come from approaching point Q from region (a) in
Fig.~\ref{fig:regions}.  The critical adiabatic index $\gamma_{\rm cr}$
for four values of the parameter $a$, namely, $a=1.001$,
$a=1.100$, $a=1.200$, and $a=\frac43=1.333$, is shown as a function of
the radius $\frac{r_0}{R}$.  Each line starts at
$\frac{r_0}{R}=0.9999$ and ends at $\frac{r_0}{R}=0.9999999$.  The
limit $\frac{r_0}{R}=1$ represents QBH configurations.
The critical adiabatic index $\gamma_{\rm cr}$ diverges in this limit.
}
\label{fig:GCMaxQBHNS}
\end{figure}

In Fig.~\ref{fig:GCMaxQBHNS} we show the numerical results for the
critical adiabatic index $\gamma_{\rm cr}$ as a function of the radius
$\frac{r_0}{R}$ for four values of the parameter $a$,
namely, $a=1.001$, $a=1.100$, $a=1.200$, and $a=\frac43=1.333$, as
indicated in the figure.  The solid curved line in each of the four
plots is for the vanishing fundamental oscillation frequency squared,
i.e., for $\omega_{0}^{2}=0$, which means that $\omega_{0}^{2}$
changes sign across such a curve.  All configurations represented by
points located above the $\omega_{0}^{2}=0$ line are stable
pressure stars,
i.e., all $\omega_{n}^{2}$ are positive, all configurations
represented by points located below the $\omega_{0}^{2}=0$ line are
unstable pressure stars.  For $\frac{r_0}{R}$ close to 1,
i.e., on the QBH limit,
one finds that to be
stable to radial perturbations the adiabatic index has to be
arbitrarily large. Thus, the QBH configurations which are obtained by
approaching the point $Q$ from region (a) are unstable unless $\gamma$
assumes arbitrarily large values.

\vskip 5cm
\newpage

In Table~\ref{tab:QBHUPS} we give details of the numerical
results 
\begin{table}[h]
\begin{ruledtabular}
\begin{tabular}{c c c c}
\textrm{$a$}& 
\textrm{$\frac{q^2}{R^2}$}&
\textrm{$\frac{m}{R}$}&
\textrm{$\gamma_{\rm cr}$}\\
\colrule
 1.001 & 0.999105 & 0.999552 & 3859.97\\
 1.100 & 0.999062 & 0.999531 & 4329.59\\
 1.200 & 0.999020 & 0.999510 & 4865.11\\
 1.333 & 0.998967 & 0.999484 & 5689.66\\
\end{tabular}
\end{ruledtabular}
\caption{
The critical adiabatic index $\gamma_{\rm cr}$ for radial
perturbations of undercharged pressure stars, i.e., stars in region
(a) of Fig.~\ref{fig:regions}, for four values of the parameter
$a$ close to the QBH configuration, i.e., for $\frac{r_0}{R}=
0.9999999$. In the QBH limit, $\frac{r_0}{R}= 1$,
$\gamma_{\rm cr}$ diverges.
}
\label{tab:QBHUPS}
\end{table}
for the stability of QBHs from regular undercharged pressure
stars.  The behavior of
$\gamma_{\rm cr}$ as a function of $a$,
$\frac{q^2}{R^2}$, and
$\frac{m}{R}$ is given for the configuration
approaching the QBH limit.
One sees that $\gamma_{\rm cr}$ is high
and in the limit diverges.

\subsubsection{Quasiblack holes from extremal dust stars}

QBHs from extremal dust stars are obtained by approaching the point
$Q$ along the curve $C_0$ of Fig.~\ref{fig:regions}, obey $a=1$,
$q^2=m^2$, and also obey $\frac{q^2}{R^2}=1$.

The resulting objects have charge density equal to mass density,
$\rho_e=\rho$, the pressure is zero and satisfy all the energy
conditions and the causality condition \cite{lemosluz2021}, see also
\cite{LemosZanchin2010}.  These are extremal dust QBHs.

As we have seen in Eq.~(\ref{stabilityextremalstars}) the condition
for stability of the stars along the $C_0$ curve is $\omega^2 \rho(r)
A^{\frac32}(r)\xi(r)=0$.  This means that for nonzero $A$ one has
$\omega^2=0$ and the corresponding stars are neutrally stable.  Now,
for a QBH $A$ is zero at one radius, the gravitational
radius $r_+$, which obeys $A(r_+)=0$.  So $A(r_+)$ is zero at $r_+$
and nonzero for all other points.  So $\omega^2=0$ for all radii
except conceivably at $r_+$. But by continuity we must infer that
$\omega^2=0$ for all radii.  This result confirms what is otherwise
known, namely, that these QBHs from extremal dust stars are
topological objects
\cite{lzjmp}, and so stable to perturbations, in particular are
neutrally stable against radial perturbations which is a remarkable
result.

\subsubsection{Quasiblack holes from overcharged tension stars}

QBHs from regular  overcharged tension stars are obtained by
approaching the point $Q$ from region (b) of Fig.~\ref{fig:regions},
obey
$0<a<1$, 
$q^2=m^2$, and also obey $\frac{q^2}{R^2}=1$.

The resulting objects are tension QBHs and satisfy all
the energy conditions.

\begin{figure}[h] 
\centering
\includegraphics[width=0.235\textwidth]{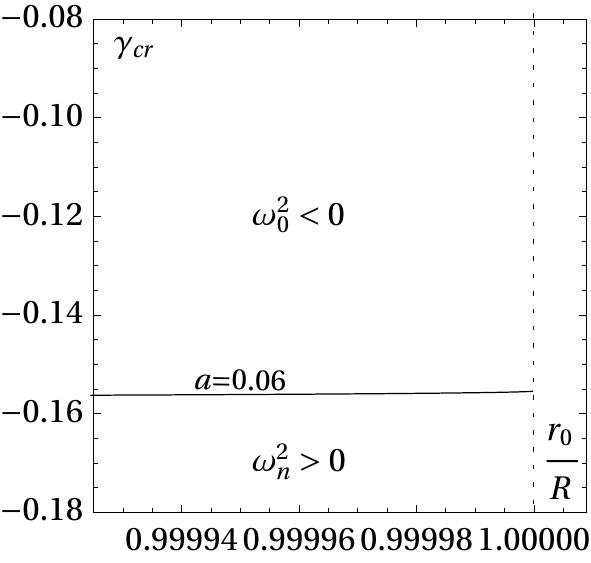}
\includegraphics[width=0.235\textwidth]{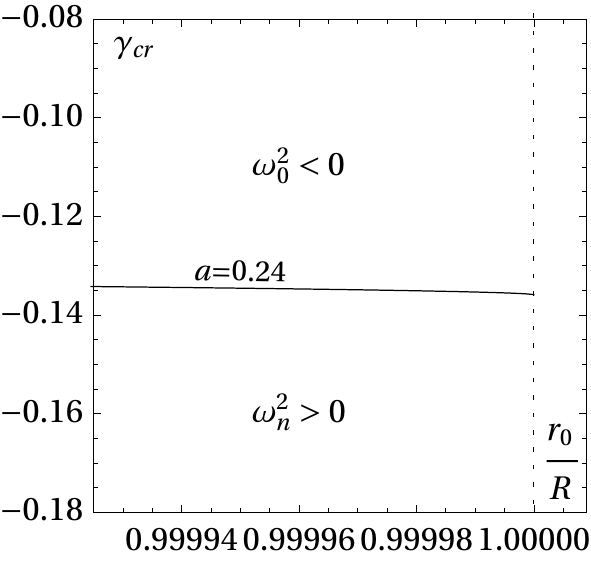}
\includegraphics[width=0.235\textwidth]{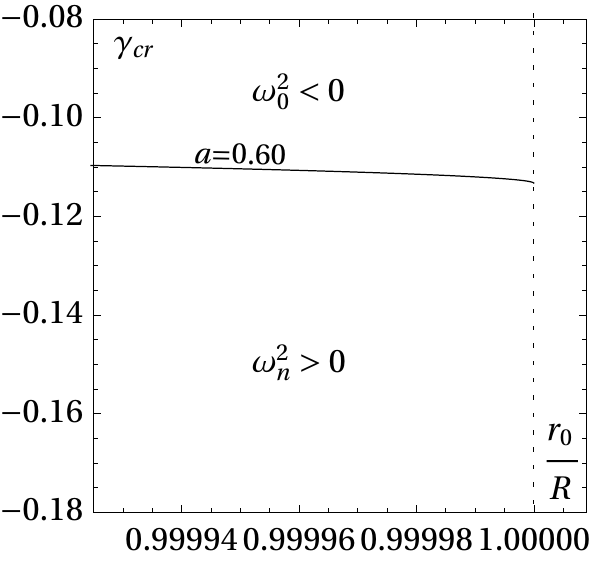}
\centering
\includegraphics[width=0.235\textwidth]{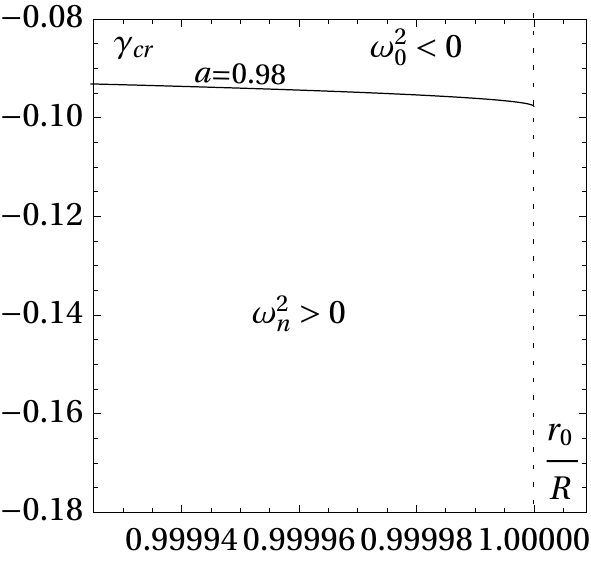}
\caption{
Stability of QBHs from regular overcharged tension stars.
These QBHs come from approaching point $Q$ from region (b) in
Fig.~\ref{fig:regions}.
The critical adiabatic index $\gamma_{\rm cr}$
for four values of the
parameter $a$, namely, $a=0.06$,
$a=0.24$, $a=0.60$, and $a=0.98$, is shown as a function of
the radius $\frac{r_0}{R}$.  Each line starts at
$\frac{r_0}{R}=0.9999$ and end sat $\frac{r_0}{R}=0.9999999$.  The
limit $\frac{r_0}{R}=1$ represents QBH configurations.
The critical adiabatic index $\gamma_{\rm cr}$ is negative
and for a sufficiently high $|\gamma_{\rm cr}|$ these QBHs
are stable against radial perturbations.
}
\label{fig:GCMinQBHTS}
\end{figure}

In Fig.~\ref{fig:GCMinQBHTS} we show the numerical results for the
critical adiabatic index $\gamma_{\rm cr}$ as a function of the radius
$\frac{r_0}{R}$ for four values of the parameter $a$,
namely, $a=0.06$, $a=0.24$, $a=0.60$, and $a=0.98$, as indicated in
the figure.
The solid curved line in each of the four plots is for
the vanishing fundamental oscillation frequency squared, i.e., for
$\omega_{0}^{2}=0$, which means that $\omega_{0}^{2}$ changes sign
across such a curve.  All configurations represented by points located
below the $\omega_{0}^{2}= 0$ line are stable tension stars, i.e., all
$\omega_{n}^{2}$ are positive, all configurations represented by
points located above the $\omega_{0}^{2}= 0$ line are unstable tension
stars.  For $\frac{r_0}{R}$ close to 1, one sees that to be stable to
radial perturbations the critical adiabatic index, $\gamma_{\rm cr}$,
is negative, and has an almost constant value.
In the limit of $\frac{r_0}{R}=1$, with $0<a<1$,
i.e., at the QBH limit from
tension stars, all $|\gamma_{\rm cr}|$ are finite. Thus, the QBH
configurations which are obtained by approaching the point $Q$ from
region (b) are stable to radial perturbations for a sufficiently
moderate $|\gamma_{\rm cr}|$.

\vskip 5cm
\newpage

In Table~\ref{tab:QBHOTS} we give details of the numerical results for
\begin{table}[h]
\begin{ruledtabular}
\begin{tabular}{c c c c}
\textrm{$a$}& 
\textrm{$\frac{q^2}{R^2}$}&
\textrm{$\frac{m}{R}$}&
\textrm{$\gamma_{\rm cr}$}\\
\colrule
 0.06 & 0.999781 & 0.999890 & -0.165394\\
 0.24 & 0.999562 & 0.999781 & -0.135874\\
 0.60 & 0.999307 & 0.999653 & -0.113167\\
 0.98 & 0.999115 & 0.999557 & -0.0975539\\
\end{tabular}
\end{ruledtabular}
\caption{
The critical adiabatic index $\gamma_{\rm cr}$ for radial
perturbations of overcharged tension stars, i.e., stars in region
(b) of of Fig.~\ref{fig:regions}, for four values of the parameter
$a$ close to the QBH configuration, i.e., for $\frac{r_0}{R}=
0.9999999$.
In the QBH limit, $\frac{r_0}{R}=1$,
$\gamma_{\rm cr}$ is negative and finite, and the
object is stable against radial perturbations for
sufficiently high negative
adiabatic index.
}
\label{tab:QBHOTS}
\end{table}
the stability of QBHs from regular overcharged tension stars.  The
behavior of $\gamma_{\rm cr}$ as a function of $a$, $\frac{q^2}{R^2}$,
and $\frac{m}{R}$ is given, i.e., for the configurations approaching
the QBH limit.  One sees that $\gamma_{\rm cr}$ is negative and finite
at the limit.

%\newpage
\subsubsection{Quasinonblack holes from regular black holes}

QNBHs from regular black holes  are obtained by
approaching the point $Q$ from region (d2) and (e1)
 of
Fig.~\ref{fig:regions}
and (e2) of
Fig.~\ref{fig:regionszoom}, this latter being
an amplification of
Fig.~\ref{fig:regions}
in order that region (e2) pops out.
Now, all regular black holes
from region (d2) and (e1) are unstable, so we do not
need to treat their approach to point $Q$. 
On the other hand, some black holes
from region (e2) 
are stable to radial perturbations, so
it is of great interest to
treat their approach to point $Q$. These QNBHs
obey $a>4$, $q^2<m^2$, and also obey $\frac{q^2}{R^2}=1$.

The resulting objects are tension QNBHs,
which have an additional
important property, namely, they satisfy the dominant
and weak energy conditions.

\begin{figure}[h] 
\centering
\includegraphics[width=0.235\textwidth]{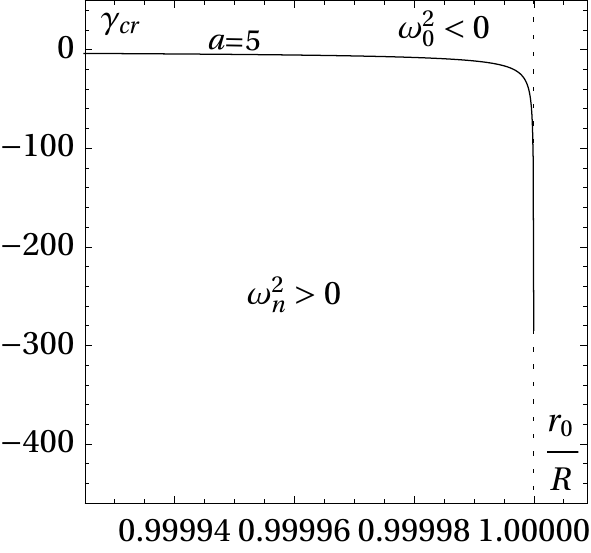}
\includegraphics[width=0.235\textwidth]{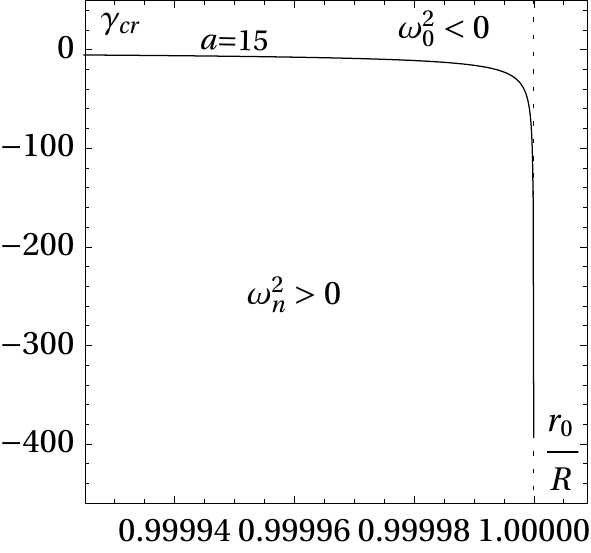}
\includegraphics[width=0.235\textwidth]{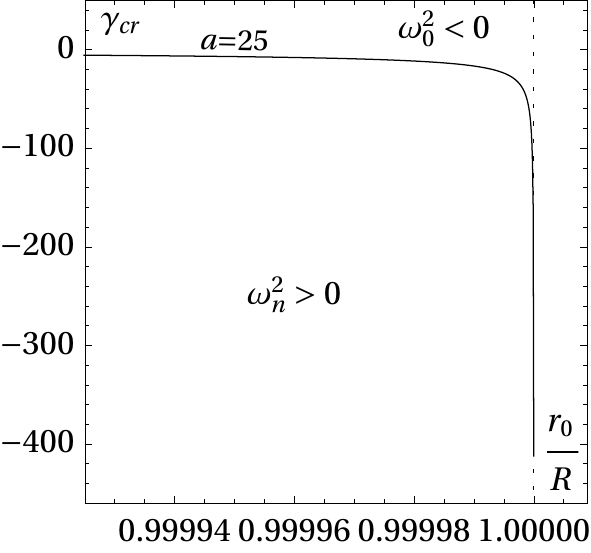}
\centering
\includegraphics[width=0.24\textwidth]{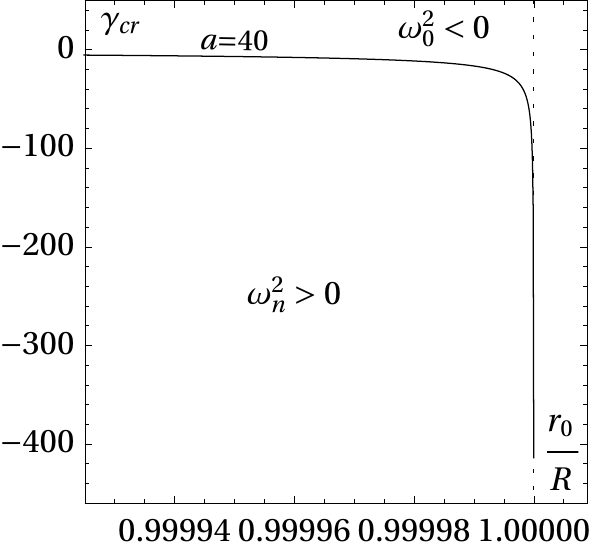}
\caption{
Stability of QNBHs from regular black holes.  These QNBHs come from
approaching point Q from region (e2) in Fig.~\ref{fig:regionszoom}
which is an ampliation of Fig.~\ref{fig:regions}.  The critical
adiabatic index $\gamma_{\rm cr}$ for four values of the Guilfoyle
parameter, $a=5$, $a=15$, $a=25$, and $a=40$, is shown as a function
of the radius $\frac{r_0}{R}$.  Each line starts at
$\frac{r_0}{R}=0.9999$ and ends at $\frac{r_0}{R}=0.9999999$.  The
limit $\frac{r_0}{R}=1$ represents QNBH configurations, and in this
limit $|\gamma_{\rm cr}|$ is high, of the order of $400$, so 
QNBHs with higher
$\gamma$ are stable
against radial perturbations for high $|\gamma_{\rm cr}|$.  }
\label{fig:GCMinQBHRBHSE} 
\end{figure}

In Fig.~\ref{fig:GCMinQBHRBHSE}, we show the numerical results for the
critical adiabatic index $\gamma_{\rm cr}$ as a function of the
radius $\frac{r_0}{R}$ for four values of the parameter $a$,
namely, $a=5$, $a=15$, $a=25$, and $a=40$, as
indicated in the figure.  The solid curved line in each of the four
plots is for the vanishing fundamental oscillation frequency squared,
i.e., for $\omega_{0}^{2}=0$, which means that $\omega_{0}^{2}$
changes sign across such a curve.
All configurations represented by
points located below the $\omega_{0}^{2}=0$ line are stable
regular black holes,
i.e., all $\omega_{n}^{2}$ are positive, all configurations
represented by points located above the $\omega_{0}^{2}=0$ line are
unstable regular black holes.
For $\frac{r_0}{R}$ close to 1, one sees that to be
stable to radial perturbations, the critical
adiabatic index, $\gamma_{\rm cr}$, has to be negative, and
that in the QNBH limit
$|\gamma_{\rm cr}|$ assumes arbitrarily large
values, and so the objects are
effectively unstable to radial perturbations
in this limit.

In Table~\ref{tab:QBHRTBH}, we give details of the numerical results
for the stability of QNBHs from regular black holes.  The behavior of
$\gamma_{\rm cr}$ as a function of $a$, $\frac{q^2}{R^2}$, and
$\frac{m}{R}$ is given, i.e., for the configurations approaching the
QNBH limit.

\begin{table}[h]
%\caption{\label{tab:QBHRTBH3} Behavior of the critical adiabatic index
%$\gamma_{\rm cr}$ in region (e2) for four values of the Guilfoyle
%parameter $a$ close the QNBH configuration, i.e., $\frac{r_0}{R}=0.9999999$.
%}
\begin{ruledtabular}
\begin{tabular}{c c c c}
\textrm{$a$}& 
\textrm{$\frac{q^2}{R^2}$}&
\textrm{$\frac{m}{R}$}&
\textrm{$\gamma_{\rm cr}$}\\
\colrule
 5 & 1.00200 & 1.00100 & -119.107\\
 15 & 1.00347 & 1.00174 & -164.433\\
 25 & 1.00448 & 1.00224 & -172.479\\
 40 & 1.00567 & 1.00284 & -173.115\\
\end{tabular}
\end{ruledtabular}
\caption{ The critical adiabatic index $\gamma_{\rm cr}$ for radial
perturbations of regular black holes, i.e., black holes in region (e2)
of Fig.~\ref{fig:regions}, for four values of the parameter $a$
close to the QNBH configuration, i.e., for $\frac{r_0}{R}=
0.9999999$. In the QBH limit, $\frac{r_0}{R}= 1$, $\gamma_{\rm cr}$
diverges.}
\label{tab:QBHRTBH}
\end{table}

\section{Conclusions
\label{sec:final}}

\subsection{ Main results
\label{sec:finalmainresults}}

We have studied the stability of several types of electrically charged
objects in general relativity, namely, the stability to radial
perturbations for stars, regular black holes, QBHs, and QNBHs has been
performed.

We have combined the theorems regarding the eigenvalues of the SL
problem to find that the stability of these compact objects depend on
the sign of the pressure $p(r)$ and on the sign of the enthalpy
density $h(r)=\rho(r)+p(r)$, which in turn have implications to the
sign of the adiabatic index $\gamma$.  Using all this and two powerful
numerical methods we were able to discover that there are objects that
can be stable against radial perturbations, in which case we have
determined the critical adiabatic index $\gamma_{\rm cr}$, and there
are objects that are always unstable.  The index $\gamma_{\rm cr}$,
when it exists, is such that objects with an adiabatic index $\gamma$
in modulus higher than $\gamma_{\rm cr}$ in modulus are stable against
radial perturbations.

Zero charge stars are the best objects to start the analysis because
they are the simplest and one can have a solid ground of direct
comparison with Chandrasekhar's analysis. We have found that stars
with small radius, and thus smaller mass, and so do not have too much
gravitation, are stable configurations against radial perturbations
for adiabatic indices with moderate values. These stars are far from
the Buchdahl limit.  Stars with large radius, and thus bigger mass,
have too much gravitation, and are generically unstable
configurations. These stars are near the Buchdahl limit. Our results
for these zero charge stars conform qualitatively and quantitatively
with Chandrasekhar's results.

Undercharged stars are supported by pressure.  Stars with small
radius, and thus smaller mass, and so do not have too much
gravitation, are stable configurations against radial perturbations
for adiabatic indices with moderate values.  These stars are far from
the Buchdahl-Andr\'easson limit.  Stars with large radius, and thus
bigger mass, have too much gravitation, and so are unstable
configurations.  These stars are near the Buchdahl-Andr\'easson limit.

Extremal charged stars are made of dust and, with small or large
radius, are neutrally stable. By itself the star neither expands nor
collapses.

Overcharged stars are supported by tension and the matter
perturbations are characterized by a negative adiabatic index.  Stars
with small radius, and thus smaller mass, are strongly repelled by the
electric charge, and so are essentially unstable.  Stars with large
radius, and thus higher mass, have enough gravitation counterbalancing
the electric repulsion, and so are stable against radial perturbations
for adiabatic indices with moderate values in modulus.

Regular black holes with negative energy density and with phantom
matter can be stable against radial perturbations if the adiabatic
index is sufficiently high, i.e., higher than $\gamma_{\rm cr}$ and as
such are of interest.  These regular black holes have a rich structure
in the stability analysis.

Regular black holes with positive energy density at the center and
with phantom matter are unstable for any values of the adiabatic
indices.

Regular black holes with positive enthalpy, and so not phantom black
holes, are stable for radial perturbations with negative adiabatic
index with sufficiently high values in modulus, i.e., for
$|\gamma|\geq |\gamma_{\rm cr}|$.

QBHs stability was one of the main motivations of this work.  We have
found that the results in the case of QBHs configurations depend on
how the QBH limit is reached because this type of solution is
degenerate. The results yield that the QBHs from undercharged stars
are unstable, unless the adiabatic index is arbitrarily large, QBHs from
extremal stars seem to be stable against radial perturbations, and
QBHS from overcharged stars are stable for reasonable negative
adiabatic indices.  QNBHSs are a recently new type of object, and we
have been able to perform a stability analyses, which indicates that
they are stable against radial perturbations for reasonable negative
adiabatic indices.
\vskip -5cm

\subsection{Summary tables}

In Tables \ref{tablesummary1} and \ref{tablesummary2}
we give a summary of the results about the stability of stars,
regular black holes, QBHs, and QNBHs.

In Table \ref{tablesummary1} (a), i.e., top part, we show the various
intervals numerically found for which there is stability for stars and
regular black holes.  The table is set as follows.
The first column represents the type of studied object.
The second column yields the interval of $\frac{r_0}{R}$ for which
there is stability, in case there is neutral stability for any
$\frac{r_0}{R}$ in the interval $0\leq\frac{r_0}{R}\leq1$ it is written
the word neutral and in case there is no stability in the interval
$0\leq\frac{r_0}{R}\leq1$ it is written the word no, noting that
the equalities should be read as numerically found, i.e, they are
approximate equalities.
The third column indicates the value of $\frac{q^2}{R^2}$ used to
calculate the interval of $\frac{r_0}{R}$ when there is stability, and
in the case of regular black holes with a phantom matter core for
which there is no stability, the interval of values of
$\frac{q^2}{R^2}$ where this region exists is indicated.
The fourth column gives the value of the adiabatic index $\gamma$ used
in the numerical calculation.
In Table \ref{tablesummary1} (b), i.e., bottom part, we present again
stability for stars and regular black holes but now in qualitative
terms giving the general features.  The table is set as the one before
with respect to the rows.
The first column represents the type of  object studied.
The second column yields whether there is stability or not with the
words yes, neutral, or no, for $\frac{q^2}{R^2}$ and $\gamma$ without
showing the intervals of $\frac{r_0}{R}$.
The third column indicates the interval of $\frac{q^2}{R^2}$ for which
there are stable solutions, with the regular black holes with
negative energy densities showing that we can find regions of
stability for $\frac{q^2}{R^2}>0$.
The fourth column gives the sign of the adiabatic index $\gamma$,
either positive or negative, which gives stability, without giving
numerical values.

In Table \ref{tablesummary2}, we present the
stability of QBHs and QNBHs
in qualitative terms giving the general features.
The table is set as follows.  The first column
represents the type of studied object.  The second column expresses
whether there is stability or not with the words yes, neutral, or no.
The third column indicates the interval of the parameter $a$ for which
there are stability solutions, in the case of QBHs from regular
undercharged pressure stars there are no stable solutions for $a>1$.  The
fourth column gives the sign of the adiabatic index $\gamma$, either
positive or negative, which gives stability, without giving numerical
values.

%\vskip 5cm

\begin{widetext}

\onecolumngrid
\begin{table}[hbtp]
\begin{ruledtabular}
\begin{tabular}{l c c c}
\textrm{Configurations}& 
\textrm{Stability}&
\textrm{$\frac{q^2}{R^2}$}&
\textrm{$\gamma$}\\
\colrule
% Zero electric charged stars  & $< \frac{r_0}{R} < $ &  & \\
 Undercharged pressure stars & $0.915703 \leq \frac{r_0}{R}
 \leq 0.942943 $ & $0.3$ & $4$\\
 Extremally charged dust stars & neutral & $0$ -- $1$ & all\\
 Overcharged tension stars & $0.938387  \leq \frac{r_0}{R}
 \leq 0.978111 $ & $0.6$ &
$-0.06$\\
 Regular black holes with negative energy densities &
 $0 \leq  \frac{r_0}{R} 
 \leq  0.866025 $ & $1.6875 $ & $4$\\
 Regular black holes with a phantom matter core & no & $1$ -- $3$ & all\\
 Regular tension black holes with positive enthalpy
density & $0.998842 \leq  \frac{r_0}{R}  \leq  0.999999 $ &
$2.2$ & $-0.4$\\
Regular de Sitter black hole & $\frac{r_0}{R} =\frac{r_-}{R} =1$ & $3$ & all\\
%\end{tabular}
%\end{ruledtabular}
%\end{table}
%
%\begin{table}[hbtp]
%\caption{Summary of the results about stability of stars and regular
%black holes: Generic cases.}
%\begin{ruledtabular}
%\begin{tabular}{l c c c}
\colrule
\textrm{Configurations}& 
\textrm{Stability}&
\textrm{$\frac{q^2}{R^2}$}&
\textrm{$\gamma$}\\
\colrule
% Zero electric charged stars  & $< \frac{r_0}{R} < $ &  & \\
 Undercharged pressure stars & yes & $0$ -- $1$ & positive\\
 Extremally charged dust stars & neutral & $0$ -- $1$ & all\\
 Overcharged tension stars & yes & $0$ -- $1$ & negative\\
 Regular black holes with negative energy densities & yes & $ \neq 0 $ &
positive\\
 Regular black holes with a phantom matter core & no & $1$ -- $3$ & all\\
 Regular tension black holes with positive enthalpy
density & yes & $1$ -- $3$ & negative\\
Regular de Sitter black hole & yes & $3$ & all\\
\end{tabular}
\end{ruledtabular}
\caption{Summary of the results on the stability of stars and regular
black holes: (a) Specific results, (b) Generic results.}
\label{tablesummary1}
\end{table}
%
%\end{widetext}

%\begin{widetext}

\begin{table}[hbtp]
\begin{ruledtabular}
\begin{tabular}{l c c c}
\textrm{Configurations}& 
\textrm{Stability}&
\textrm{$a$}&
\textrm{$\gamma$}\\
\colrule
 QBHs from regular undercharged pressure stars & no & $>1$ & $<\infty$\\
 QBHs from extremal dust stars & neutral & $1$ & all\\
 QBHs from overcharged tension stars & yes & $0$ -- $1$ & negative\\
 QNBHs from regular black holes & yes & $>4$ & negative\\
\end{tabular}
\end{ruledtabular}
\caption{Summary of the results on the stability of QBHs and QNBHs.
 }
% \vskip 0.5cm
\label{tablesummary2}
\end{table}

\end{widetext}

%\newpage
\centerline{}
\newpage
\centerline{}
\newpage

\section*{Acknowledgments}
ADDM was financed  by Coordena\c c\~ao de Aperfei\c coamento de
Pessoal de N\'ivel Superior (CAPES), Brazil, Finance Code 001.
JPSL acknowledges Funda\c{c}\~{a}o para a Ci\^{e}ncia e Tecnologia -
FCT, Portugal, for financial support through Project
No.~UIDB/00099/2020.
VTZ thanks CAPES, Brazil, Grant No. 88887.310351/2018-00, and
Conselho Nacional de Desenvolvimento Cient\'ifico e Tecnol\'ogico
(CNPq), Brazil, Grant No.~309609/2018-6.

\appendix

\section{Consistency of the Einstein-Maxwell-electric matter
system of equations}
\label{sec:c}

The Einstein-Maxwell equations with electrically charged matter
presented in Sec.~\ref{sec:basic}, specifically, in
Sec.~\ref{basicbasic}, Eqs.~\eqref{eq:Einstein}-\eqref{eq:tensorj}, is
a consistent system of equations.  We now show this consistency.  To
be self-contained and for ease of referencing the equations in the
deduction of the consistency we repeat the full set of equations.

The
two
Einstein-Maxwell equations with electrically charged matter
are
\begin{equation}\label{eq:EinsteinApp}
G_{\mu\nu}=8\pi T_{\mu\nu},
\end{equation}
\begin{equation}\label{eq:MaxwellApp}
\nabla_{\nu}F^{\mu\nu}=4\pi J^{\mu},
\end{equation}
where
$G_{\mu\nu}$ is the Einstein tensor, $T_{\mu\nu}$ is the
energy-momentum tensor, $\nabla_{\mu}$ represents the covariant
derivative, $F_{\mu\nu}$ is the Faraday-Maxwell
electromagnetic tensor, $J^{\mu}$
is the charge current density,
and Greek indices range from $0$ to $3$, $0$ corresponding
to a timelike coordinate $t$, and $1,2,3$ to spatial coordinates.
The Einstein tensor
$G_{\mu\nu}$ is a function of the metric
$g_{\mu\nu}$ and its first two derivatives,
not needed to be written explicitly here.
There are two distinct contributions to
the energy-momentum tensor
$T_{\mu\nu}$. One contribution comes from
the matter and its energy-momentum
tensor is denoted by $M_{\mu\nu}$.
The other contribution comes from the
electromagnetic field and  its energy-momentum
tensor is denoted by $E_{\mu\nu}$.
So, $T_{\mu\nu}$ can be written as
\begin{equation}\label{eq:tensortotApp}
T_{\mu\nu}=M_{\mu\nu}+E_{\mu\nu}\,.
\end{equation}
The matter energy-momentum
tensor $M_{\mu\nu}$
is assumed to be a perfect fluid
 energy-momentum
tensor, so that
\begin{equation}\label{eq:tensormatterApp}
M_{\mu\nu}=\left(\rho+p\right)u_{\mu}u_{\nu}+pg_{\mu\nu},
\end{equation}
where $\rho$ is the fluid matter energy density,
$p$ is the isotropic
fluid pressure, and $u_{\mu}$ is
the fluid's four-velocity.
The electromagnetic energy-momentum
tensor $E_{\mu\nu}$
has the expression
\begin{equation}\label{eq:tensorchargeApp}
E_{\mu\nu}=\frac{1}{4\pi}\left({F_{\mu}}^{\gamma}F_{\nu\gamma} 
-\frac{1}{4}g_{\mu\nu}F_{\gamma\beta}F^{\gamma\beta}\right)\,.
\end{equation}
The covariant derivative $\nabla_{\mu}$
is defined through the Levi-Civita connection. 
The Faraday-Maxwell tensor
$F_{\mu\nu}$ is defined in terms of a
vector potential $\mathcal{A}_{\mu}$ by 
\begin{equation}\label{eq:tensormaxApp}
F_{\mu\nu}=\nabla_{\mu}\mathcal{A}_{\nu}-
\nabla_{\nu}\mathcal{A}_{\mu}\,.
\end{equation}
With this definition one can see that
$F_{\mu\nu}$ obeys 
the internal Maxwell
equations $F_{[\mu\nu;\rho]}=0$, with all
the three indices being antisymmetrized.
The
current density of an electrically charged fluis
has the expression
\begin{equation}\label{eq:tensorjApp}
J^{\mu}=\rho_{e}u^{\mu}, 
\end{equation}
where $\rho_{e}$ is the electric charge density.

To show that the system of equations given in
Eqs.~(\ref{eq:EinsteinApp})-(\ref{eq:tensorjApp}) is consistent
we start by using 
the contracted Bianchi identities $\nabla_\nu G^{\mu\nu}=0$.
This amounts to $\nabla_\nu T^{\mu\nu}=0$, with
$T^{\mu\nu}=M^{\mu\nu}+E^{\mu\nu}$, see Eq.~(\ref{eq:tensortotApp}).
Now, $\nabla_\nu M^{\mu\nu}= [\nabla_\nu (\rho+p)]u^\mu u^\nu
+(\rho+p)[ (\nabla_\nu u^\mu) u^\nu+u^\mu(\nabla_\nu u^\nu) ]
+(\nabla_\nu p) g^{\mu\nu}$ and $\nabla_\nu E^{\mu\nu}=-J_\nu
F^{\mu\nu}$ where in the latter equation
full use of all the Maxwell equations has been made.
Then, we cross $\nabla_\nu T^{\mu\nu}=0$ with $u_\mu$.  Crossing
$\nabla_\nu M^{\mu\nu}$ with $u_\mu$, and using $u_\mu u^\mu=-1$ and
so $u_\mu\nabla_\nu u^\mu=0$, one obtains
$u_{\mu}\nabla_{\nu}M^{\mu\nu}=
 u^{\nu}\nabla_{\nu}\rho+ (\rho+p)\nabla_{\nu}u^{\nu}$.
Crossing $\nabla_\nu E^{\mu\nu}$ with
$u_\mu$, one obtains $u_\mu\nabla_\nu E^{\mu\nu}=-u_\mu J_\nu
F^{\mu\nu}=0$ since $u_\mu J_\nu$ is symmetric in $\mu\nu$ and
$F^{\mu\nu}$ antisymmetric. So $u_\mu\nabla_\nu T^{\mu\nu}=0$ implies
$\nabla_\nu( \rho u^\nu)
+p\nabla_{\nu}u^{\nu}
= 0$, 
which is the energy conservation equation for the
matter, and when $p=0$ turns into  the continuity equation,
i.e.,
$\nabla_\nu( \rho u^\nu)=0$.
Now we use the projection tensor $P_{\rho\sigma}=g_{\rho\sigma}+
u_\rho u_\sigma$ to act on $\nabla_\nu T^{\mu\nu}=0$.  First, we have
$P_{\rho\mu}\nabla_\nu M^{\mu\nu}= P_{\rho\mu}(\rho+p)(\nabla_\nu
u^\mu) u^\nu+ P_{\rho\mu}(\nabla_\nu p)g^{\mu\nu}$, i.e.,
$P_{\rho\mu}\nabla_\nu M^{\mu\nu}= (\rho+p)u^\nu(\nabla_\nu u_\rho)+
\nabla_\rho p+ u_\rho u^\nu(\nabla_\nu p)$,
where we have used that
$P_{\rho\mu}u^\mu u^\nu= (g_{\rho\mu}+ u_\rho u_\mu)u^\mu u^\nu=u_\rho
u^\nu-u_\rho u^\nu=0$, and $P_{\rho\mu}(\rho+p)u^\mu(\nabla_\nu u^\nu)
= (g_{\rho\mu}+ u_\rho u_\mu)(\rho+p)u^\mu(\nabla_\nu u^\nu)
=(\rho+p)[u_\rho(\nabla_\nu u^\nu)-u_\rho(\nabla_\nu u^\nu) ]=0 $, and
the other identities.  Second, we have $P_{\rho\mu}\nabla_\nu
E^{\mu\nu} =-P_{\rho\mu} J_\nu F^{\mu\nu}=-J_\nu {F_\rho}^\nu$,
where we
again used $u_\mu J_\nu$ is symmetric in $\mu\nu$ and $F^{\mu\nu}$
antisymmetric.  Thus, $P_{\rho\mu}\nabla_\nu T^{\mu\nu}=0$ implies
that
$(\rho+p)u^\nu(\nabla_\nu u_\rho)+ \nabla_\rho p+
u^\nu(\nabla_\nu p) u_\rho -J_\nu {F_\rho}^\nu=0$, which is the
relativistic Euler equation with a Lorentz force term as it should.
Also, clearly, one has from Eq.~(\ref{eq:MaxwellApp})
that $\nabla_\mu J^\mu=0$, which is the continuity
equation for the electric current. So
the whole setup presented in Sec.~\ref{sec:basic}, specifically,
in Sec.~\ref{basicbasic}
is consistent.

\section{Derivation of the full set of the dynamical
perturbation equations}
\label{manipulation}

Here we give the derivation of the full set of perturbation equations
of Sec.~\ref{sec:basic}, specifically, of
Sec.~\ref{theperturbationequations}.  So we have to derive
Eqs.~\eqref{eq:deltaA}-\eqref{eq:deltaQ}.

We proceed as follows.
Integrating Eq.~\eqref{eq:eint01} we arrive at $\delta A=-8\pi
r(\rho_{i}+p_{i})A_{i}^{2}\xi$ which upon using the $tt$ component of
the Einstein-Maxwell equations yields $\delta A=-A_{i}
\left(\dfrac{A_{i}^{\prime}}{A_{i}}+
\dfrac{B_{i}^{\prime}}{B_{i}}\right)\xi$, which is
Eq.~(\ref{eq:deltaA}).
From Eqs.~\eqref{eq:eint00} and \eqref{eq:eint11}, and with the help
of Eq.~(\ref{eq:deltaQL}), we can find an expression for the
perturbation $\delta B$ as $ \left(\dfrac{\delta
B}{B_{i}}\right)^{\prime}=8\pi
A_{i}\left[2rp_{i}^{\prime}-\left(\rho_{i}+p_{i}\right)\right]\xi\
+8\pi A_{i}r\delta p-\dfrac{2A_{i}Q_{i}Q_{i}^{\prime}\xi}{r^{3}} $,
which is Eq.~(\ref{eq:deltaB}).
From Eq.~\eqref{eq:eint00} and the equation resulting from the
integration of Eq.~\eqref{eq:eint01}, we can find an expression for
the perturbations $\delta\rho$, namely,
$\delta\rho=-\rho_{i}^{\prime}\xi-
(\rho_i+p_i)\dfrac{B_{i}^{\frac12}}{r^{2}}
\left(r^{2}B_{i}^{-{\frac12}}\xi\right)^{\prime}$, which is
Eq.~(\ref{eq:deltarho}).  The Lagrangian perturbation $\Delta \rho$ is
obtained by using relation~\eqref{eq:deltarho} and the fact that the
Lagrangian and the Eulerian perturbations are linked by the
relationship given in Eq.~\eqref{eq:LagEul}, resulting in
$\Delta\rho=-(\rho_i+p_i)\dfrac{B_{i}^{{\frac12}}}{r^{2}}
\left(r^{2}B_{i}^{-{\frac12}}\xi\right)^{\prime}$.
Using the definition for $\gamma$ given in Eq.~\eqref{eq:adindex} and
the equation for $\Delta\rho$ just derived, one gets $\Delta
p=-\gamma\dfrac{p_{i}B_{i}^{{\frac12}}}{r^{2}}
\left(r^{2}B_{i}^{-{\frac12}}\xi\right)^{\prime}$, or, in terms of the
Eulerian perturbation, $\delta
p=-p_{i}^{\prime}\xi-\gamma\dfrac{p_{i}B_{i}^{{\frac12}}}{r^{2}}
\left(r^{2}B_{i}^{-{\frac12}}\xi\right)^{\prime}$, which is
Eq.~(\ref{eq:deltaP}).
The equation of motion for $\xi$, Eq.~(\ref{eq:motion2}), is simply
Eq.~(\ref{eq:motion}) written more appropriately to the perturbation
problem.
The equation for the perturbed charge, Eq.~(\ref{eq:deltaQ}), is
essentially Eq.~(\ref{eq:deltaQL}), and it is worth noting that it
implies directly that $ \Delta Q=0 $, i.e., the electric charge is
conserved when a Lagrangian perturbation is performed, a fact that
comes out directly from the conservation of electric charge, i.e.,
$\nabla_\mu J^\mu=0$.

\section{Sturm-Liouville problem}
\label{sec:SLP}

Here we comment on the Sturm-Liouville problem, 
see Sec.~\ref{sec:basic}, sepcifically,
Sec.~\ref{convenientSturm-Liouvilleform}.

Standard manipulation of the perturbation equation for the
electrically charged fluid under study, Eq.~\eqref{eq:eigenvalue},
leads to a
second order ordinary homogeneous differential
equation
for the displacement $\zeta$,
Eq.~\eqref{eq:SLP}, which is again displayed here as
\begin{equation}\label{eq:SLPappendix}
F(r)\zeta^{\prime\prime}(r)+
F^{\prime}(r)
\zeta^{\prime}(r)+\left[H(r)+\omega^2
W(r)\right]\zeta(r)=0,
\end{equation}
where we have used $\zeta(r)=r^2 B_{i}^{-{\frac12}}\xi(r)$. The coefficients
$F(r)$,  $H(r)$, and $W(r)$ are given by
\begin{equation}
F(r)=\dfrac{\gamma p_i B_{i}^{{\frac32}}A_{i}^{{\frac12}}}{r^2},
\end{equation}
\begin{equation}
\begin{split}
H(r)=& \dfrac{B_{i}^{{\frac32}}A_{i}^{{\frac12}}}{r^{2}}
\left[\dfrac{1}{(\rho_{i}+p_{i})}\left(\dfrac{Q_{i}Q_{i}^{\prime}}
{4\pi r^{4}}-p_{i}^{\prime} \right)^{2}\right.\\
 &\left.- 
\dfrac{4p_{i}^{\prime}}{r} -8\pi A_{i}(\rho_{i}+p_{i})
\left( p_{i}+\dfrac{Q_{i}^{2}}{8\pi r^{4}}\right)
\right], 
\end{split}
\end{equation}
\begin{equation}
W(r)=\dfrac{(\rho_{i}+p_i) B_{i}^{{\frac12}}A_{i}^{{\frac32}}}{r^2}.
\end{equation}
Equation~\eqref{eq:SLPappendix} defines a homogeneous
Sturm-Liouville problem
or SL problem for short. 

Here we state some known theorems regarding the eigenvalues of
the SL problem that are important for our work,
see e.g.~\cite{KongZettl1996,Moller1999,ZettlBook}.
Consider the differential equation
\begin{equation}\label{SLP:1}
(F\zeta^{\prime})^{\prime}+H\zeta+\lambda W\zeta=0,\,\,
\textrm{in}\,\,I=(a,\,b)\,,
\end{equation}
with $-\infty<a<b<\infty$ and the boundary conditions
\begin{equation}\label{SLP:2}
\alpha_1\zeta(a)+\alpha_2 F(a)\zeta^{\prime}(a)=0,
\end{equation}
\begin{equation}\label{SLP:3}
\beta_1\zeta(b)+\beta_2 F(b)\zeta^{\prime}(b)=0,
\end{equation}
where $\alpha_1$ and $\alpha_2$ are not both zero, similarly for
$\beta_1$ and $\beta_2$, and with the coefficients satisfying
\begin{equation}\label{SLP:4}
F,\,H,\,W:\,(a,\,b)\rightarrow\mathbb{R},\,\,\,
\frac1F,\,H,\,W\,
\in\,L(I,\,\mathbb{R})\,,
\end{equation}
where $\mathbb{R}$ denotes the set of real numbers, and
$L(I,\,\mathbb{R})$
denotes the space of real valued Lebesgue integrable
functions in $I$.
Let \eqref{SLP:1}-\eqref{SLP:3} hold in $I$, and take the following
considerations~\cite{KongZettl1996,Moller1999,ZettlBook}:
\begin{itemize}
\item[(A)] Assume that $W>0$ and $F>0$ almost everywhere in $I$.
Then, the boundary value problem \eqref{SLP:1}-\eqref{SLP:3} has only
real and simple eigenvalues. There are an infinite but countable
number of eigenvalues that are bounded from below and can be ordered
to satisfy the inequalities
\begin{equation}
-\infty<\lambda_0<\lambda_1<\lambda_2<\lambda_3<\cdots,
\end{equation}
with $\lambda_n\rightarrow\infty$ as $n\rightarrow\infty$. If
$\zeta_n$ is
an eigenfunction of $\lambda_n$, then $\zeta_n$
has exactly $n$ zeros in
the open interval $(a,\,b)$.

\item[(B)] Assume that $W>0$ and $F< 0 $ almost everywhere in $I$.
Then, the boundary value problem \eqref{SLP:1}-\eqref{SLP:3} has only
real and simple eigenvalues. There are an infinite but countable
number of eigenvalues that are bounded from above and can be ordered
to satisfy the inequalities
\begin{equation}
\cdots<\lambda_{-2}<\lambda_{-1}<\lambda_0<\infty,
\end{equation}
with $\lambda_{-n} \rightarrow -\infty$ as $n\rightarrow\infty$.  

\item[(C)] Assume that  $W>0$ and that $F$ changes sign in $I$.
Then, the boundary value problem \eqref{SLP:1}-\eqref{SLP:3} has only
real and simple eigenvalues. There are an infinite but countable
number of eigenvalues that are unbounded from below and from above and
can be ordered to satisfy
\begin{equation}\label{SLPEV:2}
\cdots<\lambda_{-2}<\lambda_{-1}<\lambda_0<\lambda_1<\lambda_2<\cdots,
\end{equation}
with $\lambda_n\rightarrow\infty$ as $n\rightarrow\infty$, and
$\lambda_n\rightarrow-\infty$ as $n\rightarrow-\infty$. If
$\zeta_n$ is an
eigenfunction of $\lambda_n$, then
$\zeta_n$ has exactly $|n|$ zeros in
the open interval $(a,\,b)$. And $\lambda_0$ is chosen as the first
nonnegative eigenvalue in \eqref{SLPEV:2}.

\item[(D)] Assume that $F>0$ and $W$ that changes sign in $I$.
Then, the boundary value problem\eqref{SLP:1}-\eqref{SLP:3} has only
real and simple eigenvalues. There are an infinite but countable
number of eigenvalues that are unbounded from below and from above and
can be ordered to satisfy
\begin{equation}\label{SLPEV:3}
\cdots<\lambda_{-2}<\lambda_{-1}<\lambda_0<\lambda_1<\lambda_2<\cdots,
\end{equation}
with $\lambda_n\rightarrow\infty$ as $n\rightarrow\infty$, and
$\lambda_n\rightarrow-\infty$ as $n\rightarrow-\infty$.
\end{itemize}

The above theorems may be applied to the stability problems considered
in the main text by noting that the finite interval $[a, b]$
translates into the interval $[0,r_0]$ in the radial coordinate
$r$, where $r_0$ is the radius of
the boundary of the matter.

\section{Numerical methods}
\label{sec:numeric}

\subsection{Shooting method}

The pulsation equation, being an eigenvalue SL problem, can be solved
using the shooting method
\cite{PressBook1992,KongZettl1996,Moller1999,ZettlBook}.
This is one of the methods mentioned in Sec.~\ref{sec:basic},
specifically, Sec.~\ref{Subec:Numericalmethods}.
This method is implemented to find the eigenvalues of the equation
which in this case are the normal frequencies of the normal modes.
The shooting method is based on the reduction of a boundary value
problem to the solution of an initial value problem.  In concrete, the
idea of the method is to solve the differential pulsation equation
given in Eq. (32) by performing its integration from the center at
$r =
0$ toward the surface at $r_ 0$ using a Runge-Kutta integration with an
adaptive stepsize for a succession of trial
values of $\omega^2$, see, e.g., ~\cite{PressBook1992}.

To apply
in practice
this method it is advisable to transform
Eq.~\eqref{eq:eigenvalue} into two first order differential
equations. For that we have to return to the full set of perturbed
equations given in Eqs.~\eqref{eq:deltaA}-\eqref{eq:deltaQ}.
To simplify the whole scheme,
we substitute $\xi$ for a dimensionless variable
$\chi(r)$ defined by $\chi(r)=\dfrac{\xi(r)}{r}$. Then, using
Eq.~\eqref{eq:deltaP} for $\delta p$ and recalling that the Lagrangian
variation is $\Delta p=\delta p+p^\prime \xi$ we can write from the
very
same Eq.~\eqref{eq:deltaP} $\chi^{\prime}$ as
\begin{equation}\label{eq:chiL}
\chi^{\prime}=-\dfrac{3\chi}{r}-\dfrac{\Delta p}{\gamma \,r \,
p}+\dfrac{\chi}{(\rho+p)}\left(\dfrac{QQ^{\prime}}{4\pi 
r^{4}}-p^{\prime}\right),
\end{equation}
where Eq.~\eqref{eq:equiHidro} was also used to eliminate $B'$
in terms of the fluid quantities.
Now, using $\Delta p=\delta p+p^\prime \xi$ in 
Eq.~\eqref{eq:motion2}
one has the following equation for $\Delta p^{\prime}$
\begin{equation}
\begin{array}{ccl}
\Delta p^{\prime}&=& \omega^{2}r\dfrac{A}{B}(\rho+p)\chi
 -8\pi rA(\rho+p)p\chi\\
\\ &&- 4p^{\prime}\chi +\dfrac{r 
}{(\rho+p)}\left(\dfrac{QQ^{\prime}}{4\pi r^{4}}-
p^{\prime}\right)^{2}\chi\\
\\&&-(\rho+p)\dfrac{A Q^{2}\chi}{r^{3}}-4\pi
rA(\rho+p)\Delta p\\
\\&&-\dfrac{\Delta p}{(\rho+p)}\left(\dfrac{QQ^{\prime}}{4\pi 
r^{4}}-p^{\prime}\right)\,,
\end{array} \label{eq:deltap}
\end{equation}
where
Eqs.~\eqref{eq:deltaA}-\eqref{eq:deltaQ}
have also been used.
For a given $\omega^2$,
Equations~\eqref{eq:chiL} and \eqref{eq:deltap} form a first order
differential system of two equations for the
two unknowns $\chi$ and $\Delta p$.

To guarantee a regular solution
the imposition of
regular boundary conditions is mandatory.
Since $\xi(r)=\chi(r)r$,
the boundary condition $\xi(r=0)=0$,
turns into $\chi(r=0)\times0=0$, which is automatically
satisfied if $\chi(r=0)$ is finite. One can choose
any finite number and we choose 
\begin{equation}\label{chi0}
\chi(r=0) = 1\,.
\end{equation}
In many concrete problems one has that
at $r=0$, $p^\prime=0$ and $Q^\prime=0$. Imposing
also that $\chi^\prime(r)=0$ at $r=0$, which
one can always do, one finds
from Eq.~\eqref{eq:chiL}
that  Eq.~\eqref{chi0} is then equivalent to
$\Delta p(r=0)=-3\,\gamma\, p$ at $r=0$,
which is a helping equation
to start the numerical calculations.
The boundary condition at the
boundary $r=r_0$ is the same as before, i.e.,
\begin{equation}
\Delta p(r=r_0)=0\,.
\label{b2}
\end{equation}

With the two boundary conditions of 
Eqs~\eqref{chi0} and \eqref{b2}, 
Eqs~\eqref{eq:chiL} and \eqref{eq:deltap} form a first order
differential system of two equations that can 
be now solved numerically for the
two unknowns $\chi$ and $\Delta p$, when
the correct $\omega^2$ is found.
We note that in the
uncharged case, for specific
neutron star models, this strategy has been employed
in \cite{Vaeth1992,Gondek1997}.
We use Fortran 77 to implement the shooting
method, see, e.g.,~\cite{PressBook1992}.

\subsection{Chebyshev finite difference method}

The pulsation equation, as an eigenvalue SL problem, can be solved
using other methods besides the shooting method.  The other method
that we use here is the Chebyshev finite difference
method~\cite{Elgendi1969,Boyd19892013,Elbar2003,TMM2013,jansen17}
which is an instance of generic pseudospectral methods.
This is the other method mentioned in Sec.~\ref{sec:basic},
specifically, Sec.~\ref{Subec:Numericalmethods}.

The pseudospectral methods are powerful tools which represent an
efficient discretization technique for obtaining approximate numerical
solutions of differential, integral, and integro-differential
equations~\cite{Elgendi1969}.  The basic idea is considering that the
unknown solution $\zeta(r)$ of the Sturm-Liouville (SL) boundary value
problem,
as given in Eq.~\eqref{eq:SLP} 
can be approximated as a sum of a
finite set of known basis functions.
The basic functions to choose are
important because they depend on the properties of the system under
study~\cite{Boyd19892013}. A good choice is the Chebyshev functions
of the first kind defined by
\begin{equation}
    T_n(x)=\cos(n\,\arccos{x}),
\end{equation}
with $n$ running over the natural numbers, as these 
present excellent properties to approximate smooth functions.
The Chebyshev functions
are a well-known family of orthogonal polynomials in the
interval $x\in [-1,\,1]$, which can be rescaled and shifted to any
other interval.  Given this property, it is convenient to map the
domain of the radial coordinate $r$ of our problem to the domain of
these polynomials, i.e., we want to rescale the interval $[0,\,r_0]$
to $[-1,\,1]$. For this we do $r=\frac12\,(x+1) r_0$. i.e.,
\begin{equation}
x=\dfrac{2r}{r_0}-1 \,,
\end{equation}
and so $r\in [0,\,r_0]$
is mapped into $x\in [-1,\,1]$.

The formal  solution
to the perturbation problem
can be put in the form
of an infinite sum  of
Chebyshev functions
\begin{equation}\label{zeta0}
\zeta(x)=
\sum_{n=0}^{\infty}a_n
    T_n(x)\,,
\end{equation}
where
the $a_n$ are given by
$a_n=\int_{-1}^1 \zeta(x) T_n(x) dx$.
To solve it numerically
one has to approximate the infinite
sum in Eq.~\eqref{zeta0}
by a finite sum $\zeta_N (x)$
defined up to a number $N$.
So 
\begin{equation}\label{zeta1}
\zeta_N (x)=
\sum_{n=0}^{N}\theta_n
\,a_n
    T_n(x),
\end{equation}
where now 
the $a_n$ are given by
$a_n=\int_{-1}^1 \theta_n\,\zeta_N(x) T_n(x) dx$
and in these truncated Chebyshev
sums it is understood
that the first and
the last terms in the series are multiplied by the factor $\frac12$,
so that the auxiliary variable $\theta_n$ was created so that
$\theta_0=\theta_N=\frac12$ and the other $\theta_n$ are
given by 
$\theta_n=1$. 
This
numerical
approach
further
requires the definition of a grid which is a
discretization of the domain in which the problem is to be
solved. This means that the continuous independent variable
$x$ is
replaced by a discrete set of points called Chebyshev-Gauss-Lobatto
points, and are such that
\begin{equation}
x_k=\cos\left(\dfrac{k\pi}{N}\right),\quad k=0,\,1,\,2,...,N.
\end{equation}
Thus, at each $x_k$ we can write from Eq.~\eqref{zeta1}
\begin{equation}\label{zeta2}
\zeta_N(x_k)=
\sum_{n=0}^{N}\theta_n\,a_n
    T_n(x_k),
\end{equation}
where now 
$\displaystyle{a_{n}=\dfrac{2}{N}
\sum_{k=0}^{N}
\theta_n\,\zeta_N(x_{k})T_{n}(x_{k})}$.
The derivatives of $\zeta(x)$ that enter into
the problem are then expanded as a linear
combination from the values of the function $\zeta_N(x)$ at the
Chebyshev-Gauss-Lobatto points
$\zeta_N(x_k)$. Thus, the calculation process to obtain the
value of the $m$-th order derivative of $\zeta_N(x)$
at a given grid
point $x_k$ reduces to a matrix operation given by
\begin{equation}
    \zeta_N^{(m)}(x_k)=\sum_{j=0}^{N} C_{kj}^{(m)}\zeta_N(x_j)\,,
\end{equation}
where the first and
second of the coefficients $ C_{kj}^{(m)}$ are given by
\begin{equation}
C_{kj}^{(1)}=\dfrac{4\theta_j}{N}\sum_{n=0}^{N}
   \sum_{\substack{l=0\\(n+l)\,\textrm{odd}}}^{n-1
   }\dfrac{n\theta_n}{\alpha_l}T_n(x_j)T_l(x_k)
\end{equation}
\begin{equation}
C_{kj}^{(2)}=\dfrac{2\theta_j}{N}\sum_{n=0}^{N}
   \sum_{\substack{l=0\\(n+l)\,\textrm{even}}}^{n-2}
   \dfrac{n(n^2-l^2)\theta_n}{\alpha_l}T_n(x_j)T_l(x_k)
\end{equation}
where the subscripts $j,\,k$ run from $0$ to
$N$,
$\theta_0=\theta_N=\frac12$,
$\theta_n=1$ for $n=1,...,N-1$, 
$\alpha_0=2$,
$\alpha_l=1$ for $l=1,...,N-1$, see~\cite{Elbar2003}.
The general expressions for the
derivatives $\zeta_N^{(m)}$ can be found in~\cite{TMM2013}.

This procedure allows us to discretize the initial differential
problem into a system of algebraic equations that
the set of the expansion
coefficients must satisfy.
Since the pulsation equation,
see Eq.~\eqref{eq:SLP},
is linear, these
algebraic equations can be cast as a matrix equation generically of
the form of a generalized eigenvalue problem,
\begin{equation}
(\mathbf{F}+\omega^{2}\,\mathbf{W})\,\mathbf{Z}=0\,,
\end{equation}
where, $(\mathbf{F})_{kl}=F(x_k)\, C_{kl}^{(2)}+G(x_k)\,
C_{kl}^{(1)}+H(x_k)\,\delta_{kl}$,
$(\mathbf{W})_{kl}=W(x_k)\,\delta_{kl}$ are two purely numerical
square matrices constructed from the coefficients of
Eq.~\eqref{eq:SLP}, and $(\mathbf{Z})_k=\zeta_N(x_k)$ is the
vector with the unknown values of the eigenfunction at the $N+1$ grid
points. The last two rows of the coefficients matrix of the algebraic
system are replaced by a suitable formulation of the boundary
conditions in terms of the polynomial approximation and its
derivatives~\cite{TMM2013}.
This can be solved numerically using, for instance,
Mathematica's built-in function Eigenvalues, or Eigensystem to get the
eigenfunctions as well, see
\cite{jansen17} for an application.
We use Mathematica packages to implement the Chebyshev method.

\vskip -0.7cm
\centerline{}

\section{
More tables and comments} 
\label{appendixB}
\vskip -0.4cm

To complete the text on regular undercharged stars,
i.e., stars with $0<q^2<m^2$, see
Sec.~\ref{Sec:UPS}, we present the
Table~\ref{tab:underchargeappendixB}
for stars with $\frac{q^2}{R^2}=0.3$,
which completes
Table~\ref{tab:undercharge1}.  The two first columns of
Table~\ref{tab:underchargeappendixB} are also given in
Table~\ref{tab:undercharge1}, whereas columns third and fourth are new
and give the fundamental frequency squared $\omega_{0}^{2}$ and the
first overtone frequency squared $\omega_{1}^{2}$ for a matter fluid
with $\gamma=4$. 
\begin{table}[h]
\begin{ruledtabular}
\begin{tabular}{c c c c}
\textrm{$\frac{r_0}{R}$}& 
\textrm{$\gamma_{\rm cr}$}& 
\textrm{$\omega_{0}^{2}$}&
\textrm{$\omega_{1}^{2}$}\\
\colrule
$0.915704$ & $2.95794$ & $1.64116\times10^{-6}$ &
$4.35847\times10^{-5}$\\
$0.924784$ & $3.32947$ & $7.72620\times10^{-3}$ & $0.334711$\\
$0.933863$ & $3.86695$ & $2.34262\times10^{-3}$ & $0.541389$\\
$0.942943$ & $4.70936$ & $-0.0135501$ & $0.628574$\\
$0.952022$ & $6.20193$ & $-0.0371488$ & $0.602781$\\
$0.961102$ & $9.47855$ & $-0.0650874$ & $0.469199$\\
$0.970181$ & $21.1295$ & $-0.0924624$ & $0.237346$\\
$0.979261$ & $440359$ & $-0.111286$ & $3.17674\times10^{-6}$\\
\end{tabular}
\end{ruledtabular}
\caption{
For regular undercharged stars, $0<q^2<m^2$, with
$\frac{q^2}{R^2}=0.3$, in columns one and two, several $\frac{r_0}{R}$
are given along with their own $\gamma_{\rm cr}$. In columns three and
four, the eigenfrequencies $\omega_{0}^{2}$ and $\omega_{1}^{2}$ are
given for $\gamma=4$. The transition from stability to
instability of the star goes when $\omega_{0}^{2}$ goes from positive
to negative.
}
\label{tab:underchargeappendixB}
\vskip -0.2cm
\end{table}
One clearly sees from the table that there is a
change from stability to instability when $\frac{r_0}{R}$ goes from
$\frac{r_0}{R}=0.933863$ to $\frac{r_0}{R}=0.942943$.
One also sees that for small relative radii, $\frac{r_0}{R}$, the stars are
electrically charged stars with very small pressure, they are near the curve
$C_0$ of electrically charged dust stars, and  both $ \omega_0^{2}$ and
$\omega_1^{2}$ take values close to zero, as it should be from our discussion
in the main text.

%\vskip 0.3cm

To complete the text on regular overcharged tension stars,
i.e., stars with $m^2<q^2$, see
Sec.~\ref{Sec:OTS}, we present the
Tables~\ref{tab:overcharge1appendixB}
and~\ref{tab:overcharge2appendixB}.
Table~\ref{tab:overcharge1appendixB}, for
stars with $\frac{q^2}{R^2}=0.6$, presents in the first column the
radii $\frac{r_0}{R}$
for regular
overcharged stars.
\begin{table}[h]
\begin{ruledtabular}
\begin{tabular}{c c c c c}
\textrm{$\frac{r_0}{R}$}& 
\textrm{$\gamma_{\rm cr}$}&
\textrm{$\omega_{0}^{2}$}&
\textrm{$\omega_{1}^{2}$}\\
\colrule
0.880113 & 1.65801 & -1.32771 & -13.2864\\
0.894113 & 1.73187 & -0.974978 & -10.7426\\
0.908113 & 1.84207 & -0.650766 & -8.07796\\
0.922113 & 2.00277 & -0.387085 & -5.60955\\
0.936112 & 2.24897 & -0.193301 & -3.49922\\
0.950112 & 2.66600 & -0.0687836 & -1.83197\\
0.964112 & 3.51731 & $-7.57518\times10^{-3}$ & -0.652796\\
0.978111 & 6.16196 & $2.88863\times10^{-6}$ &
$-7.18147\times10^{-5}$ \\
\end{tabular}
\end{ruledtabular}
\caption{
For regular overcharged tension
stars, $m^2<q^2$, with $\frac{q^2}{R^2}=0.6$,
in columns one and two, several $\frac{r_0}{R}$ are given along with
their own $\gamma_{\rm cr}$. In columns three and four, the
eigenfrequencies $\omega_{0}^{2}$ and $\omega_{1}^{2}$ are given for
$\gamma=4$. There are no stable solutions for $\gamma=4$ and indeed
for any positive finite $\gamma$, the eigenfrequencies have a tower of
negative values.
}
\label{tab:overcharge1appendixB}
\vskip -0.7cm
\end{table}
The solutions for these overcharged stars have
radii extending from approximately $\frac{r_0}{R}= 0.880112$ to
approximately $\frac{r_0}{R}= 0.978113$. The endpoints
for the radius $\frac{r_0}{R}$ shown in the first column of the table
represent the minimum and the maximum values where the numerical
methods are in agreement to six decimal places.  In the second column
of the table the $\gamma_{\rm cr}$ corresponding to the given
$\frac{r_0}{R}$ radius of the star is shown, where $\gamma_{\rm cr}$ is
the $\gamma$ for which $\omega_0^2$ is zero.  The third and fourth
columns give the fundamental frequency squared $\omega_0^2$ and the
first excited frequency squared $\omega_1^2$ for each $\frac{r_0}{R}$
considering that the $\gamma$ of the fluid has the value $\gamma=4$.
The two eigenvalues start to be negative for small $\frac{r_0}{R}$ and 
become less negative for larger
$\frac{r_0}{R}$. Indeed, $\omega_0^2$
is negative in the approximate range $0.880112\leq\frac{r_0}{R}
\leq 0.964112$ where $\gamma>\gamma_{\rm cr}$, and turns
up positive for
approximately $0.964112\leq \frac{r_0}{R}\leq 0.978111$ with
$\gamma<\gamma_{\rm cr}$. But $\omega_1^2$ still remains negative. This
behavior suggests that all overcharged tension stars, configurations
belonging to the region (b) between the lines $C_0$ and $C_1$ of
Fig.~\ref{fig:regions}, are dynamical unstable against small radial
perturbation for positive $\gamma$, unless $\gamma=\infty$.
As explained in the main text this is expected on physical
grounds for stars that are held up by tension rather than
pressure.
In Table~\ref{tab:overcharge2appendixB} for stars with
$\frac{q^2}{R^2}=0.6$ the two first columns for $\frac{r_0}{R}$ and
$\gamma_{\rm cr}$ are also given in Table~\ref{tab:overcharge2}.  The
values of the critical adiabatic index $\gamma_{\rm cr}$ only depend
on the pair $\frac{q^2}{R^2}$ and $\frac{r_0}{R}$, and $\gamma_{\rm
cr}$ decreases in negative value when $\frac{r_0}{R}$ grows.
\begin{table} [h]
\begin{ruledtabular}
\begin{tabular}{c c c c c}
\textrm{$\frac{r_0}{R}$}& 
\textrm{$\gamma_{\rm cr}$}&
\textrm{$\omega_{0}^{2}$}&
\textrm{$\omega_{1}^{2}$}\\
\colrule
0.880113 & -0.125874 & -0.112415 & 0.107096\\
0.894113 & -0.113132 & -0.0799539 & 0.118102\\
0.908113 & -0.0952036 & -0.0446134 & 0.119851\\
0.922113 & -0.0779790 & -0.0179504 & 0.106768\\
0.936112 & -0.0623795 & -$1.69590\times10^{-3}$ & 0.0827208\\
0.950112 & -0.0483227 & $5.03599\times10^{-3}$ & 0.0529323\\
0.964112 & -0.0352687 & $4.47316\times10^{-3}$ & 0.0231310\\
0.978111 & -0.0220528 & $9.30989\times10^{-7}$ &
$3.26101\times10^{-6}$ \\
\end{tabular}
\end{ruledtabular}
\caption{
For regular overcharged tension stars, $m^2<q^2$, with
$\frac{q^2}{R^2}=0.6$, in columns one and two, several $\frac{r_0}{R}$
are given along with their own $\gamma_{\rm cr}$,  which
has negative values. In columns three and
four, the eigenfrequencies $\omega_{0}^{2}$ and $\omega_{1}^{2}$ are
given for $\gamma=-0.06$. The transition from instability to
stability of these stars goes when $\omega_{0}^{2}$ goes from negative
to positive.}
\label{tab:overcharge2appendixB}
%\vskip -0.6cm
\end{table}
Columns
third and fourth are new and give the fundamental frequency squared
$\omega_{0}^{2}$ and the first overtone frequency squared
$\omega_{1}^{2}$, respectively, for a matter fluid with
$\gamma=-0.06$.  The fundamental frequency squared $\omega_{0}^{2}$
has negative values for small $\frac{r_0}{R}$ where
$|\gamma|<|\gamma_{\rm cr}|$, and it has positive values for large
$\frac{r_0}{R}$ where $|\gamma|>|\gamma_{\rm cr}|$.  For large
relative radii, $\frac{r_0}{R}$, i.e., for stars that are almost
electrically charged dust stars with very small tension, and so they
are near the curve $ C_0 $, it is clear from the table that both $
\omega_0^{2} $ and $ \omega_1^{2} $ take values close to zero, as it
should be from our discussion in the main text.  One clearly sees from
the table that there is a change from instability to stability when
$\frac{r_0}{R}$ goes from approximately
$\frac{r_0}{R}=0.936112$ to approximately
$\frac{r_0}{R}=0.950112$, which means that stars with more mass and
less electric charge, and so less tension, become stable
against
radial perturbations.

%\vskip 0.3cm

To complete the text on regular black holes with negative energy
densities, see Sec.~\ref{rbhsned}, we present the
Table~\ref{tab:regularbhneappendixB} for regular black holes with
\begin{table}[h]
\begin{ruledtabular}
\begin{tabular}{c c c c}
\textrm{$\frac{r_0}{R}$}& 
\textrm{$\gamma_{\rm cr}$}&
\textrm{$\omega_{0}^{2}$}&
\textrm{$\omega_{1}^{2}$}\\
\colrule
 0.0186989 & 2.59470 & $8.6867\times10^{9}$ &
 $3.31562\times10^{10}$ \\
 0.139572 & 1.30905 & 206173 & 696030 \\
 0.260445 & 0.916685 & 9970.58 & 32291.6\\
 0.381318 & 0.766649 & 1569.94 & 5123.63 \\
 0.502191 & 0.701179 & 378.774 & 1268.46 \\
 0.623064 & 0.670727 & 106.670 & 364.258 \\
 0.743937 & 0.659659 & 25.9222 & 87.7674 \\
 0.864810 & 0.666663 & $6.86986\times10^{-3}$ &
 $2.18738\times10^{-2}$ \\
\end{tabular}
\end{ruledtabular}
\caption{
For regular black holes with negative energy densities
with
$\frac{q^2}{R^2}=\frac{27}{16}=1.6875$,
in columns one and two, several $\frac{r_0}{R}$
are given along with their own $\gamma_{\rm cr}$. In columns three and
four, the eigenfrequencies $\omega_{0}^{2}$ and $\omega_{1}^{2}$ are
given for $\gamma=4$. All systems are stable
against
radial perturbations, there are no
negative frequency squares. The systems with
$\frac{r_0}{R}=0.864810$, approximately,
are neutrally stable, indeed they are systems
with mass equal to charge
or almost.}
\label{tab:regularbhneappendixB}
\vskip -0.3cm
\end{table}
$\frac{q^2}{R^2}=\frac{27}{16}=1.6875$, which completes
Table~\ref{tab:regularbhne}.  The two first columns of
Table~\ref{tab:regularbhneappendixB} are also given in
Table~\ref{tab:regularbhne}, whereas columns third and fourth are new
and give the fundamental frequency squared $\omega_{0}^{2}$ and the
first overtone frequency squared $\omega_{1}^{2}$ for a matter fluid
with $\gamma=4$. The values of the critical adiabatic index
$\gamma_{\rm cr}$ are obtained respectively from the pseudospectral and
the shooting methods, and are in agreement to each other up to six
decimal places.
The radii of these
regular black holes extend from just above
$\frac{r_0}{R}=0$ to approximately
$\frac{r_0}{R}=0.866025$.
The critical adiabatic index $\gamma_{\rm cr}$ decreases
with growing $\frac{r_0}{R}$, and for $\frac{r_0}{R}$
larger, $\gamma_{\rm cr}$ rises again, the heuristic physical
reason for this behavior is not clear.
All eigenfrequencies are positive,
i.e., $\omega_{0}^{2}$ and
$\omega_{1}^{2}$ are positive, 
and so all these regular black holes
with $\gamma=4$ are stable
against
radial perturbations.
 In addition, when $\frac{r_0}{R}$
is large enough to yield an object with mass equal to charge, i.e.,
when it is near $C_2$, the two frequencies become zero and the system
is neutrally stable. This latter behavior
possibly holds for any $\gamma$.

%\vskip 1.0cm

To complete the text on regular black holes with
a phantom matter core,
see Sec.~\ref{regbhspmc}, we present
the Table~\ref{tab:RBHPMappendixB}
which gives 
some more detail for such regular black holes.
Regular black holes with a central core made by a charged fluid of
phantom matter are configurations whose parameters belong to the
regions (d2) and (e1) above the curve $C_{31}$
plus $C_{31}C_{32}$,
and below the line
$C_{33}$ of Figs.~\ref{fig:regions}
and \ref{fig:regionszoom}, in the region (d2), the energy density is
positive and finite at the center of the distribution of matter and
changes to negative values close to the surface,
and in the region (e1), the energy
density is positive everywhere inside matter and the pressure is
negative.
\begin{table}[h]
\begin{ruledtabular}
\begin{tabular}{c c c c c c c}
\textrm{$\frac{r_0}{R}$}& 
\textrm{$\frac{q^2}{R^2}$}&
\textrm{$\gamma$}&
\textrm{$\omega_{0}^{2(+)}$}&
\textrm{$\omega_{0}^{2(-)}$}&
\textrm{$\omega_{1}^{2(+)}$}&
\textrm{$\omega_{1}^{2(-)}$}\\
\colrule
0.910& 1.8 & 0.60 &  6.11397 & -3.89608 & 27.9524 & -6.23978\\
0.940 & 2.0 & 0.65 & 1.51244 & -7.81232 & 29.9170 & -96.1171\\
0.970 & 2.5 & 0.55 & 2.95108 & -44.2682 & 42.4200 & -516.103\\
\end{tabular}
\end{ruledtabular}
\vskip 0.5cm
\begin{ruledtabular}
\begin{tabular}{c c c c c c c}
\textrm{$\frac{r_0}{R}$}& 
\textrm{$\frac{q^2}{R^2}$}&
\textrm{$\gamma$}&
\textrm{$\omega_{0}^{2(+)}$}&
\textrm{$\omega_{0}^{2(-)}$}&
\textrm{$\omega_{1}^{2(+)}$}&
\textrm{$\omega_{1}^{2(-)}$}\\
\colrule
0.910 & 1.8 & -0.50 & 11.2388 & -15.0966 & 65.4825  & -36.5027\\
0.940 & 2.0 & -0.45 & 13.3861 & -25.7676 & 75.5721 & -57.0211\\
0.970 & 2.5 & -0.30 & 42.9441 & -19.185 & 300.872  & -43.2299\\
\end{tabular}
\end{ruledtabular}
\caption{The top table shows for some values of $\frac{r_0}{R}$,
$\frac{q^2}{R^2}$, and $\gamma$ positive, that $\omega_{0}^{2}$ and
$\omega_{1}^{2}$ are degenerated, i.e., there is one positive and one
negative corresponding eigenvalue.  The bottom table shows for the
same values of $\frac{r_0}{R}$, $\frac{q^2}{R^2}$, and $\gamma$
negative, that $\omega_{0}^{2}$ and $\omega_{1}^{2}$ are degenerated,
i.e., there is one positive and one negative corresponding
eigenvalue.}
\label{tab:RBHPMappendixB}
\vskip -0.3cm
\end{table}
In both regions, the pressure is larger in absolute value
than the energy density at the center of the distribution, and
it goes to zero at the surface $r_0$. Thus, for a finite region inside
the matter one finds $\rho+p<0$. As a consequence, the coefficient
$W(r)$ in the SL problem is a negative function in
$0\leq r\leq r_d$ for some $r_d<r_0$, and it is positive in
$r_d\leq r\leq r_0$, while the coefficient
$F(r)$ is a negative function on the whole interval
$0\leq r\leq r_0$
if
$\gamma$ is a positive number, or conversely the coefficient $F(r)$ is
a positive function on the whole interval
$0\leq r\leq r_0$ if $\gamma$ is a
negative number.  As pointed out in Appendix~\ref{sec:SLP}, case (D),
in such a case the behavior of the eigenvalues of the SL problem is as
follows.
There are two simple ground states $\omega_{0}^{2(+)}>0$ and
$\omega_{0}^{2(-)}<0$, where the frequencies of excited states
accumulate at, from above and from below, respectively, i.e., there
are exactly one positive eigenvalue $\omega_{n}^{2(+)}$ and one
negative eigenvalue $\omega_{n}^{2(-)}$ associated to which there
exist respectively two branches of eigenvalues each one within the
corresponding intervals $(\omega_{0}^{2(+)},\,\infty)$ and $(-\infty,
\,\omega_{0}^{2(-)})$, i.e.,
the eigenvalues of the SL problem for regular black holes with a
phantom matter central core belonging to regions (d2) and (e1) are
unbounded from below and above.
Thus, in these regions the solutions are unstable due to the double valued,
negative and positive, of the squared frequencies for the same adiabatic
index. In the table,  these
results are presented.
For some values of $\frac{r_0}{R}$, $\frac{q^2}{R^2}$, and $\gamma$
positive and negative, it is shown that $\omega_{0}^{2}$ and $\omega_{1}^{2}$
are degenerated, i.e., there exist one positive and one negative corresponding
eigenvalues.

%\vskip 1.0cm

\begin{figure}[h] 
\centering
\includegraphics[width=0.235\textwidth]
{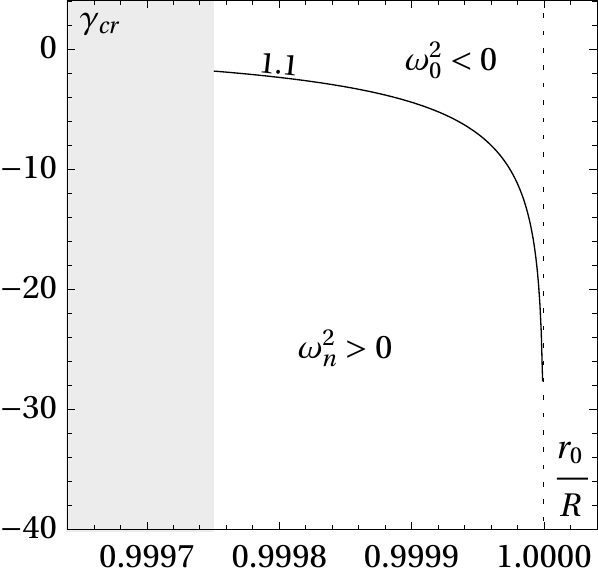}
\includegraphics[width=0.235\textwidth]
{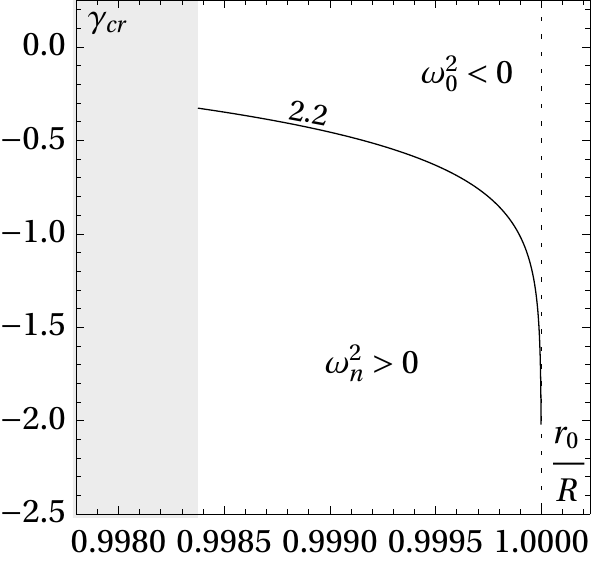}
\centering
\includegraphics[width=0.235\textwidth]
{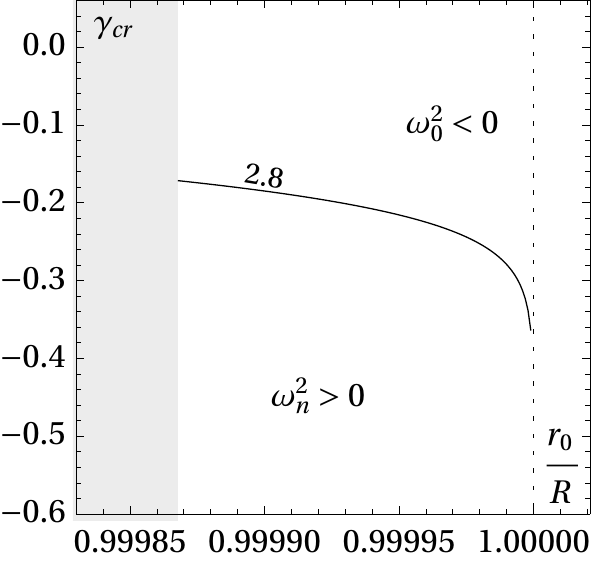}
\includegraphics[width=0.235\textwidth]
{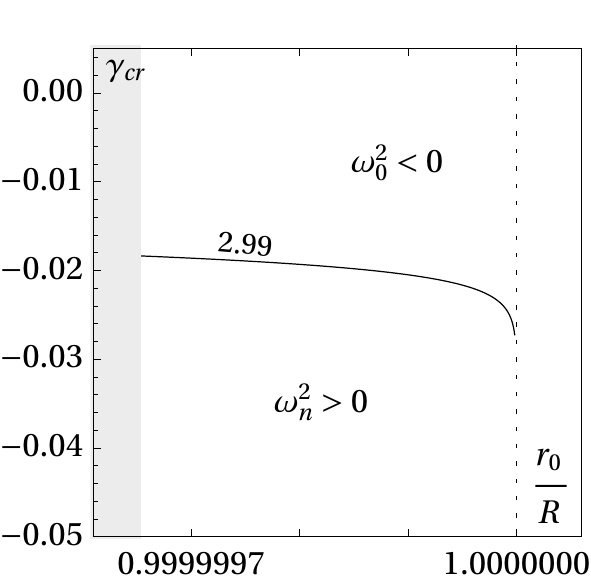}
\caption{
adiabatic index $\gamma_{\rm cr}$ as a function of the radius for four
values of the electric charge, namely, $\frac{q^2}{R^2}=1.1$,
$\frac{q^2}{R^2}=2.2$, $\frac{q^2}{R^2}=2.8$, and
$\frac{q^2}{R^2}=2.99$, as indicated in the figure.
Stability of regular tension black holes with positive enthalpy
density. These regular black holes belong to region (e2), above the
curve $C_{33}$ of Fig.~\ref{fig:regionszoom} which is an enlargement of
Fig.~\ref{fig:regions}.  The critical adiabatic index $\gamma_{\rm cr}$
for four values of the electric charge parameter
$\frac{q^2}{R^2}=1.1$, $\frac{q^2}{R^2}=2.2$, $\frac{q^2}{R^2}=2.8$,
and $\frac{q^2}{R^2}=2.99$, is shown as a function of the radius
$\frac{r_0}{R}$.  In each of the four plots, the line starts at a
minimum radius $\frac{r_0}{R}$ on the curve $C_{33}$  for which
$\gamma_{\rm cr}$ is negative and for larger $\frac{r_0}{R}$,
$\gamma_{\rm cr}$
becomes more negative up to the line $\frac{r_0}{R}=1$.  The light
gray region on the left  side of each plot corresponds to objects
that are not regular tension black holes with positive enthalpy
density configurations. The difference of this figure to  
Fig.~\ref{fig:RTBH},
is that here 
the range of $\frac{r_0}{R}$ is widened
in each plot.}
\label{fig:RTBHappendix}
\end{figure}

To complete the text on regular black holes with positive enthalpy
density, see Sec.~\ref{rbhpositiveh}, we present some more detail for
such regular black holes.  In Fig.~\ref{fig:RTBHappendix} 
we show the
numerical results for the critical adiabatic index $\gamma_{\rm cr}$
as a function of the radius for four values of the electric charge,
namely, $\frac{q^2}{R^2}=1.1$, $\frac{q^2}{R^2}=2.2$,
$\frac{q^2}{R^2}=2.8$, and $\frac{q^2}{R^2}=2.99$, as indicated in the
figure.  The difference of this figure to Fig.~\ref{fig:RTBH}, is that
here in Fig.~\ref{fig:RTBHappendix} the range of $\frac{r_0}{R}$ is
widened in each plot to get a good portion of the plotted line,
whereas in Fig.~\ref{fig:RTBH} the range of $\frac{r_0}{R}$ is fixed
so that one sees clearly by a comparison between the four plots
themselves the range of the validity of the solutions in the axis
$\frac{r_0}{R}$.

\centerline{}

\end{document}